\documentclass[twocolumn]{jpsj2} 

\def\Tc{T_\mathrm{c}}

\def\runauthor{Kiwamu {\sc Kudo} and Kosaku {\sc Yamada} 
}

\title{Reduction of $\Tc$ due to Impurities in Cuprate Superconductors}

\author{
Kiwamu \textsc{Kudo}
\thanks{E-mail:kiwamu.kudo@toshiba.co.jp}
and Kosaku \textsc{Yamada}
}

\inst{
Department of Physics, Kyoto University, Sakyo-ku, Kyoto 606-8502
}

\recdate{\today}

\abst{
In order to explain how impurities affect the unconventional 
superconductivity, 
we study non-magnetic impurity effect on the transition temperature using 
on-site $U$ Hubbard model within 
a fluctuation exchange (FLEX) approximation. 
We find that in appearance, the reduction of $\Tc$ roughly coincides with 
the well-known Abrikosov-Gor'kov formula. 
This coincidence results from the cancellation between two effects; 
one is the reduction of attractive force due to randomness, 
and another is the reduction of the damping rate of quasi-particle 
arising from electron interaction. 
As another problem, we also study impurity effect on underdoped cuprate 
as the system showing pseudogap phenomena. 
To the aim, we adopt the pairing scenario for the pseudogap and discuss how 
pseudogap phenomena affect the reduction of $\Tc$ by impurities. 
We find that 'pseudogap breaking' by impurities plays the essential role 
in underdoped cuprate and suppresses the $\Tc$ reduction due to the 
superconducting (SC) fluctuation.
}

\kword{
high-$\Tc$ cuprates; impurity effect; fluctuation exchange; 
superconducting fluctuation; 
pseudogap
}

\begin{document}
\sloppy
\maketitle

\section{Introduction} 

Until now many experiments on the effect of impurity substitution 
in hole-doped cuprate high-$\Tc$ superconductors (HTS cuprates) have been done, and it is well known that the substitution in the copper site leads to a rapid 
$\Tc$ reduction. In their experiments, the following Abrikosov-Gor'kov (AG) 
formula~\cite{Ab-G} is usually used to analyse the dependence of 
$\Tc$ on impurity content :
\begin{equation}
\ln \left(\frac{T_c}{T_{c0}} \right)
=\psi \left(\frac{1}{2} \right)
-\psi \left(\frac{1}{2}+\frac{\Gamma}{2\pi T_c} \right),
\label{ag}
\end{equation}
where $\psi(x)$ is the digamma function, $T_\mathrm{c0}=\Tc(n_\mathrm{imp}=0)$, and $\Gamma$ is the pair breaking scattering rate. 
In most of experiments which study impurity effect in the HTS, 
fitting the measured data  by Eq.(\ref{ag}) seems to be successful, 
especially in the low impurity concentration regime 
because of its linear bahavior. For example, it is reported theoretically and 
experimentally that the $\Tc$ reduction by 
Zn substitution in HTS follows quantitatively Eq.(\ref{ag}), 
where Zn is treated as a strong non-magnetic scatterer, 
and $\Gamma$ is evaluated in the unitarity limit.~\cite{Hotta,Ishida,SunMaki}

However, it is not so obvious that AG formula can be used to estimate 
impurity effects in HTS cuprates. 
The simplicity of Eq.(\ref{ag}) is based on the treatment of HTS as the 
weak-coupling $d$-wave BCS superconductor and the calculation technique of 
taking an average on Fermi surface. 
In such a treatment, the microscopic mechanism of superconducting pairing is 
not included at all, so the pairing interaction is not affected by impurities, 
whereas HTS is the typical system in which the pairing interaction is caused
by Coulomb repulsion $U$ and the momentum dependence plays the essential 
role in the mechanism of superconductivity. Thus, the effect of impurity on
the pairing interaction may not be negligible in HTS. 
Therefore, it is necessary to study how impurities suppresses 
$\Tc$ from the microscpic view. 
To clarify the reason why Eq.(\ref{ag}) represents $\Tc$ reduction 
in HTS cuprate is one of the purpose of this paper and discussed in $\S3$. 

As another problem, it is frequently debated that the impurity effect on 
underdoped cuprate is different from that on overdoped cuprate. 
In the experiments on the $\Tc$-reduction by copper site 
substitution,~\cite{Kluge,Tallon} 
it has been noted that in underdoped regime there is 
more rapid reduction of $\Tc$ than in the overdoped regime. 
In the other kind of experiments on electron irradiated samples, 
it is reported that the variation of $\Tc$ with defect content 
in underdoped and optimally doped 
$\mathrm{Y}\mathrm{Ba}_\mathrm{2}\mathrm{Cu}_\mathrm{3}
\mathrm{O}_\mathrm{7-\delta}$ 
is quite linear and contrasts with the AG formula. 
Because one of the feature of underdoped cuprate in the clean samples 
is the pseudogap phenomena, it is natural to consider that such remarkable 
behaviors in underdoped region originate from the pseudogap. 
Therefore, we disscuss how pseudogap phenomena affect the reduction of 
$\Tc$ by impurities. 
This is the second purpose of this paper and is discussed in $\S4$. 

Through this study, we treat impurities as non-magnetic 
ones in the mind of the Zn-substitution and defects; We assume that 
the scattering is elastic and that the impurity potential is the form of 
$\delta$-function type and we take an average on the positions of impurities. 
Moreover, we take the unitarity limit.

This paper is constructed in the following way. 
In $\S2$, we consider the impurity effect in the weak-coupling $d$-wave 
BCS model. We point out that $\Tc$ reduction in the model almost coincides 
with AG formula and that the slight deviation from the AG curve 
originates from the direct momentum summation done numerically. 
In $\S3$, using the FLEX approximation, we discuss $\Tc$ reduction 
by impurities in the system where the pairing interaction is caused by $U$. 
First, we introduce the FLEX theory and our theorical treatment for 
impurities. 
In $\S3.2$, we show how impurities affect 
the main factors to determine the value of $\Tc$. 
In $\S3.3$, we discuss the coincidence between the obtained $\Tc$ reduction 
and AG formula and we explain the reason. 
In $\S3.4$, we point out the necessity of our treatment for the impurity 
effect in the HTS. 
In $\S4$, using 'pairing scenario', we discuss how pseudogap phenomena affect 
$\Tc$ reduction due to impurities 
in the mind of the disordered underdoped cuprate. 
We first review the theorical treatment in the pairing scenario. 
In $\S4.2$, we include impurity effect in the theory and 
show some features of $\Tc$ reduction in the presence of pseudogap. 
We point out that such features are caused by 'pseudogap breaking'. 
In $\S4.3$, we explain what the 'pseudogap breaking' is. 
Finally, we shortly comment on the quantum dynamics of SC fluctuation. 
In the last section $\S5$, we conclude the whole discussion.

\section{Reduction of $\Tc$ by impurities in the weak-coupling 
$d$-wave BCS model}

\begin{figure}[t]
  \begin{center}
    \begin{tabular}{cc}
      \resizebox{40mm}{!}{\includegraphics{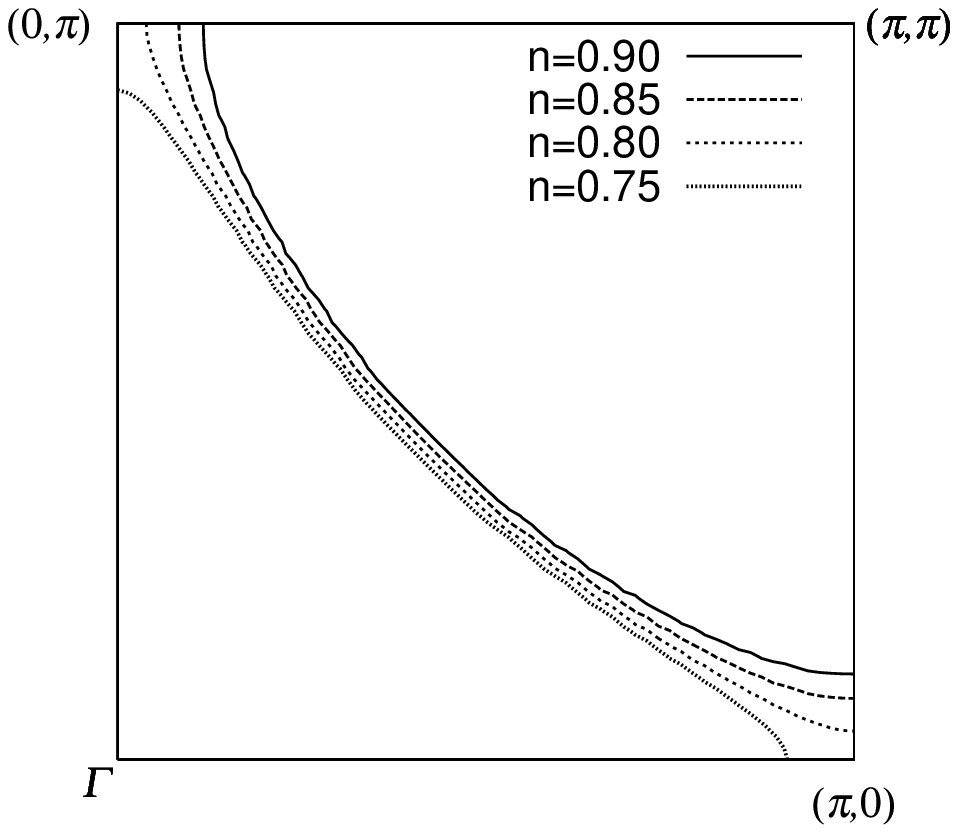}} &
      \resizebox{40mm}{!}{\includegraphics{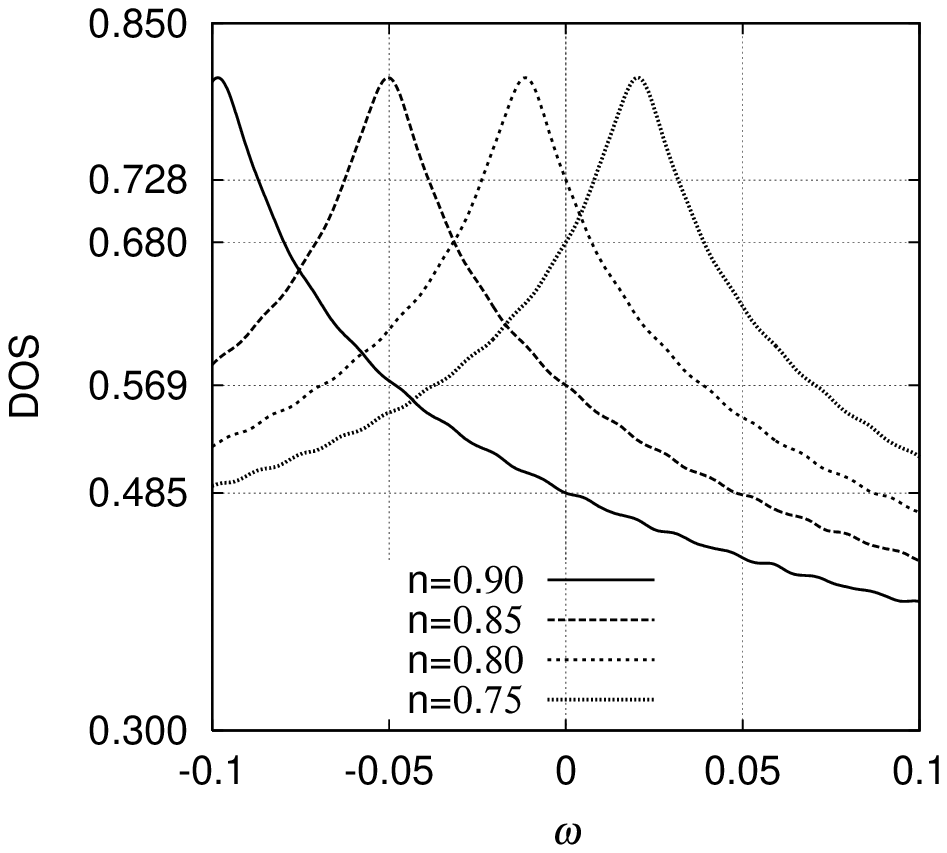}} 
    \end{tabular}
    \caption{Fermi surfaces of bare electrons with $\epsilon_\mathbf{k}$ and 
             DOS. The parameters are chosen as $t'=0.25t$, $g=0.7t$, 
             $T=0.015t$ and $n=0.75$, $0.80$, $0.85$ and $0.90$. 
            }
    \label{BCS}
  \end{center}
\end{figure}

In this section, we consider non-magnetic impurity effect on $\Tc$ reduction 
in the unitarity limit within the weak-coupling $d$-wave BCS model. 
We assume the following separable attractive interaction:
\begin{equation}
V_\mathbf{k,k'}=-g\varphi_\mathbf{k} \varphi_\mathbf{k'}, 
\label{attractive}
\end{equation}
where $g>0$ is the strength of the interaction and 
$\varphi_\mathbf{k}=\cos k_x -\cos k_y$, 
the form factor of $d_\mathrm{x^2-y^2}$-wave. 
The transition temperature is determined by 
\begin{equation}
1=-\frac{1}{\beta N}\sum_{\mathbf{k},n}V_{\mathbf{k},\mathbf{k}}
G(\mathbf{k},i\omega_n)G(-\mathbf{k},-i\omega_n),
\label{bcsTc}
\end{equation}
where 
\begin{equation}
G(\mathbf{k},i\omega_n)
=[i\omega_n-(\epsilon_\mathbf{k}-\mu)
-\Sigma_\mathrm{imp}(i\omega_n)]^{-1}
\end{equation}
with 
\begin{equation}
\Sigma_\mathrm{imp}(i\omega_n)=
-\frac{n_\mathrm{imp}}
{\frac{1}{N}\sum_\mathbf{k}G(\mathbf{k},i\omega_n)}.
\label{unisca}
\end{equation}
We numerically solve these equations. 
Here $\mu$ is the chemical potential, $\omega_n$ is the fermionic 
Matubara frequency, 
$\omega_n=(2n+1)\pi T$, 
and $n_\mathrm{imp}$ is the impurity concentration. 
We consider the 2D square lattice and assume the following 
tight-binding dispersion
\begin{equation}
\epsilon_{\mathbf{k}}=-2t(\cos k_x + \cos k_y)+4t' \cos k_x \cos k_y.
\label{dispersion}
\end{equation}
The filling of electrons is represented by $n$, and $n=1$ corresponds to 
the half filling. The parameters are chosen as $2t=1$, $t'/t=0.25$, 
$g=0.7t$ and $n=0.75 \sim 0.90$. 
The Fermi surfaces of bare electrons with dispersion Eq. (\ref{dispersion}) 
and the density of state (DOS), 
$\rho(w)=-\frac{1}{\pi}\sum_\mathbf{k}\mathrm{Im}G(\mathbf{k},\omega)$, 
are shown in Fig. \ref{BCS}. Here, $\omega=0$ is the Fermi level. 

When we assume the particle-hole symmetry 
and replace the momentum summation with the Fermi surface average 
in calculating Eqs.(\ref{bcsTc})-(\ref{unisca}), 
we obtain the well-known AG formula, Eq.(\ref{ag}), 
with the pair-breaking parameter 
$\Gamma=\frac{n_\mathrm{imp}}{\pi \rho(0)}$.~\cite{Hotta,SunMaki} 
The critical impurity concentration $n_\mathrm{imp}^\mathrm{c}$ 
is given by 
$n_\mathrm{imp}^\mathrm{c}=\frac{\pi^2 \rho(0) T_\mathrm{c0}}
{2 \gamma'}$, where $\gamma'=1.781072$. 
In the limiting case of low impurity concetration, we have from 
Eq.(\ref{ag}), 
\begin{equation}
\frac{\Tc}{T_\mathrm{c0}}
=1-\frac{\pi}{4}\frac{\Gamma}{n_\mathrm{imp} T_\mathrm{c0}}n_\mathrm{imp}.
\label{lin}
\end{equation}

The numerically calculated results of $\Tc$ reduction 
and the fitted curves by Eq.(\ref{ag}) 
are shown in Fig. \ref{agfit_weak}. 
The fitted AG curves are obtained in the following way. 
In low impurity concetration regime, the calculated reduction 
of $\Tc$ as the function of $n_\mathrm{imp}$ is almost linear. 
Thus, we can estimate the coefficient of Eq.(\ref{lin}), 
$\frac{\pi}{4}\frac{\Gamma}{n_\mathrm{imp} T_\mathrm{c0}}$, 
which is independent of $n_\mathrm{imp}$, 
obtain the value of $\Gamma$ and plot AG curves in each chart of 
Fig. \ref{agfit_weak}. 
From the figure, we can see that $\Tc=T_\mathrm{c}(n_\mathrm{imp})$ 
obtained as the solution of Eq.(\ref{bcsTc}) almost 
coincides with the AG curve. 
However, there is slight deviation of $n_\mathrm{imp}^\mathrm{c}$ from 
the value expected by AG formula. 
In the calculation of Eqs.(\ref{bcsTc})-(\ref{unisca}), 
we have done directly the momentum summation not taking the average on 
the Fermi surface. 
It results in the difference of the estimation of DOS and 
the difference leads to the deviation of $n_\mathrm{imp}^\mathrm{c}$. 
In fact, from Fig. \ref{BCS} and \ref{agfit_weak}, 
we can see that as DOS increases at Fermi level, 
the deviation becomes large. 
Therefore, the deviation is the result of dispersion 
and filling of electrons. 

\begin{fullfigure}[t]
  \begin{center}
    \begin{tabular}{cc}
      \resizebox{60mm}{!}{\includegraphics{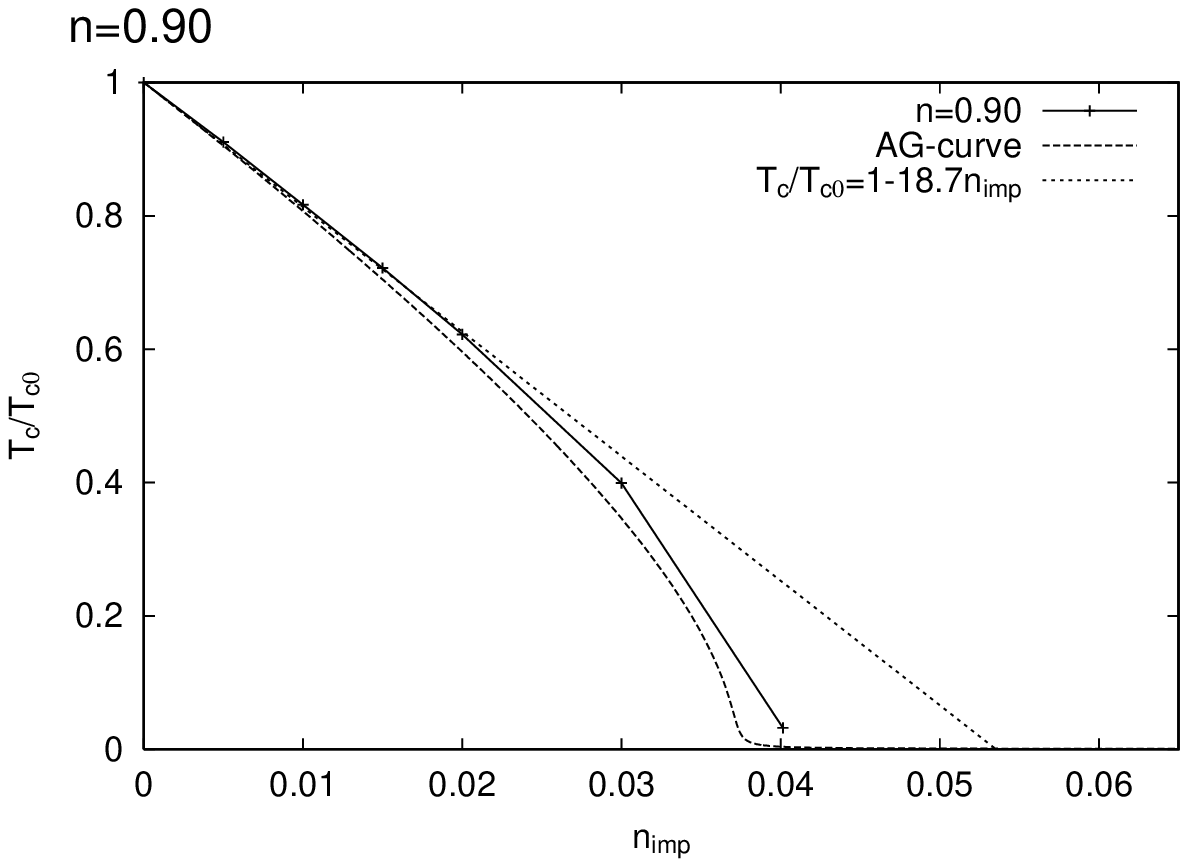}} &
      \resizebox{60mm}{!}{\includegraphics{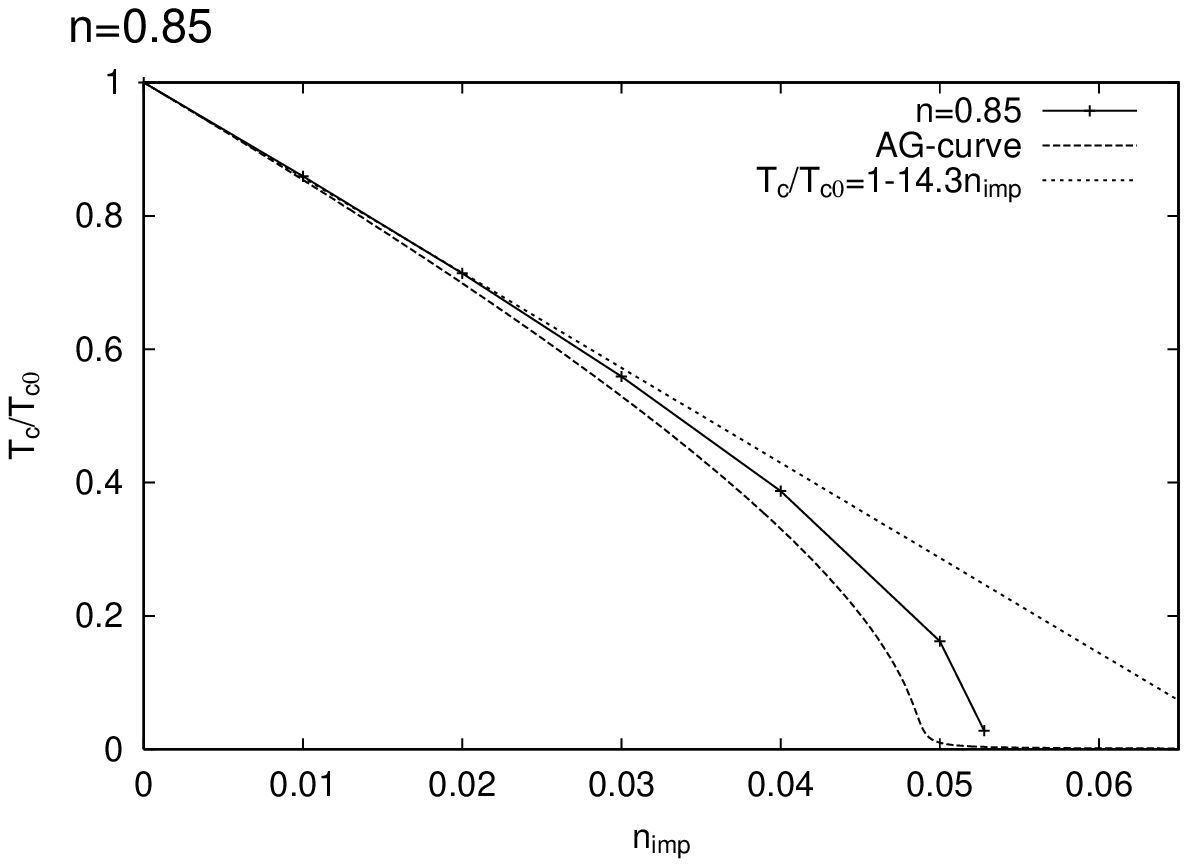}} \\
      \resizebox{60mm}{!}{\includegraphics{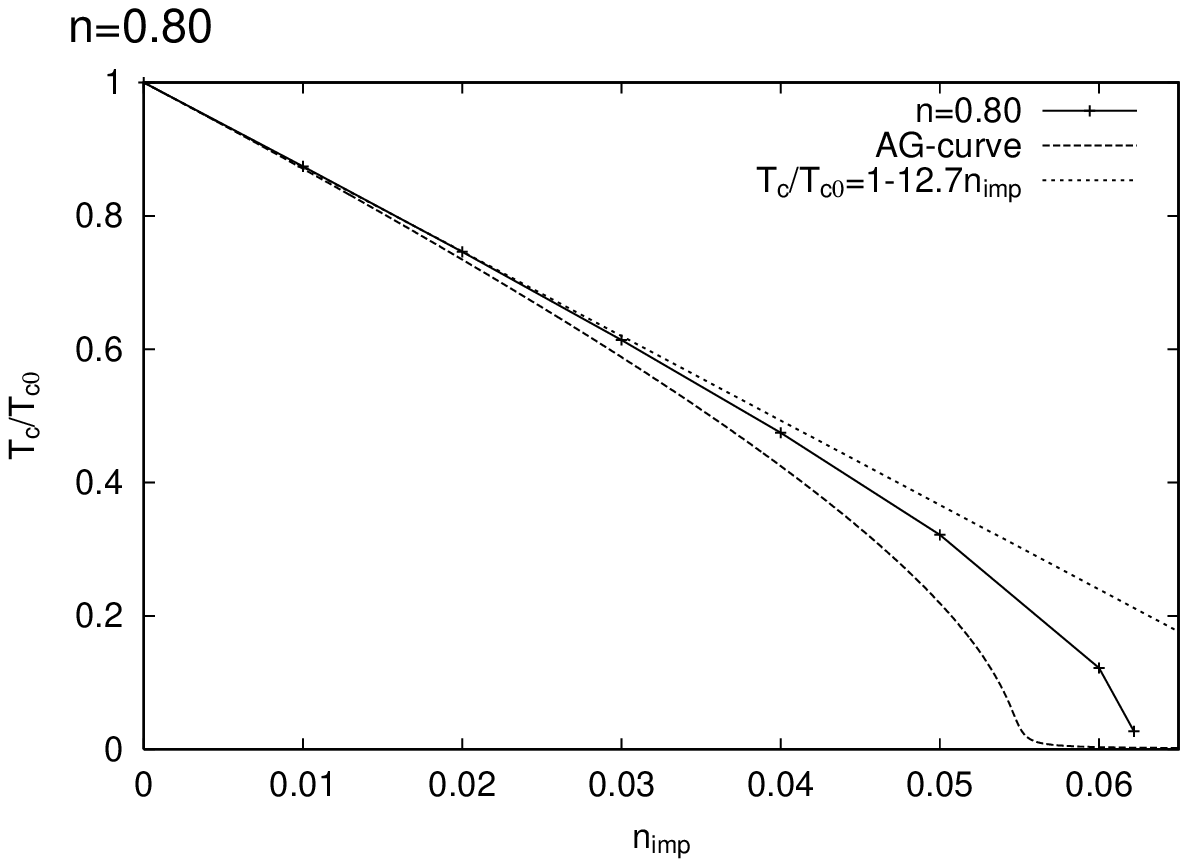}} &
      \resizebox{60mm}{!}{\includegraphics{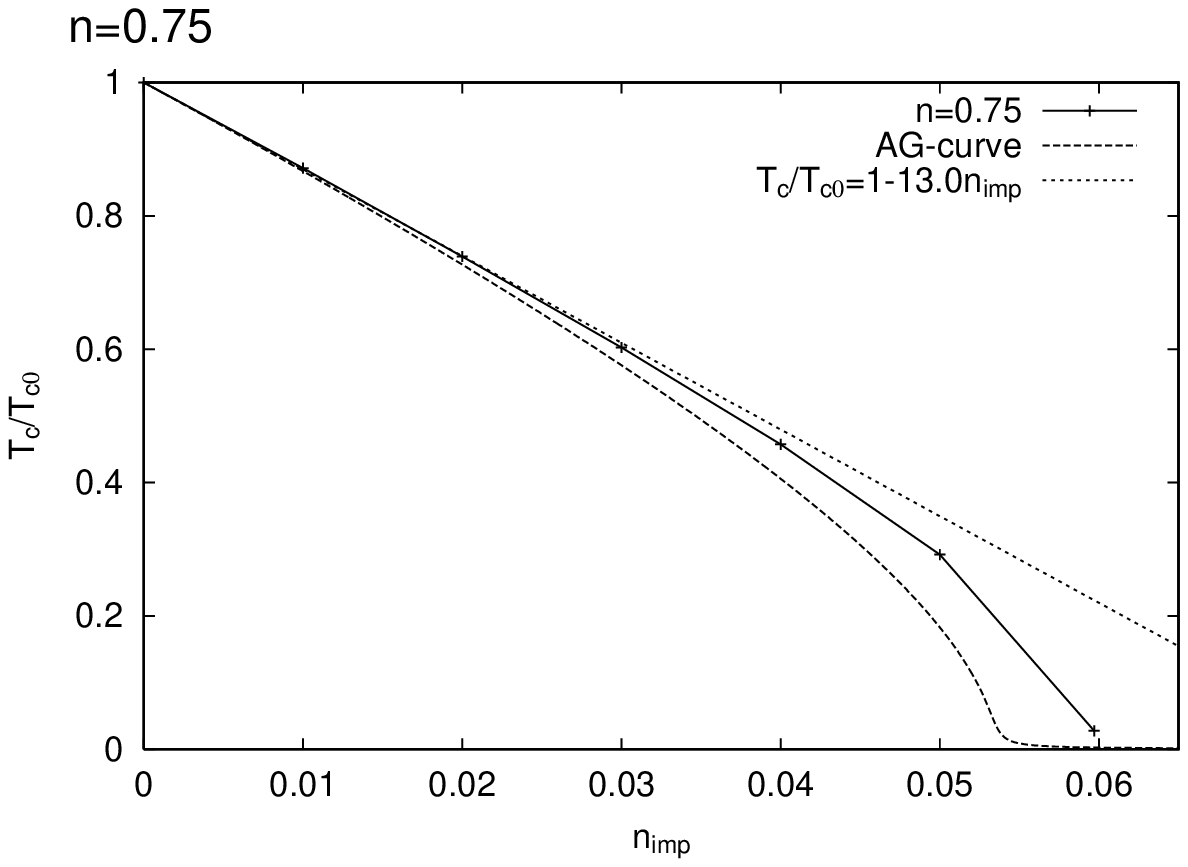}} 
    \end{tabular}
    \caption{$\Tc/T_\mathrm{c0}$ vs. $n_\mathrm{imp}$ and 
             the curves fitted by AG formula, Eq.(\ref{ag}). 
             The fitting curves are obtained in the way explained in the 
             text. 
            }
    \label{agfit_weak}
  \end{center}
\end{fullfigure}

\section{Reduction of $\Tc$ by impurities in the FLEX theory}

We start from on-site $U$ t-t' Hubbard Hamiltonian as the model of 
the HTS cuprates. 
In order to describe the superconductivity in the Hubbard model with 
the repulsive Coulomb interaction, it is necessary to derive the effective 
interaction from the many body effect. 
Here we adopt the FLEX approximation.~\cite{flex1,flex2,flex3,flex4,flex5} 
Earlier calculations using the FLEX approximation were rather successful 
in determining many physical quantities in hole-doped cuprates. 
Therefore we consider the reduction of $\Tc$ by impurities within 
the FLEX theory. 

\subsection{Model}

\begin{figure}[t]
\begin{center}
\includegraphics[width=5cm]{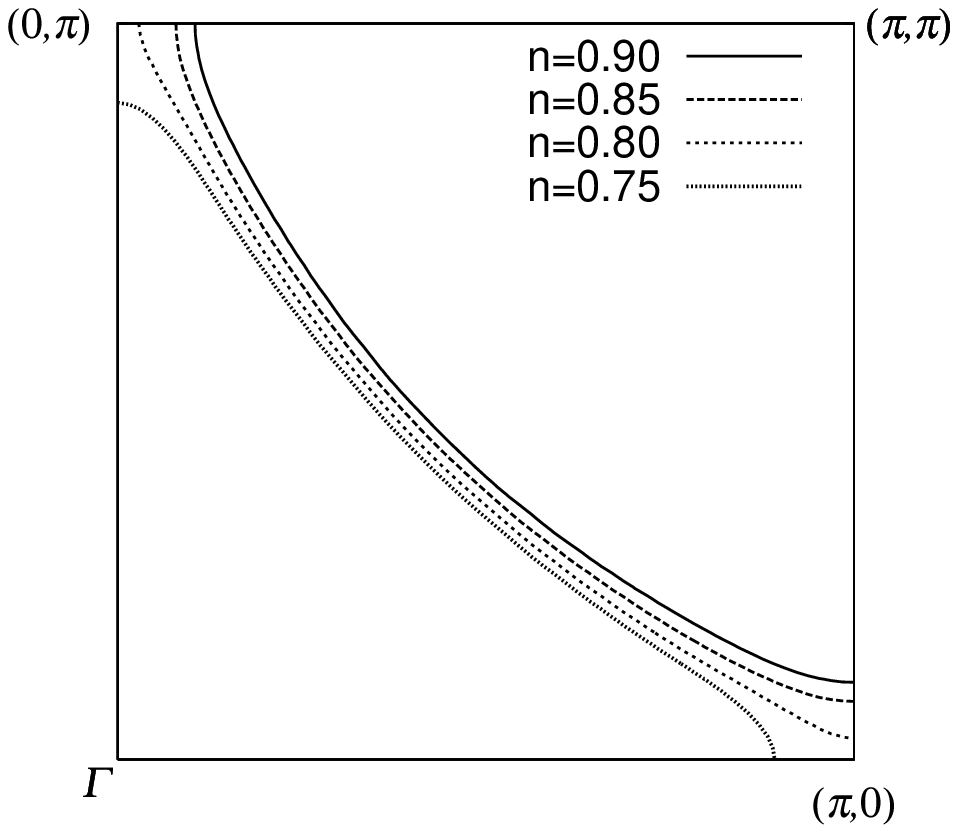}
\end{center}
\caption{Fermi surfaces calculated by the FLEX approximation without 
         impurities at T=0.015t. The change from the bare case is very slight. 
         It is recognized by comparing this figure with Fig. \ref{BCS}. 
         }
\label{fermi}
\end{figure}

Here we summarize the FLEX theory and explain our theorical treatment 
for impurities. 

We consider the t-t' Hubbard Hamiltonian 
on 2D square lattice. It is given by
\begin{equation}
H=\sum_{\mathbf{k} \sigma} (\epsilon_{\mathbf{k}} - \mu)
c_{\mathbf{k} \sigma}^{\dagger} c_{\mathbf{k} \sigma}
+\frac{U}{N}\sum_{\mathbf{k} \mathbf{k'} \mathbf{q}}
c_{\mathbf{k+q} \uparrow}^{\dagger} c_{\mathbf{k'-q} \downarrow}^{\dagger}
c_{\mathbf{k'} \downarrow} c_{\mathbf{k} \uparrow},
\label{Hamiltonian}
\end{equation}
where $c_{\mathbf{k} \sigma}^{\dagger}$ is the creation operator of an 
electron with momentum $\mathbf{k}$ and spin $\sigma$, $\mu$ is 
chemical potential, and $U$ is the on-site Coulomb repulsion. 
$\epsilon_\mathbf{k}$ is the same tight-binding dispersion 
as Eq.(\ref{dispersion}). 
The parameters are chosen as $2t=1$, $t'/t=0.25$ and $U/t=3$. 
These parameters are reasonable in the sense that
the calculated Fermi sufaces coincide with those determined by the angle 
resolved photoemission (ARPES) experiments.
Fig. \ref{fermi} shows the Fermi surfaces calculated by the FLEX approximation 
without impurity at T=0.015t. The spectrum at $(\pi,0)$ is above $\mu$ for 
$n<0.79$. Comparing Fig. \ref{fermi} with 
the ARPES data,~\cite{Uchida} we can see that 
the above set of parameters covers the optimal doped to overdoped 
region of the HTS cuprates, for example 
$\text{La}_\mathrm{2-x}\text{Sr}_\mathrm{x}\text{Cu}\text{O}_\mathrm{4}$.

In singlet pairing superconducting states, the one-particle Green's function 
is 2$\times$2 matrix in Nambu notaion, and expressed in terms of a 
self-energy matrix:
\begin{equation}
\hat{G}(k)=[\hat{G}^{-1}_0(k)-\hat{\Sigma}(k)]^{-1}.
\label{Dyson}
\end{equation}
where
\begin{equation}
\hat{G}^{-1}_0(k)=i\omega_n\hat{\tau}_0
-(\epsilon_{\mathbf{k}} - \mu)\hat{\tau}_3,
\end{equation}
and
\begin{equation}
\hat{\Sigma}(k)=i\omega_n(1-Z(k))\hat{\tau}_0+X(k)\hat{\tau}_3
+\phi(k)\hat{\tau}_1.
\end {equation}
Here $k$ is a shorthand notation as $k=(\mathbf{k},i\omega_n)$. 
$Z(k)$ is the renormalization parameter, 
$X(k)$ is the energy shift, $\phi(k)$ is the gap parameter, 
and $\hat{\tau}_i(i=1\sim3)$ and $\hat{\tau}_0$ are the Pauli matrices and 
the unit matrix, respectively.
$Z(k)$ and $X(k)$ are defined as follows:
\begin{equation}
Z(k)=1-\frac{1}{2i\omega_n}(\Sigma(k)-\Sigma(-k)),
\end{equation}
\begin{equation}
X(k)
=\frac{1}{2}(\Sigma(k)+\Sigma(-k)),
\end{equation}
where $\Sigma(k)$ is the normal self-energy. 
In these notations, the diagonal and off-diagonal one-particle Green's 
functions can be written as
\begin{equation}
G(k)=-D(k)(i\omega_n Z(k)+(\epsilon_{\mathbf{k}} - \mu)+X(k)),
\end{equation}
\begin{equation}
F(k)=-D(k)\phi(k),
\end{equation}
with
\begin{equation}
D(k)=[(\omega_nZ(k))^2+((\epsilon_{\mathbf{k}} - \mu)+X(k))^2
+\phi(k)^2]^{-1}.
\end{equation}

In the self-energy matrices, we simply neglect the mixed diagrams of 
Coulomb and impurity interaction:
\begin{equation}
\hat{\Sigma}(\mathbf{k},i\omega_n)
=\hat{\Sigma}_\mathrm{FLEX}(\mathbf{k},i\omega_n)
+\hat{\Sigma}_\mathrm{imp}(i\omega_n).
\label{self-energy}
\end {equation}

The self-energy matrix $\hat{\Sigma}_\mathrm{imp}$ is given by 
$\hat{\Sigma}_\mathrm{imp}(i\omega_n)=n_\mathrm{imp}\hat{t}(i\omega_n)$ 
where $n_\mathrm{imp}$ is impurity concentration and $\hat{t}$ is t-matrix. 
The t-matrix is the solution of the following self-consistent equation:
\begin{equation}
\hat{t}(i\omega_n)=\hat{u}
+\hat{u}\frac{1}{N}\sum_\mathbf{k}\hat{G}(\mathbf{k},i\omega_n)
\hat{t}(i\omega_n),
\end {equation}
where $\hat{u}=u\hat{\tau}_3$ and $u$ is the amplitude of impurity 
potential. In the unitarity limit, we obtain the 
diagonal and the off-diagonal components of 
$\hat{\Sigma}_\mathrm{imp}$ as follows:
\begin{equation}
\Sigma_{\mathrm{imp}}(i\omega_n)
=n_\mathrm{imp}\frac{g_{\mathrm{0}}(-i\omega_n)}
{g_{\mathrm{0}}(i\omega_n)g_{\mathrm{0}}(-i\omega_n)
+g_{\mathrm{1}}(i\omega_n)^2},
\label{imp1}
\end{equation}
\begin{equation}
\phi_{\mathrm{imp}}(i\omega_n)
=n_\mathrm{imp}\frac{g_{\mathrm{1}}(i\omega_n)}
{g_{\mathrm{0}}(i\omega_n)g_{\mathrm{0}}(-i\omega_n)
+g_{\mathrm{1}}(i\omega_n)^2},
\label{imp2}
\end{equation}
where 
\begin{equation}
g_{\mathrm{0}}(i\omega_n)
=-\frac{1}{N}\sum_\mathbf{k}G(\mathbf{k},i\omega_n),
\end{equation}
and
\begin{equation}
g_{\mathrm{1}}(i\omega_n)
=-\frac{1}{N}\sum_\mathbf{k}F(\mathbf{k},i\omega_n).
\end{equation}

$\hat{\Sigma}_\mathrm{FLEX}(\mathbf{k},i\omega_n)$ is determined by 
the following spin and charge fluctuation mediated interactions: 
\begin{equation}
V_\mathrm{s}(q)=U^2 \left[\frac{3}{2}
\frac{\chi_\mathrm{0}^\mathrm{s}(q)}
{1-U \chi_\mathrm{0}^\mathrm{s}(q)}
-\frac{1}{2}\chi_\mathrm{0}^\mathrm{s}(q) \right],
\end{equation}
\begin{equation}
V_\mathrm{c}(q)=U^2 \left[ \frac{1}{2}
\frac{\chi_\mathrm{0}^\mathrm{c}(q)}
{1+U \chi_\mathrm{0}^\mathrm{c}(q)}
-\frac{1}{2}\chi_\mathrm{0}^\mathrm{c}(q) \right].
\end{equation}
Here q is a shorthand notation as $q=(\mathbf{q},i \nu_m)$, where 
$\nu_m \equiv 2 \pi T m$ is a bosonic Matsubara frequency. 
$\chi_\mathrm{0}^\mathrm{s}$ and $\chi_\mathrm{0}^\mathrm{c}$ are the 
irreducible spin and charge susceptibilities which are written as follows: 
\begin{equation}
\chi_\mathrm{0}^\mathrm{s}(q)=-\frac{1}{\beta N}
\sum_{\mathbf{k}n}[G(k+q)G(k)+F(k+q)F(k)],
\end{equation}
\begin{equation}
\chi_\mathrm{0}^\mathrm{c}(q)=-\frac{1}{\beta N}
\sum_{\mathbf{k}n}[G(k+q)G(k)-F(k+q)F(k)].
\end{equation}
In terms of these interactions, the diagonal and off-diagonal components of 
the self-energy matrix, $\hat{\Sigma}_\mathrm{FLEX}$, are written as follows:
\begin{equation}
\Sigma_\mathrm{FLEX}(k)
=\frac{1}{\beta N}\sum_{\mathbf{q}m}
V_\mathrm{eff}(q)G(k+q),
\label{flex1}
\end{equation}
\begin{equation}
\phi_\mathrm{FLEX}(k)
=\frac{1}{\beta N}\sum_{\mathbf{k'},n'}
V_\mathrm{a}(k-k')F(k'),
\label{flex2}
\end{equation}
with
\begin{equation}
V_\mathrm{eff}(q)
=V_\mathrm{s}(q)+V_\mathrm{c}(q),
\end{equation}
\begin{equation}
V_\mathrm{a}(q)
=V_\mathrm{s}(q)-V_\mathrm{c}(q).
\end{equation}

Combining Eqs.(\ref{imp1}), (\ref{imp2}), (\ref{flex1}) and (\ref{flex2}), 
we obtain the explicit expression of Eq.(\ref{self-energy}):
\begin{equation}
\Sigma(k)=\frac{1}{\beta N}\sum_{\mathbf{q}m}
V_\mathrm{eff}(q)G(k+q)+\Sigma_\mathrm{imp}(i\omega_n),
\label{selfN}
\end{equation}
\begin{equation}
\phi(k)=-\frac{1}{\beta N}\sum_{\mathbf{k'},n'}
V_\mathrm{a}(k-k')D(k')\phi(k')+\phi_\mathrm{imp}(i\omega_n).
\label{selfA}
\end{equation}

Eq.(\ref{selfA}) is the so called gap equation. We numerically solve 
Eqs. (\ref{Dyson})-(\ref{selfA}) 
in the self-consistent way, choosing the chemical 
potential $\mu$ so as to keep the filling constant, 
\begin{equation}
n=2\frac{1}{N}\sum_{\mathbf{k}}n_{\mathbf{k}}
=2\frac{1}{\beta N}\sum_{\mathbf{k} n}G(\mathbf{k},i\omega_n)
e^{i\omega_n 0+}.
\end {equation}

The divergent point of superconduncting susceptibility, that is, 
the transition temperature $\Tc$, is estimated by the following eigenvalue 
equation: 
\begin{equation}
\lambda \tilde{\phi}(k)=-\frac{1}{\beta N}\sum_{\mathbf{k'},n'}
V_\mathrm{a}(k-k')D(k')|_{\phi=0}
\tilde{\phi}(k').
\label{eigen}
\end{equation}
The transition temperature is determined by the criterion that the maximum 
eigenvalue of Eq.(\ref{eigen}) is unity at $T=\Tc$. 
Here, we have assumed that $\phi(\mathbf{k},i\omega_n)$ is 
anisotropic, thus $\phi_\mathrm{imp}(i\omega_n)=0$, 
because in the clean case 
$\phi$ has $d_\mathrm{x^2-y^2}$-symmetry. 
It is not immediately obvious that this assumption may be reasonable 
when the calculation is done self-consistently 
in the presence of impurites. 
However, this is indeed the case as we will shortly disscuss in $\S3.5$. 
Then, within the assumption, we study the impurity effect on $\Tc$ obtained by 
solving Eq. (\ref{eigen}).

\subsection{Impurity effects on the quasi-particle damping due to electron 
correlation and on the pairing interaction }

From the earlier calculation using the FLEX theory, it is well known that 
the main factors to determine the value of $\Tc$ 
are the magnitude of $V_\mathrm{s}(\mathbf{q},\omega)$ around 
$\mathbf{q}=(\pm \pi,\pm \pi)$ and the imaginary part of self-energy around 
$\mathbf{k}=(\pm \pi,0),(0,\pm \pi)$. Therefore, we examine what is going on 
in these factors by impurities. 

In our calculation, $V_\mathrm{s}(\mathbf{q},0)$ has 
the commensurate peak at $\mathbf{q}=\mathbf{Q}=(\pi,\pi)$ for $n=0.85$ 
and the incommensurate peak for $n=0.75$. As for the discussion below, 
the essential points do not change whether $V_s$ has the commensurate 
peak or not. That is, the essential features discussed below do not 
depend on the amount of hole-doping. Then, 
we discuss mainly the commensurate case.

\begin{figure}[t]
\begin{center}
\includegraphics[width=7cm]{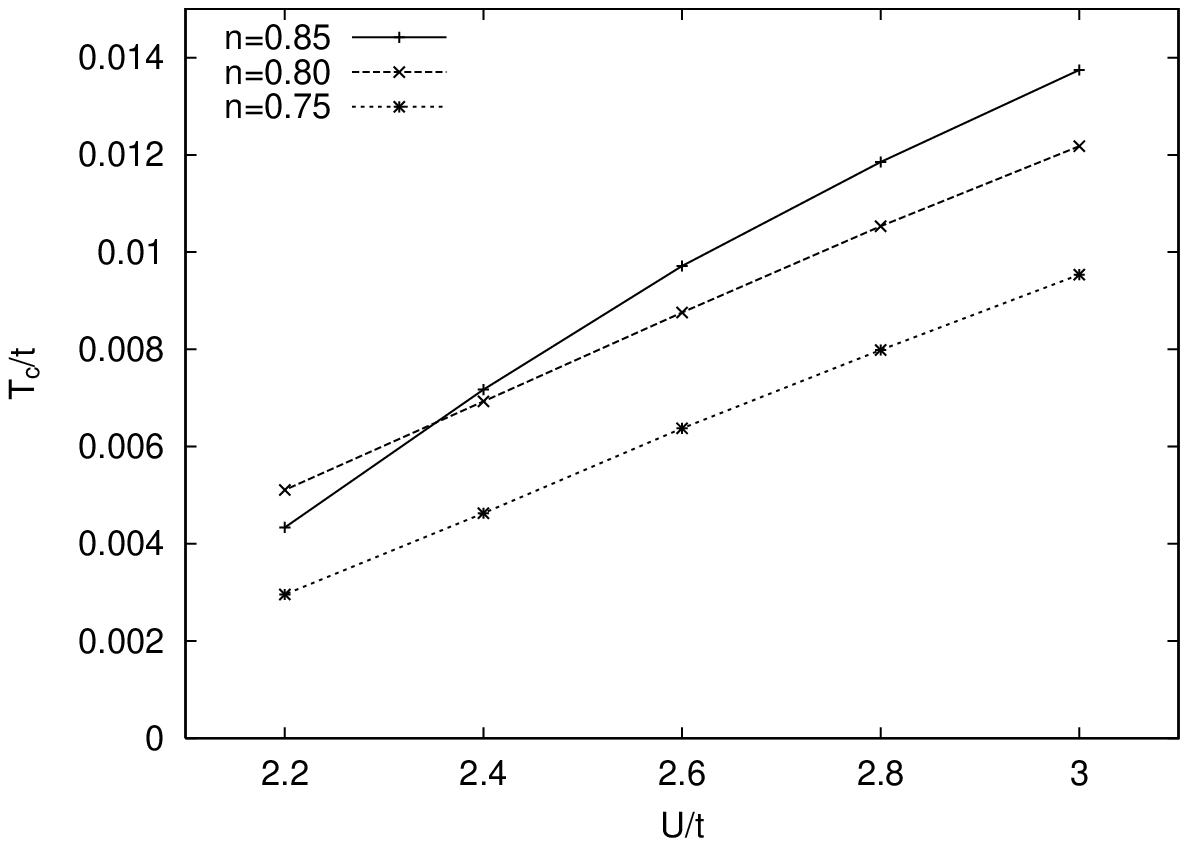}
\end{center}
\caption{$\Tc$ vs. $U$ for various fillings of electrons without impurities. 
         The feature that small $U$ leads to small $\Tc$ is common in 
         the usual FLEX theory.}
\label{U-T}
\end{figure}

Fig. \ref{ImSEVs} shows $\mathrm{Im}\Sigma_\mathrm{FLEX}$ at 
$\mathbf{k}=(0,\pi)$ and $\mathrm{Im}V_\mathrm{s}(\mathbf{q},\omega)$ at 
$\mathbf{q}=\mathbf{Q}$ as the function of $\omega$. 
The filling of electron is $n=0.85$. 
In our formulation, the normal self-energy is given by 
$\Sigma=\Sigma_\mathrm{FLEX}+\Sigma_\mathrm{imp}$. 
Then, $|\mathrm{Im}\Sigma_\mathrm{FLEX}|$ means the quasi-particle 
damping originated from the electron correlation. 
From the figure, we can see that $V_\mathrm{s}$ and 
$|\mathrm{Im}\Sigma_\mathrm{FLEX}|$ are suppressed by impurities. 
This is because the specific scattering process due to electron correlation 
is blurred by impurity scattering. 
By comparing the data of '$U=2.71t,n_\mathrm{imp}=0.00$' and 
'$U=3t,n_\mathrm{imp}=0.02$', 
we can see that 
the amount of $|\mathrm{Im}\Sigma_\mathrm{FLEX}|$ suppression by 
impurities ($n_\mathrm{imp}=0.02$) is almost identical to 
that by the reduction of $U$ from $3t$ to $2.71t$. 
In the usual FLEX theory without impurities, 
as $U$ reduces, $\Tc$ becomes small. This is shown in Fig. \ref{U-T}. 
From the data of '$n=0.85$' in Fig. \ref{U-T}, 
the change of $\Tc$ expected by the suppression can be roughly 
estimated as $\Delta \Tc \sim 0.003t$. 
It indicates that the reduction of $V_\mathrm{s}$ and 
$|\mathrm{Im}\Sigma_\mathrm{FLEX}|$ is rather large. 
Then, the pairing interaction and the pair-breaking due to the electron 
correlation are significantly suppressed by impurities.

\begin{fullfigure}[t]
  \begin{center}
    \begin{tabular}{cc}
      \resizebox{80mm}{!}{\includegraphics{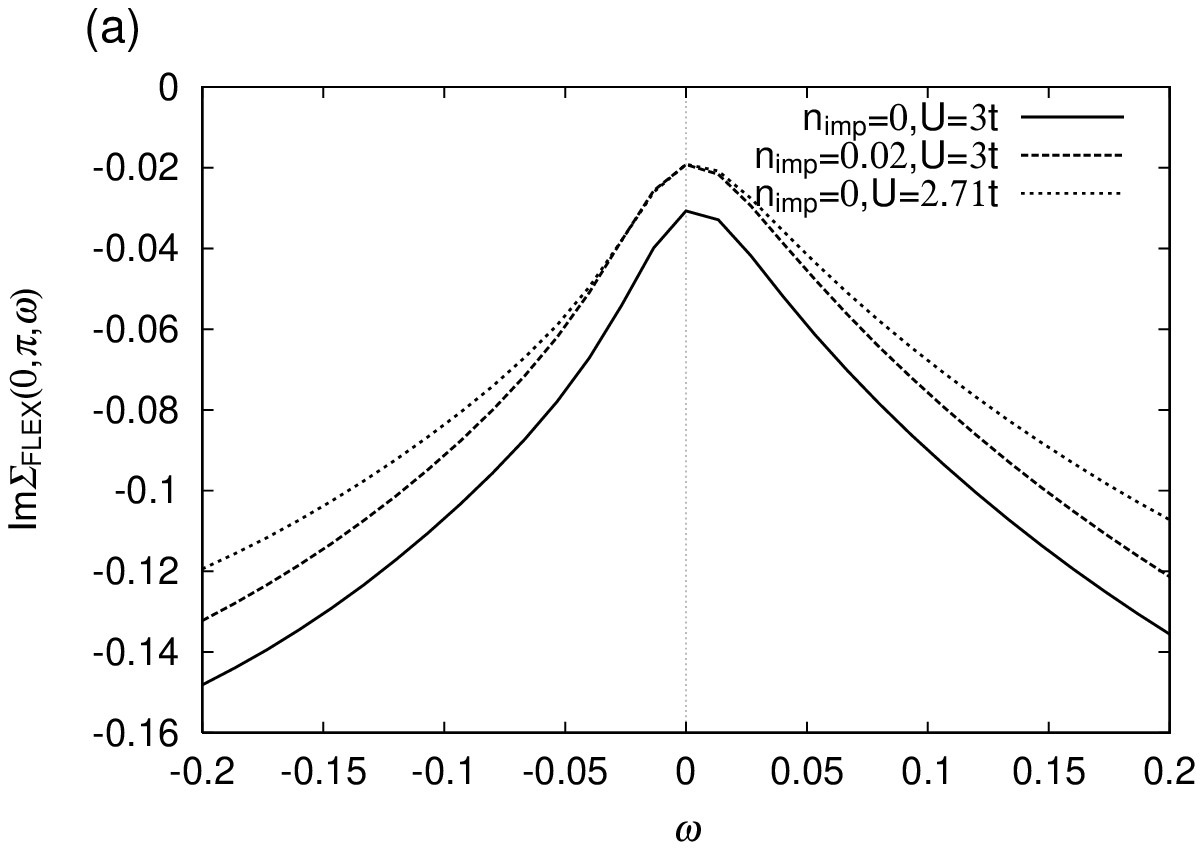}} &
      \resizebox{55mm}{!}{\includegraphics{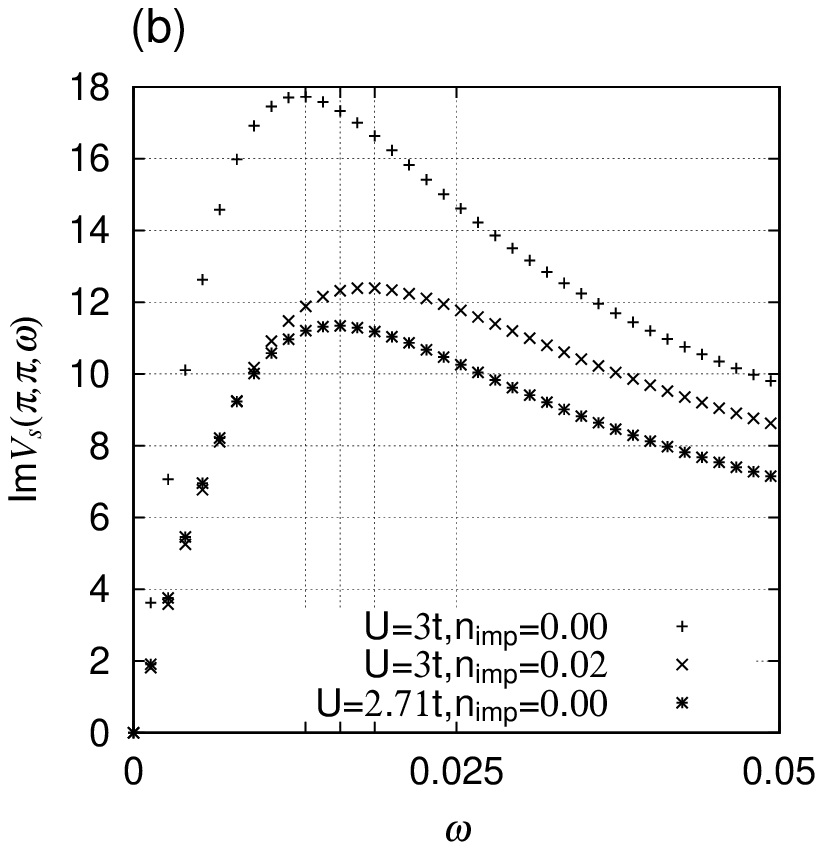}} 
    \end{tabular}
    \caption{The imaginary part of $\Sigma_\mathrm{FLEX}$ 
             at $\mathbf{k}=(0,\pi)$ (a) 
             and $V_s$ at $\mathbf{Q}=(\pi,\pi)$ (b) as the 
             function of real $\omega$. 
             The filling of electron, $n$ is 0.85. 
             The pure case corresponds to '$U=3t,n_\mathrm{imp}=0.00$' and 
             the dirty case corresponds to '$U=3t,n_\mathrm{imp}=0.02$'. 
             For the comparison, the case of the reduced $U$ without 
             impuries ('$U=2.71t,n_\mathrm{imp}=0.00$') is included. 
             The strength of the spin fluctuation interaction 
             associated with momentum $\mathbf{Q}$ can be estimated 
             by the integral, $V_\mathrm{s}(\mathbf{Q},0)
             =2\int_0^\infty \frac{\mathrm{d}\omega}{\pi}
             \mathrm{Im}V_\mathrm{s}(\mathbf{Q},\omega)/\omega$. 
            }
    \label{ImSEVs}
  \end{center}
\end{fullfigure}

\subsection{Reduction of $\Tc$ as the function of 
$n_\mathrm{imp}$ and the mechanism}

\begin{figure}[t]
\begin{center}
\includegraphics[width=7cm]{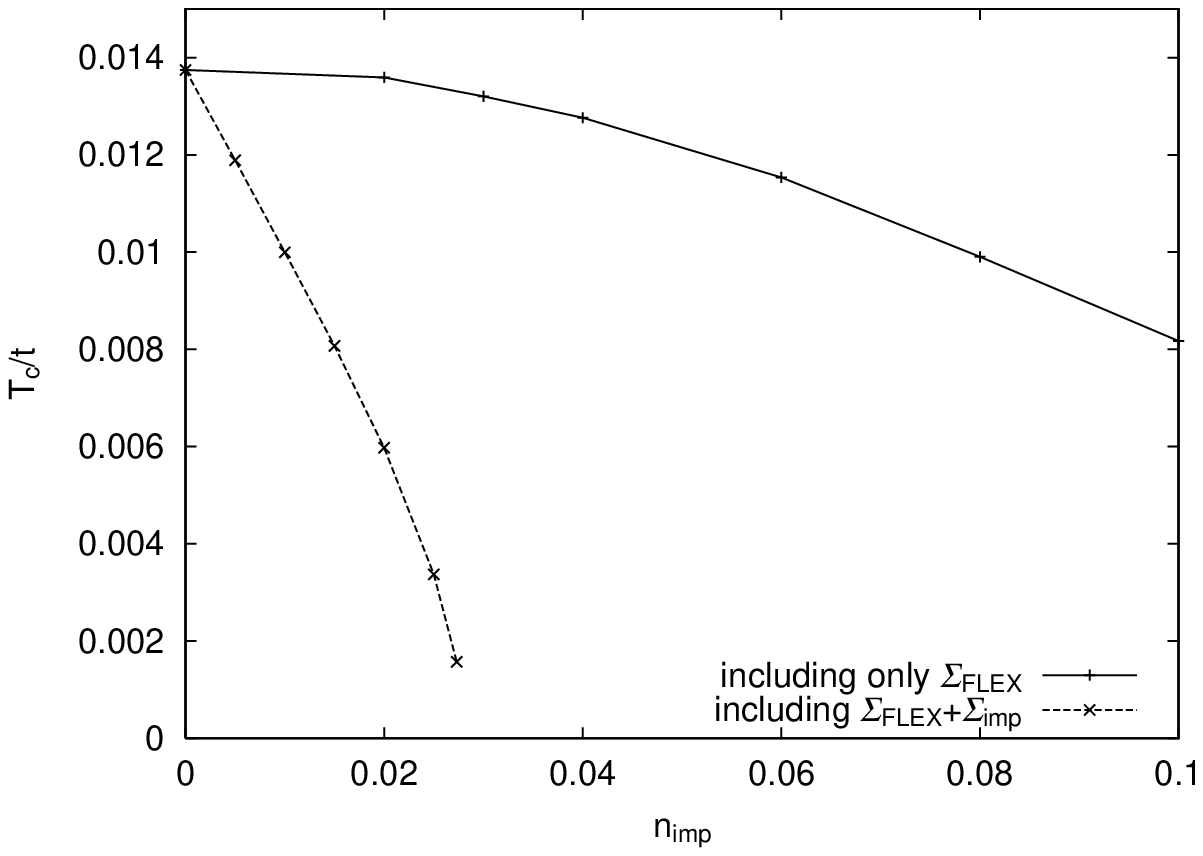}
\end{center}
\caption{Full line and dotted line show 
         $\Tc$ vs. $n_\mathrm{imp}$ in the FLEX theory 
         with impurity-suppressed $\Sigma_\mathrm{FLEX}$ 
         and $U=3t$ without and with $\Sigma_\mathrm{imp}$, respectively. 
         The filling of electrons is $0.85$. 
         Dotted line is as same as 
         the data of '$n=0.85$' chart in Fig. \ref{agfit}, 
         although here $\Tc$ is not scaled by $T_\mathrm{c0}$($=0.0137t$). 
         This is shown for comparision.}
\label{fleximp}
\end{figure}

Fig. \ref{agfit} shows $\Tc$ vs. $n_\mathrm{imp}$ obtained 
by the FLEX theory in the presence of impurities. 
We can see that the reduction of $\Tc$ almost coincides with AG curves and 
then keeps the almost same form as that in the weak-coupling case. 
This result is a surprise because 
in spite of the rather large pair-weakening and the reduction 
of the pair-breaking originating from the electron interaction by impurities 
as mensioned in the last section, those changes never seem to be 
effective on $\Tc$. 
The result 
originates from the cancellation between these two impurity effects. 
Fig. \ref{fleximp} shows $\Tc=\Tc(n_\mathrm{imp})$ 
in the FLEX theory with the impurity-suppressed pairing interaction, 
$V_\mathrm{s}$, and the impurity-suppressed FLEX part of the self-energy. 
It can be seen that 
$\Tc$ hardly changes up to $n_\mathrm{imp} \sim 0.03$. 
This indicates that the cancellation between two effects occurs. 

Now, we summarize the $\Tc$ reduction by impurities in the FLEX theory. 
First of all, the value of $\Tc$ is mainly determined by the magnitude 
of $V_\mathrm{s}(\mathbf{q},\omega)$ with 
$\mathbf{q} \sim \mathbf{Q}$ and the quasi-pariticle damping rate 
$|\mathrm{Im}\Sigma(\mathbf{k},\omega)|$ around 
$\mathbf{k}=(\pm \pi,0),(0,\pm \pi)$. 
$V_\mathrm{s}$ can be regarded as the factor to raise $\Tc$ and 
$|\mathrm{Im}\Sigma|$ as the factor to lower $\Tc$. 
Impurities significantly suppress 
both $V_\mathrm{s}$ and $|\mathrm{Im}\Sigma|$; 
the pair-weakening and the reduction of the quasi-particle 
damping due to the electron correlation. 
These two effects, however, almost cancel each other 
to give no effect on $\Tc$. 
Then, we can regard the system described by the FLEX theory with impurities 
as that with an impurity-independent pairing interaction and impurities; 
that is, the system can be described by the weak-coupling $d$-wave BCS model. 
Consequently, the reduction of $\Tc$ as the function of 
$n_\mathrm{imp}$ keeps the almost same form as that in the BCS model. 

It is notable that the coincidence between the 
reduction of $\Tc$ in the FLEX theory and AG formula is the very appearance. 
In order to analyze correctly the experimental data on impurity effect in HTS, 
we must take into account the pair-weakening and the reduction of 
$|\mathrm{Im}\Sigma_\mathrm{Coulomb}|$, because 
the superconductivity emerges out of Coulomb interaction in HTS. 
In the next section, we will pick up the example which indicates that 
we have to take the reduction of $|\mathrm{Im}\Sigma_\mathrm{Coulomb}|$ 
into consideration. 

\begin{fullfigure}[t]
  \begin{center}
    \begin{tabular}{cc}
      \resizebox{60mm}{!}{\includegraphics{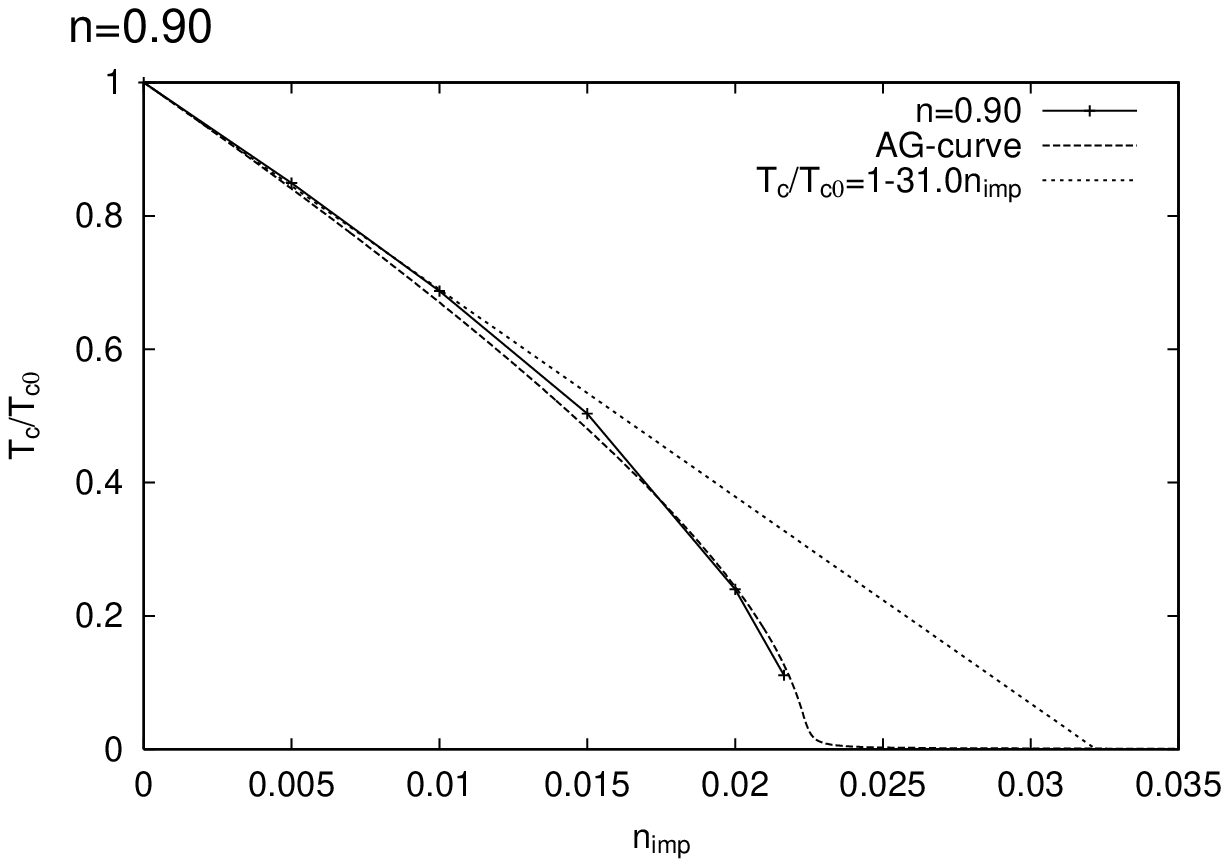}} &
      \resizebox{60mm}{!}{\includegraphics{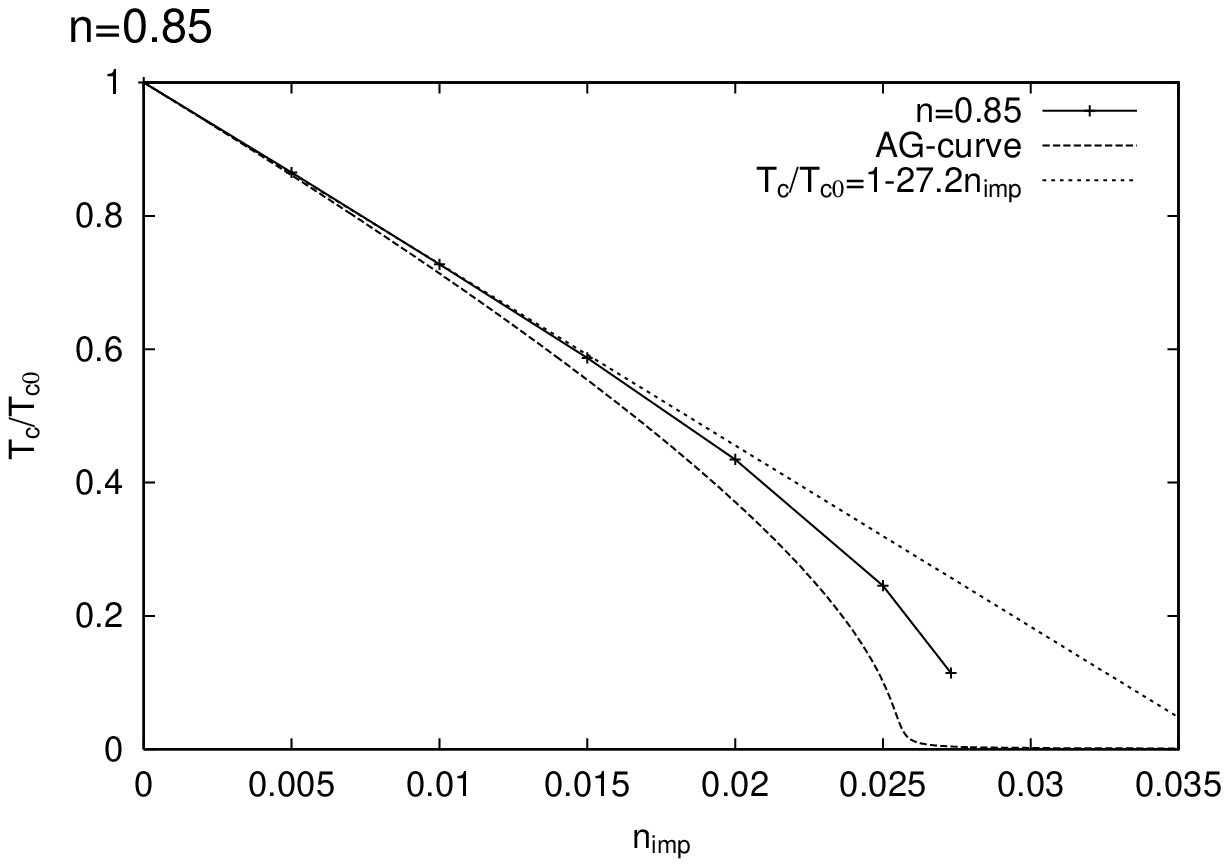}} \\
      \resizebox{60mm}{!}{\includegraphics{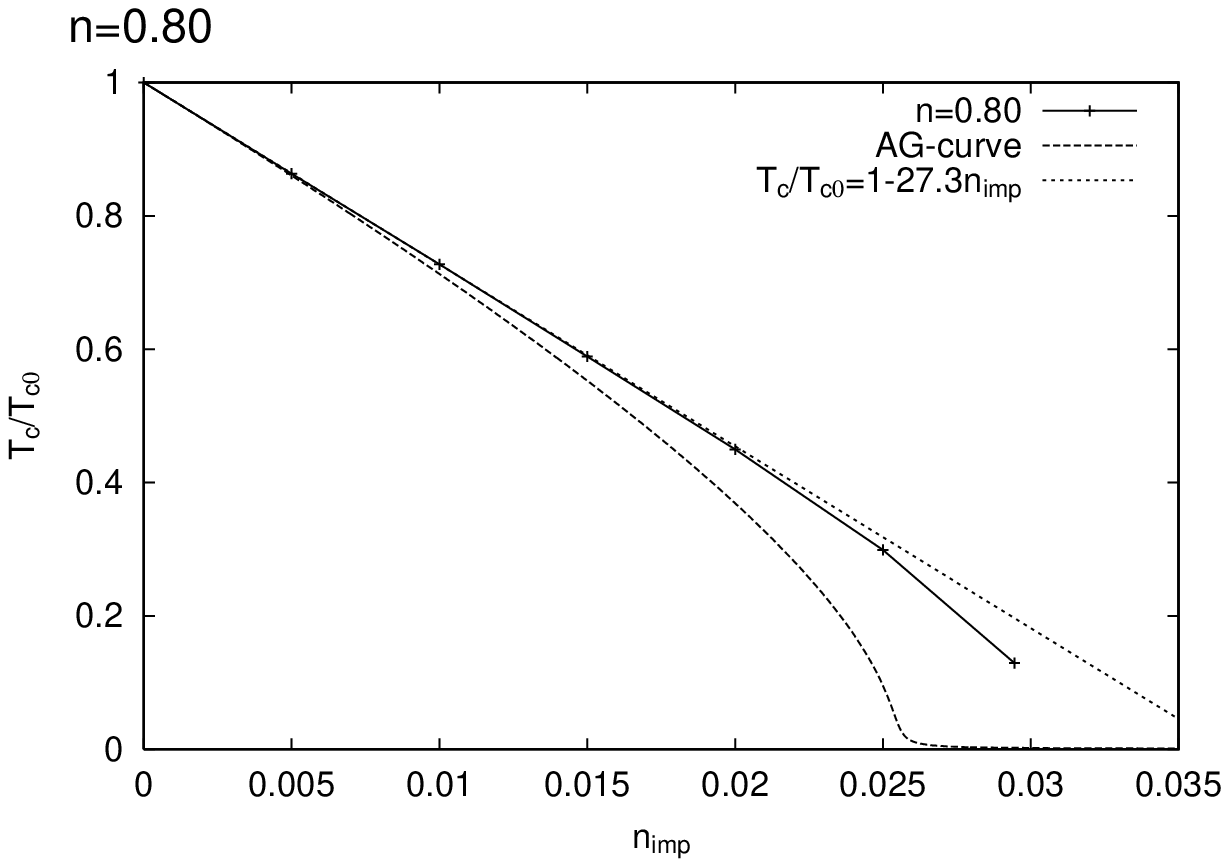}} &
      \resizebox{60mm}{!}{\includegraphics{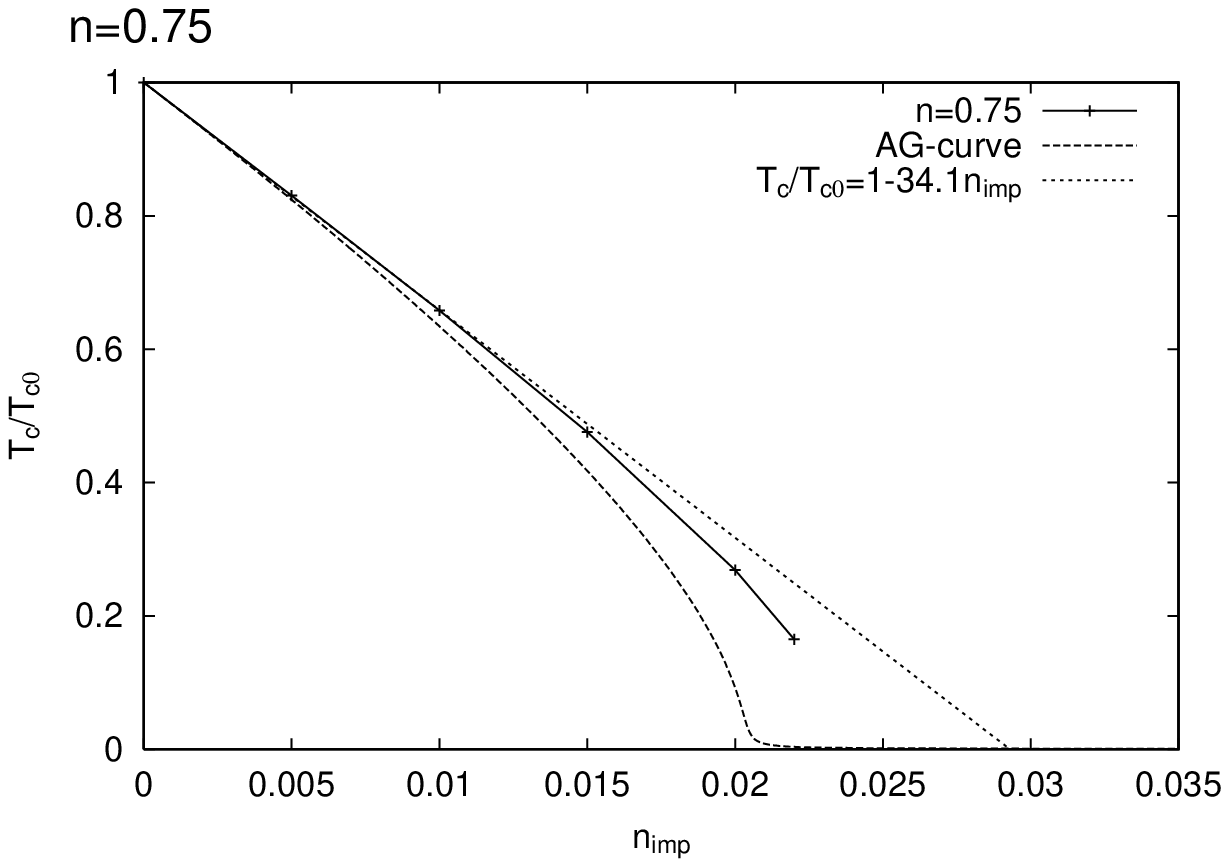}} 
    \end{tabular}
    \caption{$\Tc/T_\mathrm{c0}$ vs. $n_\mathrm{imp}$ and 
             the curves fitted by AG formula, Eq.(\ref{ag}). 
             The fitting curves are obtained in the same way 
             as we have done in Fig. \ref{agfit_weak}.}
    \label{agfit}
  \end{center}
\end{fullfigure}

\subsection{Dependence of impurity effect on $\mathbf{k}$-point}

\begin{figure}[t]
\begin{center}
\includegraphics[width=6cm]{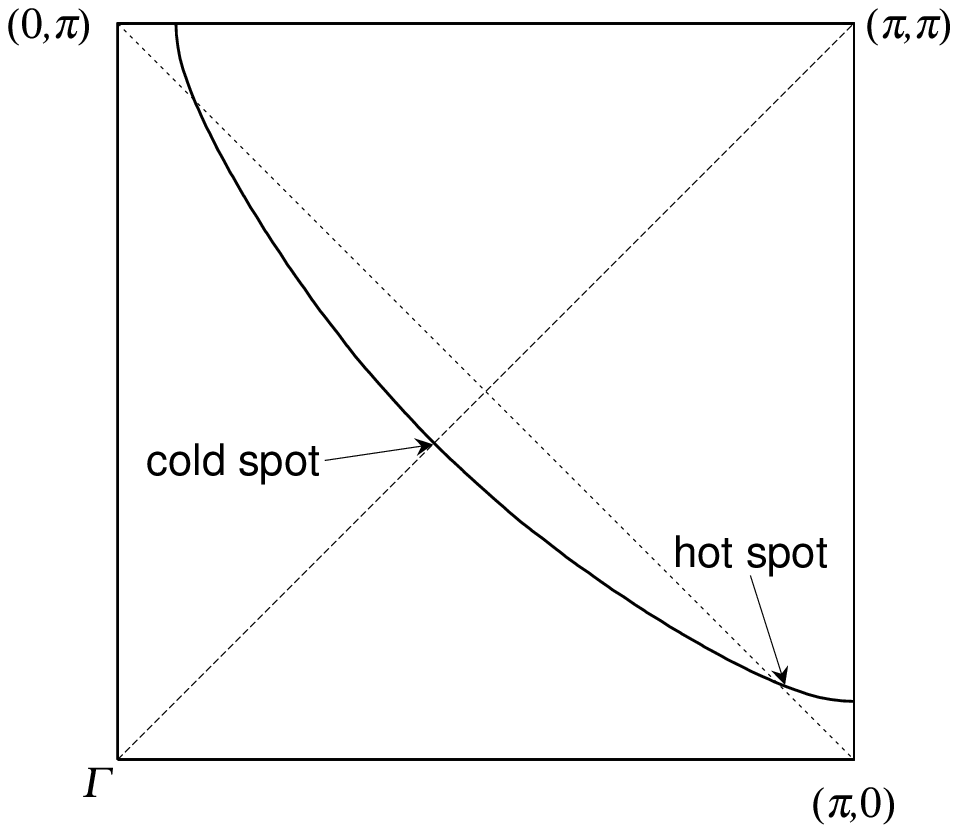}
\end{center}
\caption{The definition of 'hot spot' and 'cold spot' of the Fermi surface. 
         The Fermi surface is obtained for the parameters chosen as 
         $n=0.85$, $t'=0.25$, $U=3t$ and $T=0.015t$. 
         }
\label{spot}
\end{figure}

\begin{fullfigure}[t]
  \begin{center}
    \begin{tabular}{cc}
      \resizebox{78mm}{!}{\includegraphics{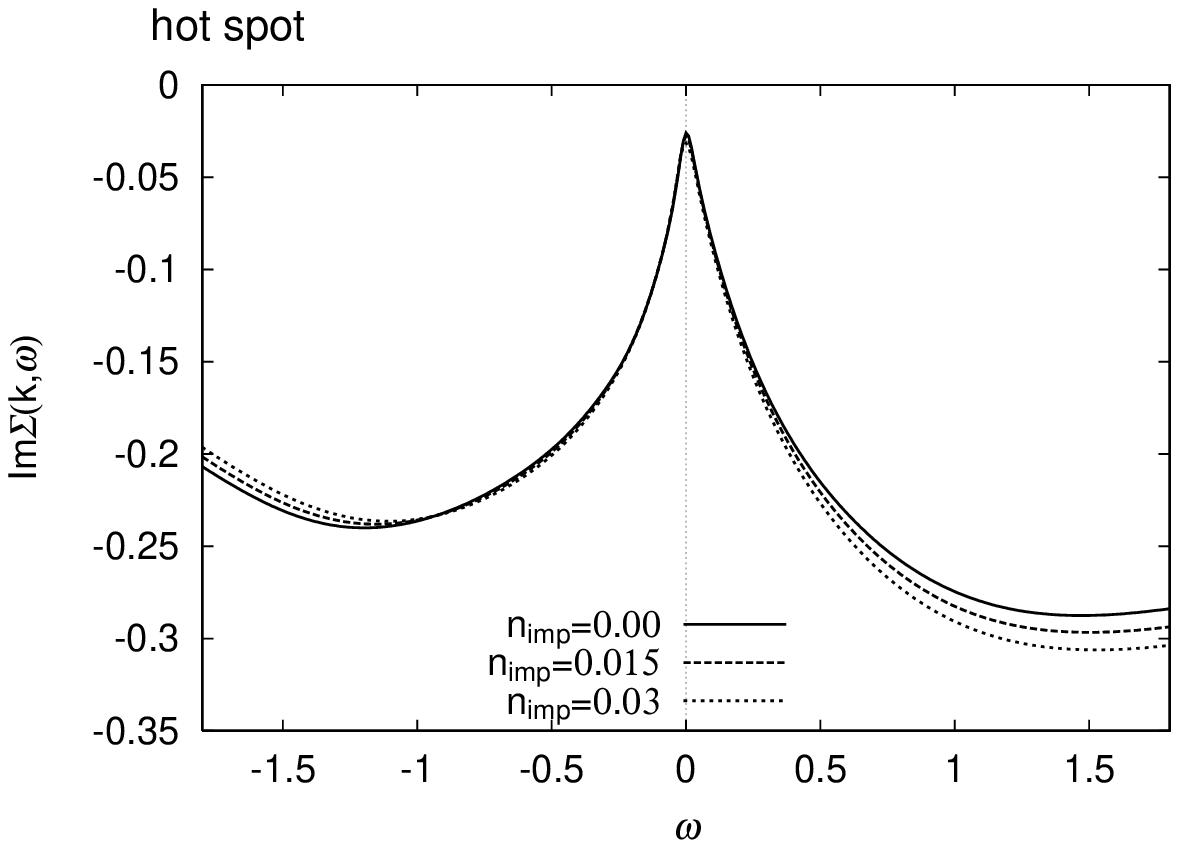}} &
      \resizebox{78mm}{!}{\includegraphics{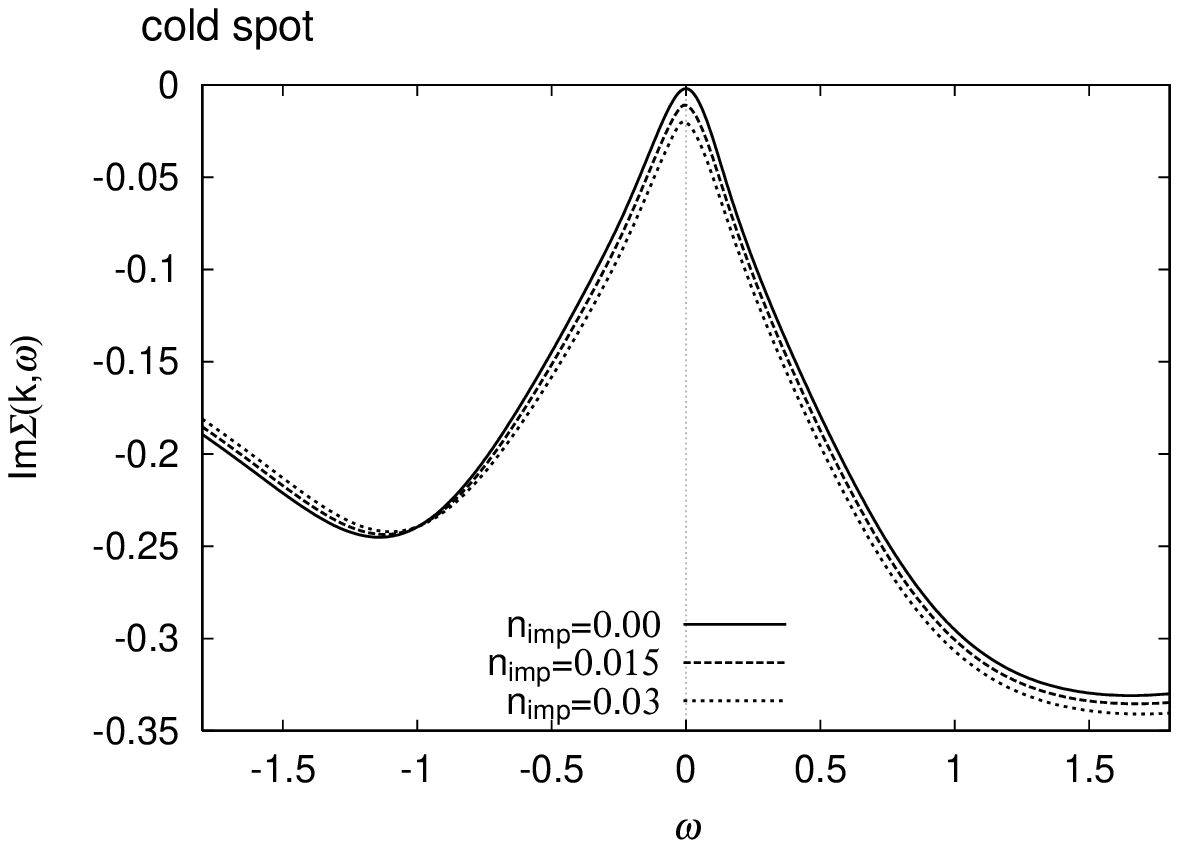}}
    \end{tabular}
    \caption{The imaginary part of the quasi-particle self-energy 
             in the presence of impurity scattering at hot spot (a) and 
             cold spot (b). The parameters are chosen as 
             $n=0.85$, $t'=0.25t$, $U=3t$, and $T=0.015t$.
            }
    \label{hotcold}
  \end{center}
\end{fullfigure}

We define 'hot spot' and 'cold spot' of the Fermi surface 
following Pines's definition for the hole-doping case. 
As we show in Fig. \ref{spot}, 'hot spot' is located in the vicinity of 
the anti-ferromagnetic Brillouin zone, and 'cold spot' is far from the 
anti-ferromagnetic Brillouin zone. We find that there is a remarkable 
difference in the impurity effect on the quasi-particle damping 
between hot spot and cold spot. At hot spot, 
the change of $|\mathrm{Im}\Sigma|$ at Fermi level is small 
for the increase of impurity concentration, but at cold spot, 
$|\mathrm{Im}\Sigma|$ becomes large; that is, 
impurities hardly affect the 
quasi-particles at hot spot but are effective at cold spot. 
This is shown in Fig. \ref{hotcold}. The feature of hot spot is 
$\epsilon_\mathbf{k}=\epsilon_\mathbf{k-Q}=0$. 
Therefore, the quasi-particles at hot spot 
exchange the strongly enhanced spin fluctuation interaction 
$V_\mathrm{s}$ with the momentum $\mathbf{Q}$, and have a large 
damping rate even in the absence of impurities. 
In the above discussion, we have found that impurities reduce 
the spin fluctuation 
part of $|\mathrm{Im}\Sigma|$, i.e. $|\mathrm{Im}\Sigma_\mathrm{FLEX}|$. 
Therefore, at hot spot, the spin fluctuation part and the impurity 
part of $|\mathrm{Im}\Sigma|$ preserve a balance, and therefore, 
$\mathrm{Im}\Sigma=\mathrm{Im}\Sigma_\mathrm{FLEX}+
\mathrm{Im}\Sigma_\mathrm{imp}$ hardly change. 
On the other hand, at cold spot, since the original quasi-paritcle damping 
rate by the spin fluctuation scattering in the absence of impurities is 
small, the reduction of $|\mathrm{Im}\Sigma_\mathrm{FLEX}|$ is very slight, 
and then impurity scattering makes the magnitude of 
$|\mathrm{Im}\Sigma|$ large. 

The following point should be noted. 
In the absence of impurities, the conductance is mainly determined by 
the contribution from cold spot region. 
As impurities increases, the momentum 
dependence of lifetime becomes small, and the conductance becomes to be 
determined by the whole $\mathbf{k}$'s on Fermi surface.

\subsection{The supplementary discussion}

\begin{figure}[t]
\begin{center}
\includegraphics[width=8cm]{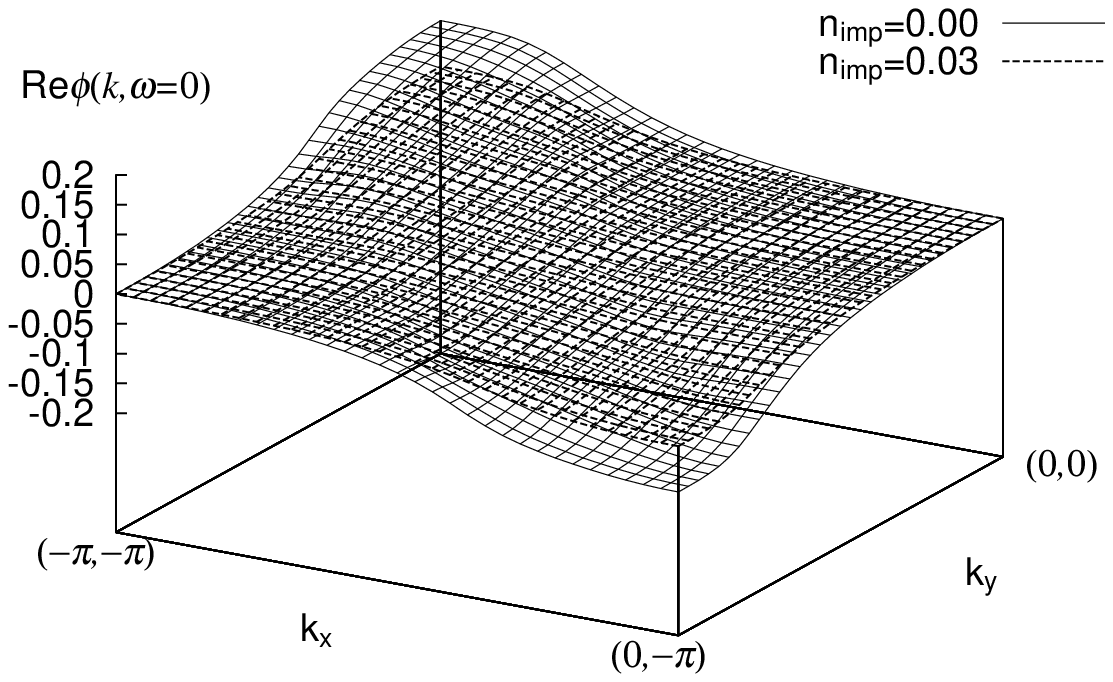}
\end{center}
\caption{The momentum dependence of the gap paremeter $\phi(\mathbf{k},0)$ 
         in the superconducting state. 
         The parameters are chosen as $t=0.5$, $t'=0.25t$, $T=0.007t$, 
         $U=4t$, and $n=0.85$. }
\label{Dgap}
\end{figure}

Here, we make sure that the assumption stated at the last of $\S3.1$. 
Fig. \ref{Dgap} shows the momentum dependence of the gap parameter for 
the clean case and the dirty case. In both cases, the 
$d_{\mathrm{x^2}-\mathrm{y^2}}$-wave arises clearly. Therefore, 
within the model we explained in $\S3.1$, impurities do not change 
the symmetry of the superconducting gap. Then, the neglect of 
$\phi_\mathrm{imp}$ in Eq.(\ref{eigen}) is justified. 

\section{Impurity effect in the presence of the pseudogap}

Pseudogap phenomena can be seen in underdoped cuprate. 
In order to describe the pseudogap, 
we adopt the "pairing scenario" 
where the pseudogap is induced by the superconducting (SC) 
fluctuation.~\cite{Yanase1,Yanase2,Yanase3,Levin1,Levin2,Rhoe} 
In the approach, pseudogap phenomena 
are understood as the resonance phenomena between the states of Cooper pairs 
and quasi-particles in the Fermi liquid; 
the resonance is described by extraordinarily
increased imaginary part of the quasi-particle self-energy. 
Y. Yanase and one of the authors (K. Y) have succeeded in 
describing the pseudogap by the 1-loop order theory 
with respect to 
SC fluctuation on the basis of the Hubbard model with $d$-wave attractive 
interaction.~\cite{Yanase1} 
Therefore, including effects of impurities into the theory, 
we discuss the $\Tc$ suppression by impurities 
in the presence of the pseudogap. 
It is notice that the similar study has already been done by 
Qijin Chen and J. R. Schrieffer.~\cite{chen} 
However, their approach for pseudogap is different from ours. 
They have used a $G_0G$ approximation scheme. 
Thus our study may serve as a test to judge what is an appropriate theory 
of pseudogap.

\subsection{Reviews on the 1-loop order theory 
and the self-consistent calculation}

\begin{figure}[t]
\begin{center}
\includegraphics[width=4cm]{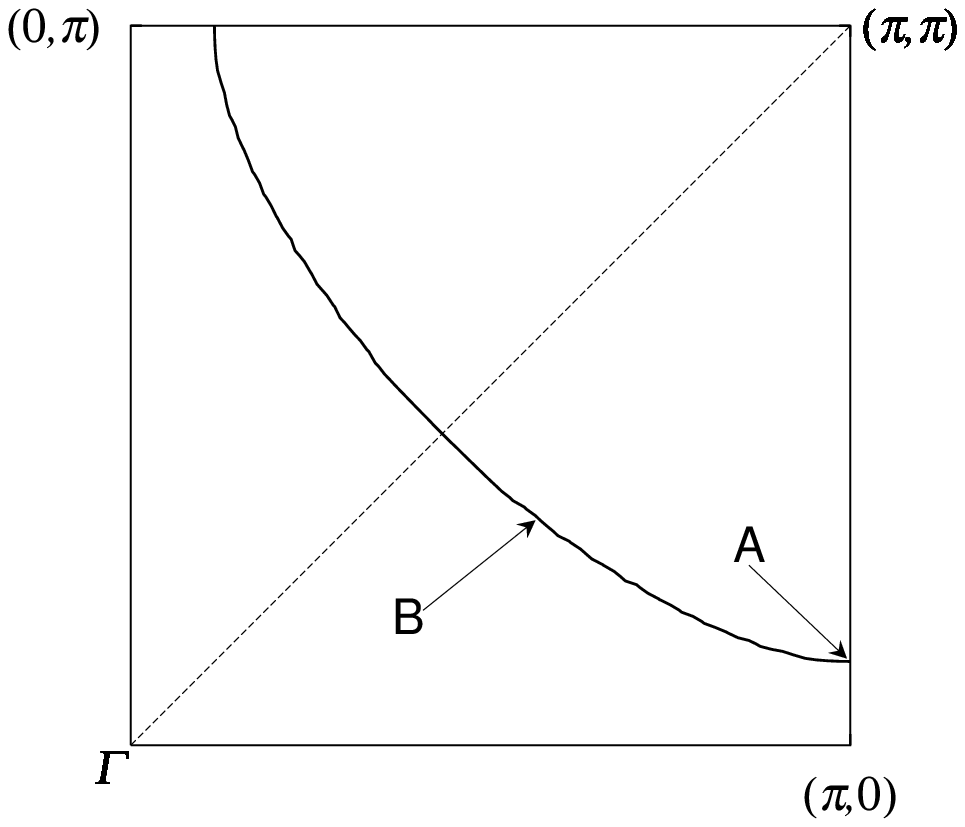}
\end{center}
\caption{The Fermi surface calculated using the parameters chosen as
         $t'=0.25t$ and $n=0.9$. 
         The point {\bf A} shows $\mathbf{k}=(\pi,0.12\pi)$ and 
         {\bf B} shows $\mathbf{k}=(0.56\pi,0.31\pi)$. 
         {\bf A} is so called antinodal point. 
         We refer these points {\bf A} and {\bf B} many times in the below 
         discussion.}
\label{AB}
\end{figure}

The attractive Hubbard model on 2D square lattice is given by 
\begin{equation}
H=\sum_{\mathbf{k}\sigma}\xi_\mathbf{k}
c_{\mathbf{k} \sigma}^{\dagger} c_{\mathbf{k} \sigma}
+\frac{1}{N}\sum_\mathbf{k,k',q}V_\mathbf{k,k'}
c_{\mathbf{k} \uparrow}^{\dagger} c_{\mathbf{q-k} \downarrow}^{\dagger}
c_{\mathbf{q-k'} \downarrow} c_{\mathbf{k'} \uparrow}. 
\end{equation}
Here, $V_\mathbf{k,k'}$ is the $d_\mathrm{x^2-y^2}$-wave separable pairing 
interaction written by the same form as Eq.(\ref{attractive}) and 
$\xi_\mathbf{k}=\epsilon_\mathbf{k}-\mu$, 
where $\epsilon_\mathbf{k}$ 
is the same dispersion as Eq.(\ref{dispersion}). 
The parameters are chosen as $t'=0.25t$, $g=0.4$, the filling number 
$n=0.9$, and $t=0.1\sim1.0$. 
Fig. \ref{AB} shows the Fermi surface. 
The physical meaning of changing the value of 
$t$ is as follows. Generally, in the high density electron system, the band 
width $W$ ($=8t$) of quasi-particle is renormalized 
by the effect of the electron correlation. 
Then, small $t$ corresponds to the high density case. 
As for hole-doped HTS cuprates, small $t$ corresponds to the underdoped region 
and large $t$ does to the overdoped region. 
Therefore, by changing the value of $t$, we can cover the whole region 
of cuprate.

\begin{figure}[t]
\begin{center}
\includegraphics[width=3.6cm]{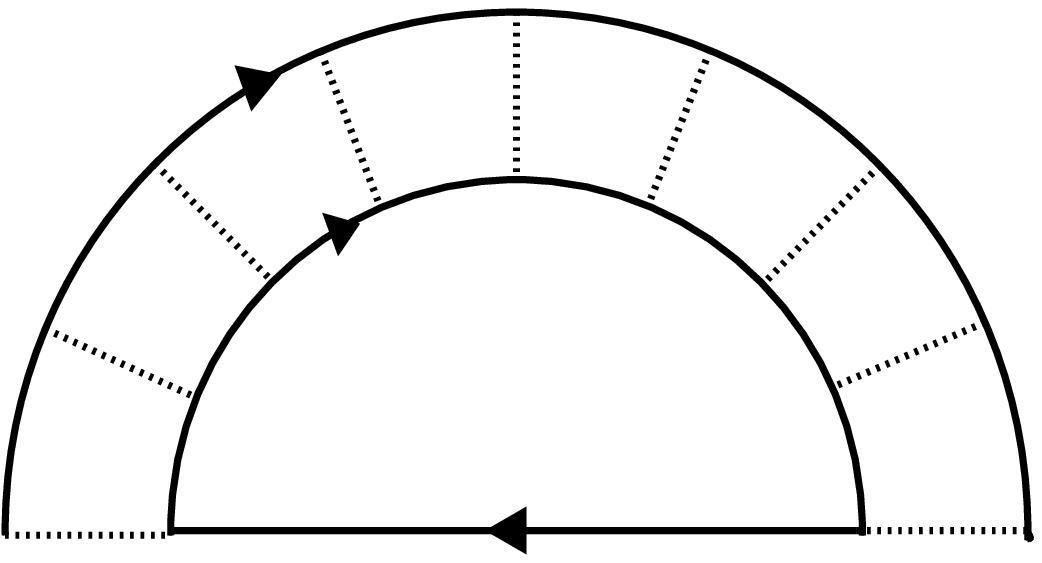}
\end{center}
\caption{Self-energy diagrams in the 1-loop order theory. 
         The dotted line means $V_\mathbf{k,k'}$.}
\label{scf}
\end{figure}

In the 1-loop order theory, 
the self-energy by SC fluctuation is given 
by the diagram in Fig. \ref{scf} and written as 
\begin{equation}
\Sigma_\mathrm{sc}(\mathbf{k},i\omega_n)=\varphi_\mathbf{k}^2\frac{1}{\beta N}
\sum_{\mathbf{q},m}T(\mathbf{q},i\nu_m)G(\mathbf{q-k},i\nu_m-i\omega_n),
\label{SEsc}
\end{equation}
with 
\begin{equation}
T(\mathbf{q},i\nu_m)=\frac{-g}{1-g \chi_\mathrm{pp}(\mathbf{q},i\nu_m)}.
\label{Tmat}
\end{equation}
Here $\nu_m$ is the bosonic Matsubara frequency, 
$\nu_m=2\pi T m$, and 
$\chi_\mathrm{pp}$ is the pairing susceptibility written as
\begin{equation}
\chi_\mathrm{pp}(\mathbf{q},i\nu_m)=\frac{1}{\beta N}\sum_{\mathbf{k},n}
\varphi_\mathbf{k}^2G(\mathbf{k},i\omega_n)G(\mathbf{q-k},i\nu_m-i\omega_n).
\label{chipp}
\end{equation}
In the 1-loop order theory, the Green functions, $G(\mathbf{k},i\omega_n)$, 
in Eq.(\ref{SEsc})$\sim$(\ref{chipp}) are given by bare ones, 
$G_\mathrm{0}(\mathbf{k},i\omega_n)=(i\omega_n-\xi_\mathbf{k})^{-1}$.

Fig. \ref{ASE} shows the real and imaginary parts of the self-energy 
and the spectral weight obtained by the 1-loop order theory. 
It can be seen that as a result 
of the resonance scattering, the damping rate 
$|\mathrm{Im}\Sigma_\mathrm{sc}^\mathrm{R}(\mathbf{k},\omega)|$ at 
the Fermi level is large and 
then the spectral weight is suppressed there. The suppression of the spectral 
weight is related to the suppression of DOS. 
Hence, as mentioned above, the 1-loop order theory well describes the gap-like 
behavior of the single-particle spectrum, that is the pseudogap phenomena. 
It also can be seen that 
$\mathrm{Re}\Sigma_\mathrm{sc}^\mathrm{R}(\mathbf{k},\omega)$ 
has a positive slope at the Fermi level. It is notable that this 
non-Fermi liquid behavior is naturally derived from the Fermi liquid state.

\begin{fullfigure}[t]
  \begin{center}
    \begin{tabular}{ccc}
      \resizebox{50mm}{!}{\includegraphics{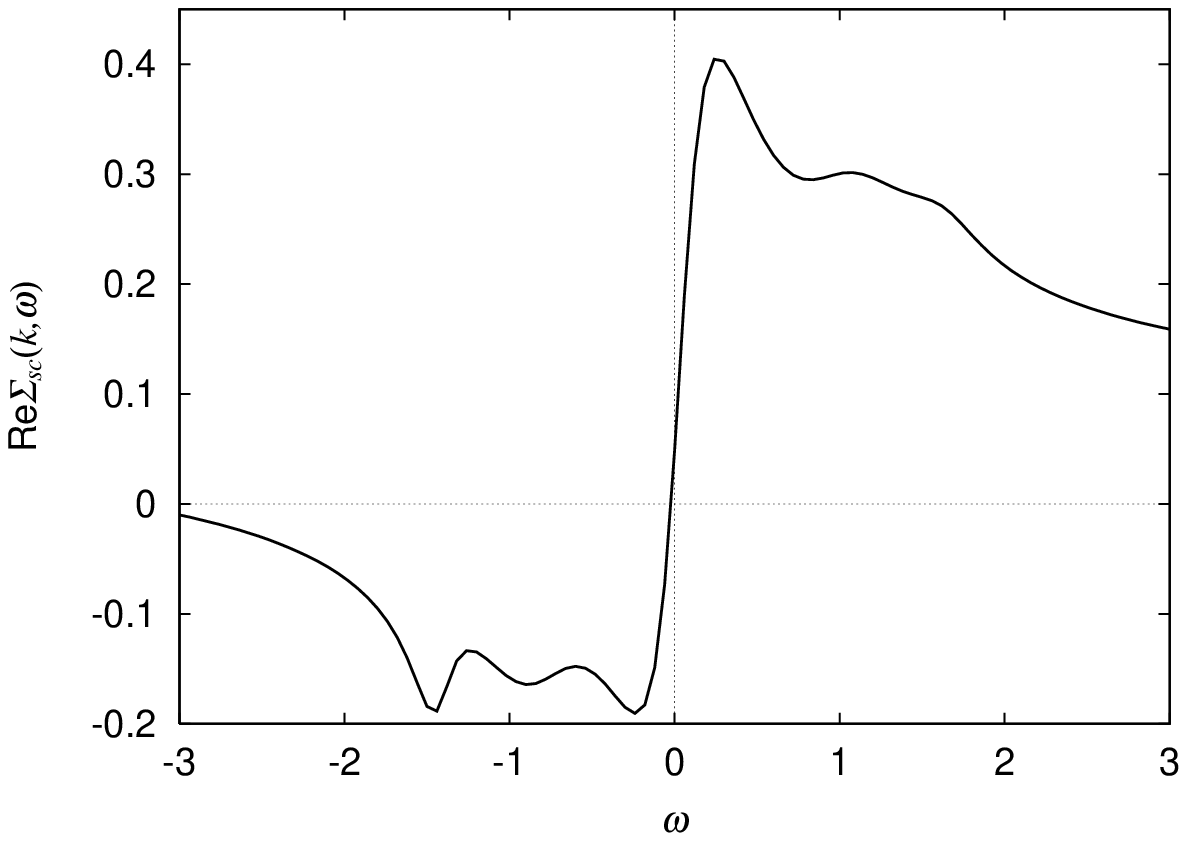}} &
      \resizebox{50mm}{!}{\includegraphics{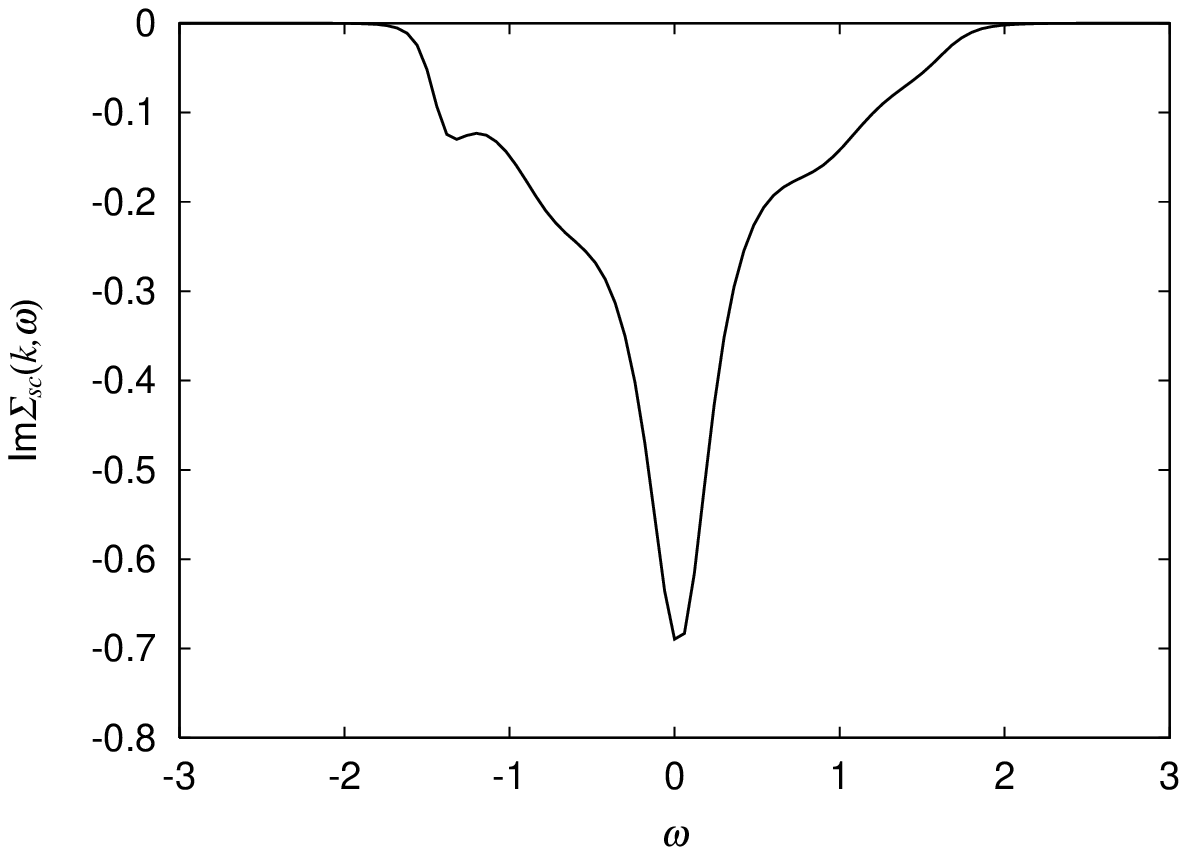}} &
      \resizebox{50mm}{!}{\includegraphics{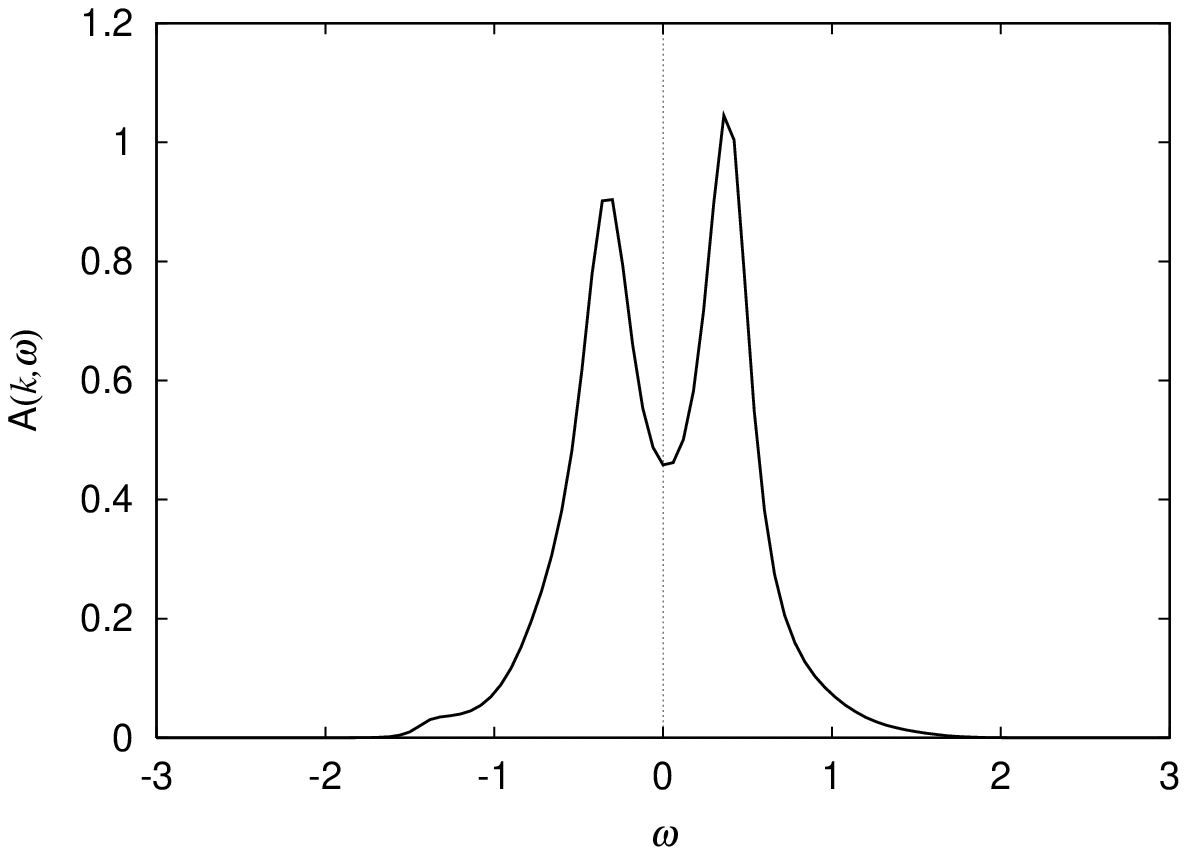}} 
    \end{tabular}
    \caption{The real and imaginary part of the single particle self-energy
             and the spectral weight obtained by the 1-loop order theory. 
             Here, $g=0.4$, $t=0.2$, and $T=0.15$.}
    \label{ASE}
  \end{center}
\end{fullfigure}

It should be paid attention that the above caluculation by the 1-loop order 
theory is applicable to the higher temperature than 
the BCS mean field transition temperature, $\Tc^\mathrm{MF}$. 
In the framework here, the transition temperature is determined by the 
so called Thouless condition:
\begin{equation}
1-g\chi_\mathrm{pp}(\mathbf{0},0)=0.
\label{thouless}
\end{equation}
For the bare Green function $G_\mathrm{0}(\mathbf{k},i\omega_n)$, 
Eq.(\ref{thouless}) is identical with the BCS mean field form: 
\begin{equation}
1=g\frac{1}{\beta N}\sum_{\mathbf{k},n}
\varphi_\mathbf{k}^2 \frac{1}{\omega_n^2+\xi_\mathbf{k}^2}.
\label{bcs}
\end{equation}
Therefore the transition temperature in the 1-loop order theory is identical 
to $\Tc^\mathrm{MF}$. As the result, the temperature region 
as to the pseudogap phenomena is $T > \Tc^\mathrm{MF}$.

To include the suppresion of $\Tc$ by the effect of the SC fluctuation, 
we have to go beyond the 1-loop order theory . In order to 
treat such effect, we carry out the self-consistent calculation 
(the self-consistent 1-loop theory) using the renormalized Green function, 
i.e., $G^{-1}=G_\mathrm{0}^{-1}-\Sigma_\mathrm{sc}$.

\begin{figure}[t]
\begin{center}
\includegraphics[width=7.4cm]{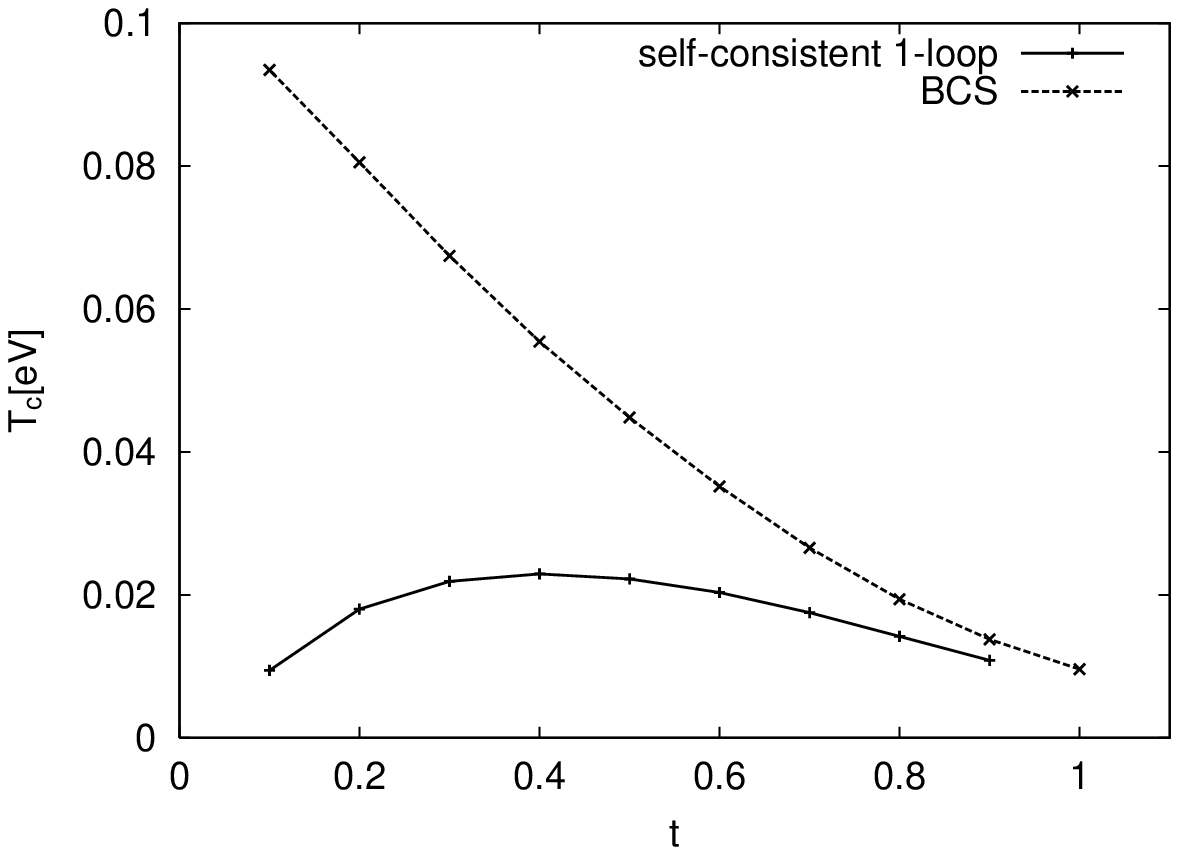}
\end{center}
\caption{$t$ vs. $\Tc$ in 
the self-consistent 1-loop theory and the BCS mean field theory. 
'BCS' means the solution of Eq.(\ref{bcs}), that is 
$\Tc^\mathrm{MF}$. The parameters are chosen as $g=0.4$ and $n=0.9$.}
\label{tT1}
\end{figure}

In the self-consistent 1-loop theory, we solve 
Eq.(\ref{SEsc})$\sim$(\ref{chipp}) self-consistently, choosing the chemical 
potential $\mu$ so as to keep the filling constant $n=0.9$. Using the 
self-consistent solution, we estimate $\Tc$ by the Thouless condition. 
It should be noticed that in 2D system the superconducting transition 
does not occur as is well known as Marmin-Wagner's theorem. Therefore, 
we phenomenologically introduce the three-dimensionality and define 
$\Tc$ as the temperature in which $1-g\chi_\mathrm{pp}(\mathbf{0},0)=0.015$. 
By the procedure, we obtain the phase diagram, Fig. \ref{tT1}, which shows 
$t$ vs. $\Tc$ in the self-consistent 1-loop theory and 
the BCS mean field theory. In the figure, as $t$ decreases, 
$\Tc$ is more suppressed from the value of the BCS transition temperature. 
The small $t$ leads to large SC fluctuation via large $g/t$; 
therefore Fig. \ref{tT1} shows the suppression of $\Tc$ by the effect 
of SC fluctuation. This suppression is mainly due to the reduced DOS. 
Actually, the smaller is $t$, the more suppressed is 
the spectral weight at Fermi level. 
This is displayed in Fig. \ref{ASEful} which 
shows the real and imaginary parts of the self-energy and the spectral 
weight obtained by the self-consistent 1-loop theory. It can be seen that 
at $t=0.05$, 
$|\mathrm{Im}\Sigma_\mathrm{sc}^\mathrm{R}(\mathbf{k},\omega)|$ 
is most enhanced and then the spectral weight near the Fermi level is 
most suppressed. 
Unlike in the 1-loop order theory, the Fermi liquid behaviors 
recover as can be seen in the negative slope of 
$\mathrm{Re}\Sigma_\mathrm{sc}^\mathrm{R}(\mathbf{k},\omega)$ 
at the Fermi level. Moreover, the resonance feature is weaken and the spectral 
peak at Fermi level is somewhat recovered. Then the gap-like behavior 
disappears. 
This is due to the partial summation included in the self-consistent 
calculation. 
However, the reduction of DOS at Fermi level as the feature of the pseudogap 
remaines. 
Yanase~\cite{Yanase3} has shown that the peak structure near the 
Fermi energy in the spectral weight is not reliable but $\Tc$ 
obtained by the self-consistent 1-loop theory is reliable, 
since the value of $\Tc$ is determined by wide structure in the spectrum. 
Therefore, we consider the impurity effect on $\Tc$ 
in the presence of the pseudogap within the self-consistent treatment.

\begin{fullfigure}[t]
  \begin{center}
    \begin{tabular}{ccc}
      \resizebox{50mm}{!}{\includegraphics{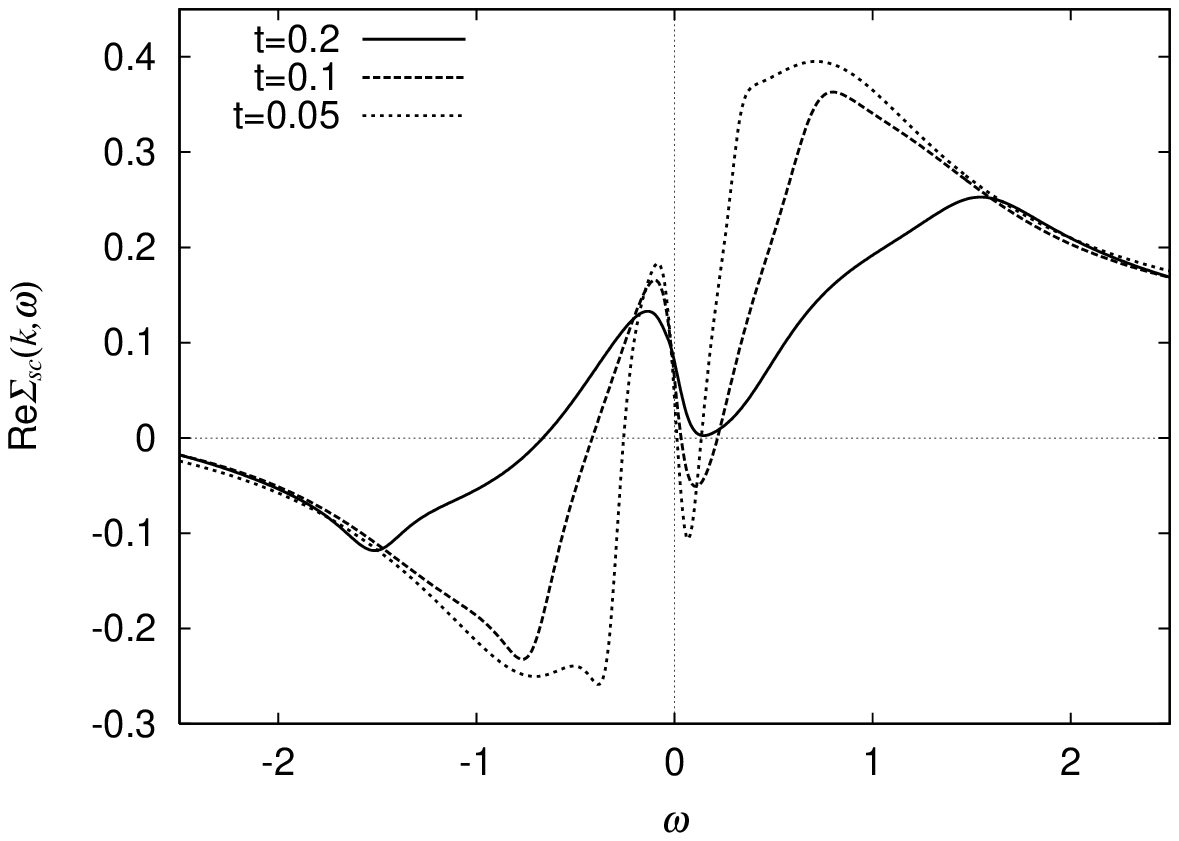}} &
      \resizebox{50mm}{!}{\includegraphics{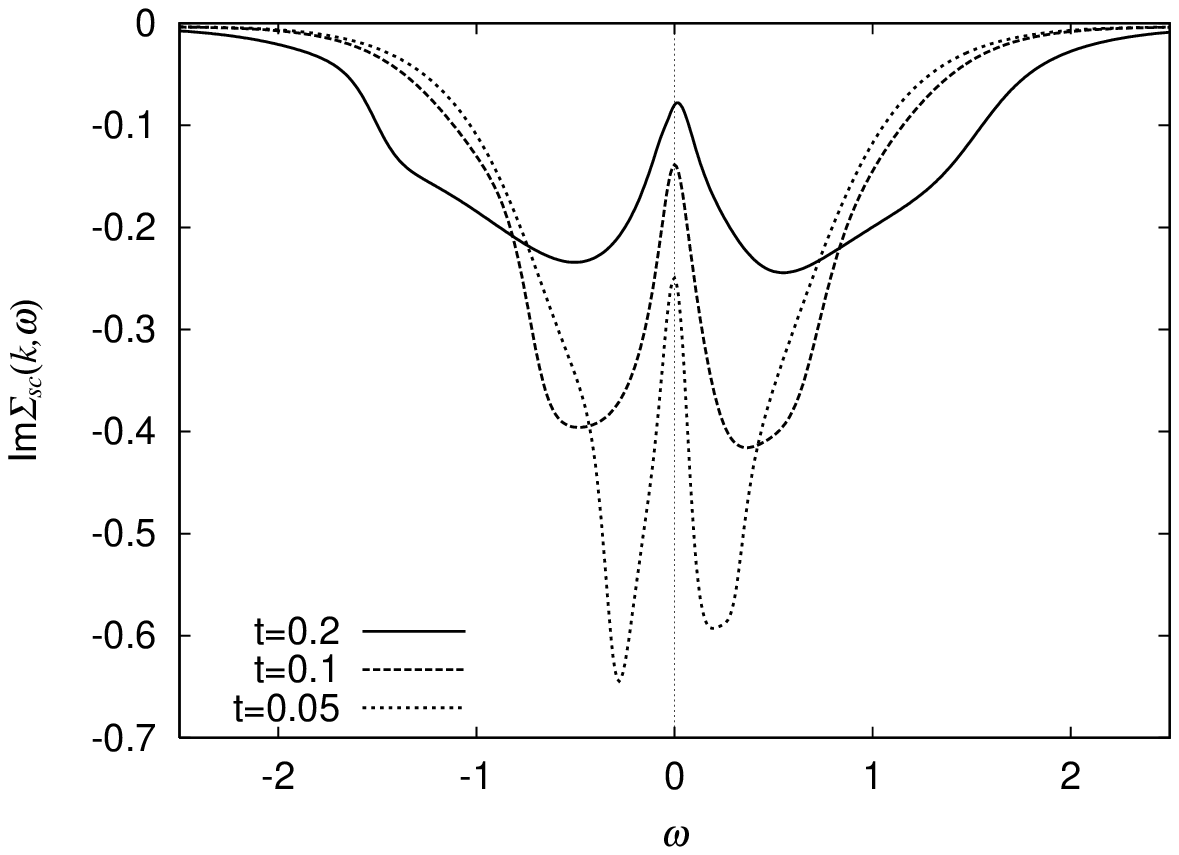}} &
      \resizebox{50mm}{!}{\includegraphics{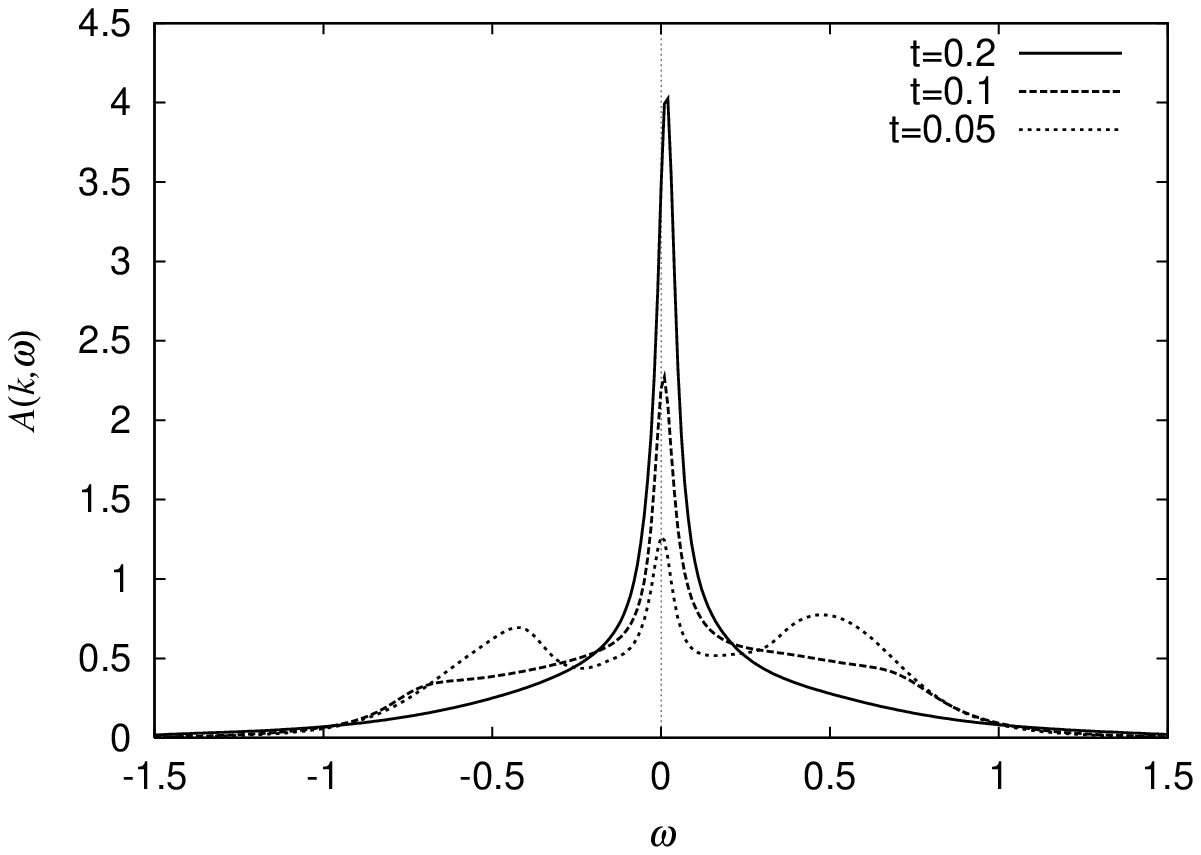}} 
    \end{tabular}
    \caption{The real and imaginary parts of the self-energy and the spectral 
             weight obtained by the self-consistent 1-loop theory. Here, 
             $g=0.4$ and $T=0.02$. The hopping parameter $t$ is chosen as
             $t=0.05$, $0.1$ and $0.2$. }
    \label{ASEful}
  \end{center}
\end{fullfigure}

\subsection{Self-consistent 1-loop theory with impurities}

\begin{figure}[t]
\begin{center}
\includegraphics[width=8.2cm]{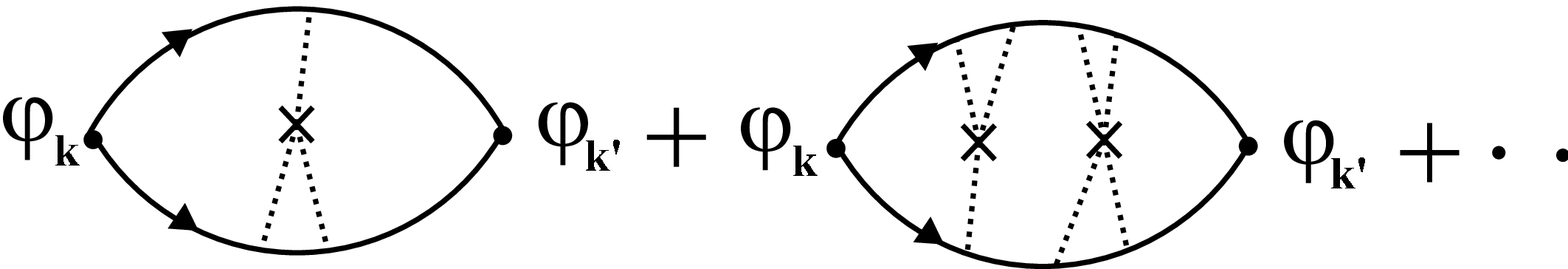}
\end{center}
\caption{Impurity contributions to $\chi_\mathrm{pp}$ which are 
         formally taken into account in our calculation.}
\label{impdiagram}
\end{figure}

Let us include impurity effects in the self-consistent 1-loop theory. 
We incorporate the impurity effect into the single-particle Green 
function via the self-energy and also take into account the diagrams 
shown in Fig. \ref{impdiagram}. However, in our $d$-wave model, the diagrams 
in Fig. \ref{impdiagram} have no contribution due to the momentum summation 
which has the form of $\frac{1}{N}\sum_\mathbf{k}\varphi_\mathbf{k}\cdots$, 
where $\varphi_\mathbf{k}=\cos k_x-\cos k_y$. Therefore, impurities enter 
just only the Green function via Dyson equation:
\begin{equation}
G(\mathbf{k},i\omega_n)^{-1}
=G_\mathrm{0}(\mathbf{k},i\omega_n)^{-1}-\Sigma(\mathbf{k},i\omega_n),
\label{DG}
\end{equation}
with
\begin{equation}
\Sigma(\mathbf{k},i\omega_n)
=\Sigma_\mathrm{sc}(\mathbf{k},i\omega_n)
+\Sigma_\mathrm{imp}(i\omega_n).
\label{SEscimp}
\end{equation}
Here 
\begin{equation}
\Sigma_\mathrm{imp}(i\omega_n)
=-\frac{n_\mathrm{imp}}{\frac{1}{N}\sum_\mathbf{k}G(\mathbf{k},i\omega_n)}.
\label{uniimp}
\end{equation}
We solve Eqs.(\ref{SEsc}), (\ref{Tmat}), (\ref{chipp}), (\ref{DG}), 
(\ref{SEscimp}) and (\ref{uniimp}) self-consistently choosing $\mu$ so as to 
keep the filling of electrons.

\begin{figure}[t]
\begin{center}
\includegraphics[width=6cm]{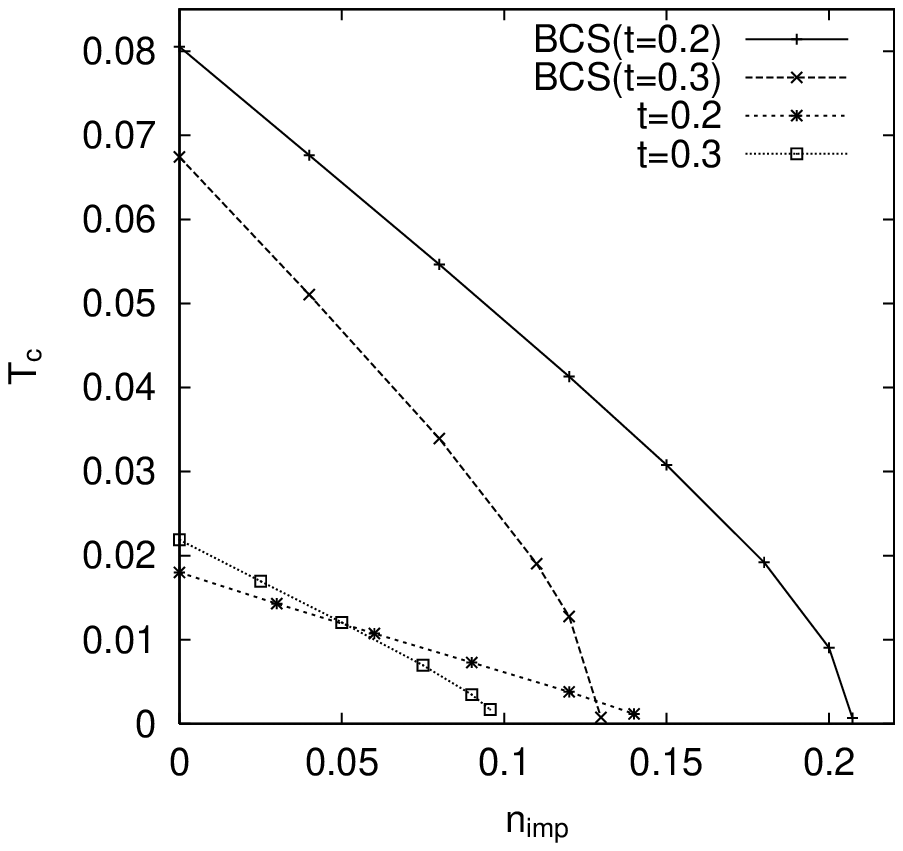}
\end{center}
\caption{$n_\mathrm{imp}$ vs. $\Tc$ in the BCS model and in the 
         self-consistent 1-loop theory for the case of $t=0.2$ and $0.3$. 
         The parameters are chosen as $g=0.4$ and $n=0.9$.}
\label{bcs_self}
\end{figure}

Fig. \ref{bcs_self} shows $n_\mathrm{imp}$ vs. $\Tc$ in the self-consistent 
1-loop theory and, for comparison, that in the BCS model. 
In the self-consistent 1-loop theory, 
the critical impurity concentration 
for $t=0.2$ is larger than that for $t=0.3$ 
in spite of smaller $T_\mathrm{c0}$ for $t=0.2$ than for $t=0.3$. 
This situation is contrary to that in the BCS model 
without SC fluctuation. 
In $\S2$, we have seen that $\Tc$ reduction in the BCS model almost coincides 
with AG curves and that $n_\mathrm{imp}^\mathrm{c}$ is 
given by 
$\frac{\pi^2 \rho(0) T_\mathrm{c0}}{2 \gamma'}$ in AG formula, that is, 
$n_\mathrm{imp}^\mathrm{c}$ depends on $T_\mathrm{c0}$. Therefore, 
we can consider that the magnitude of $n_\mathrm{imp}^\mathrm{c}$
in the self-consistent 1-loop theory is mostly determined by the 
mean-field value of $n_\mathrm{imp}^\mathrm{c}$ which reflects the 
value of $T_\mathrm{c0}^\mathrm{MF}$. 
Strictly speaking, however, there exists the clear difference between 
$n_\mathrm{imp}^\mathrm{c}$ in the mean-field theory and that in 
the self-consistent 1-loop theory, as seen in Fig. \ref{bcs_self}. 
We consider that this difference 
originates from a quantum dynamics and will comment on this issue in $\S4.4$.

\begin{figure}[t]
\begin{center}
\includegraphics[width=6cm]{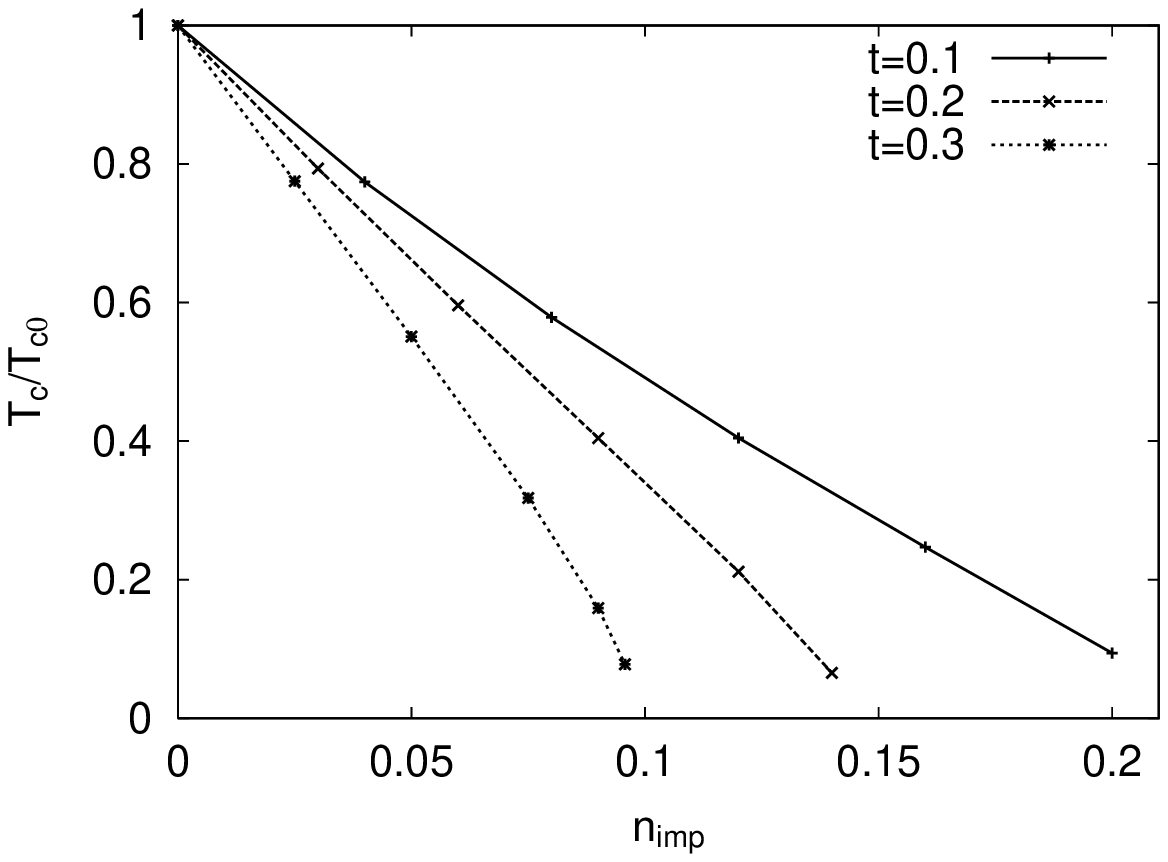}
\end{center}
\caption{The $\Tc$ scaled by $T_\mathrm{c0}$ as the function of 
         the impurity concentration. The parameters are chosen as 
         $g=0.4$, $n=0.9$ and $t=0.1$, $0.2$ and $0.3$.}
\label{dept}
\end{figure}

\begin{figure}[t]
\begin{center}
\includegraphics[width=7cm]{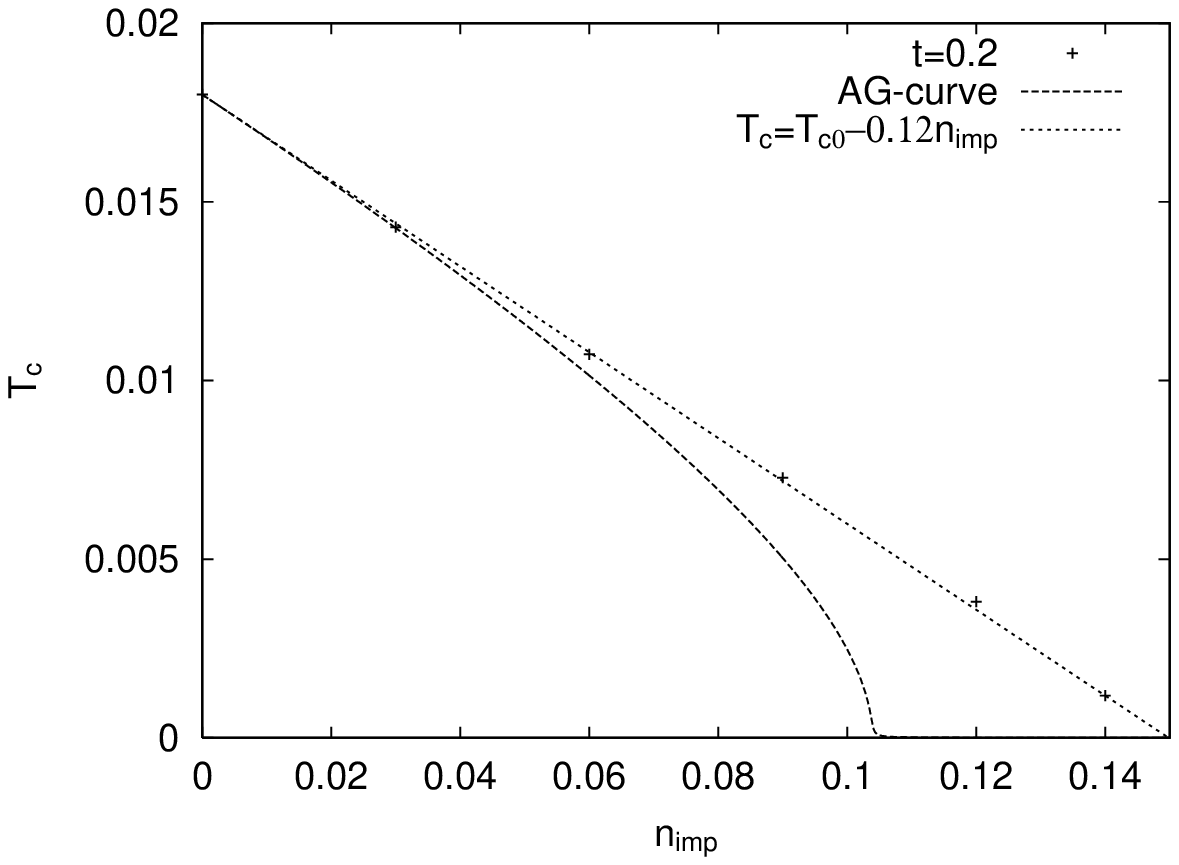}
\end{center}
\caption{The $n_\mathrm{imp}$ vs. $\Tc$ data in the self-consistent 1-loop 
         theory for the parameter as $t=0.2$, $g=0.4$, and
         $n=0.9$, and the fitting curve by Eq.(\ref{ag}). 
         The fitting is done in the same way as in $\S2$.}
\label{agfit_t02uni}
\end{figure}

The most peculiar feature of $\Tc$ reduction 
is its nearly linear variation with impurity concentration. 
This is shown in Fig. \ref{dept} and \ref{agfit_t02uni}. 
From Fig. \ref{dept}, we can see that 
the curvature of $\Tc=\Tc(n_\mathrm{imp})$ 
gradually becomes positive as $t$ is small and 
the reduction of $\Tc$ is almost linear at $t=0.2$. 
This feature originates from the strong SC fluctuation. 
In Fig. \ref{agfit_t02uni}, we compare the data for $t=0.2$ and 
AG curve. AG formula 
can not be the good fitting curve due to the absence of the negative 
curvature in the calculated data for $t=0.2$. 
Rullier-Albenque $et$ $al$. 
have reported this kind of the linear variation.~\cite{Alloul} 
They have measured $\Tc$ and $ab$ plane resistivity of electron irradiated 
$\mathrm{YBa}_\mathrm{2}\mathrm{Cu}_\mathrm{3}\mathrm{O}_\mathrm{7-\delta}$ 
crystals (underdoped and optimally doped) and mensioned that the 
variation of $\Tc$ does not follow any current prediction of pair-beraking 
theories. However, we believe that a pair-breaking theory still holds even 
in their experiment and the 'pseudogap breaking' by impurities 
or defects, which will be introduced below, 
is the reasonable keyword for an explanation of their data.

So, why the linear variation of $\Tc$ seen 
in Fig. \ref{agfit_t02uni} emerges?

Before tackling the issue, we analyze carefully Fig. \ref{SEAs1_A} and 
Fig. \ref{SEAs1_B} which show the spectral weight and the imaginary part of 
$\Sigma$ and $\Sigma_\mathrm{sc}$ at {\bf A} and {\bf B} in the 
$\mathbf{k}$-space as the funtion of $\omega$ for $n_\mathrm{imp}=0.00$, 
$0.07$, $0.14$. 
We proceed with a discussion regarding the point {\bf A} as the representative 
of the $\mathbf{k}$-points around the 'hot spot' and the point {\bf B} as 
the one around the 'cold spot', and then simply call {\bf A} 'hot spot' 
and {\bf B} 'cold spot' in the below discussion.

\begin{figure}[t]
\begin{center}
\includegraphics[width=3.6cm]{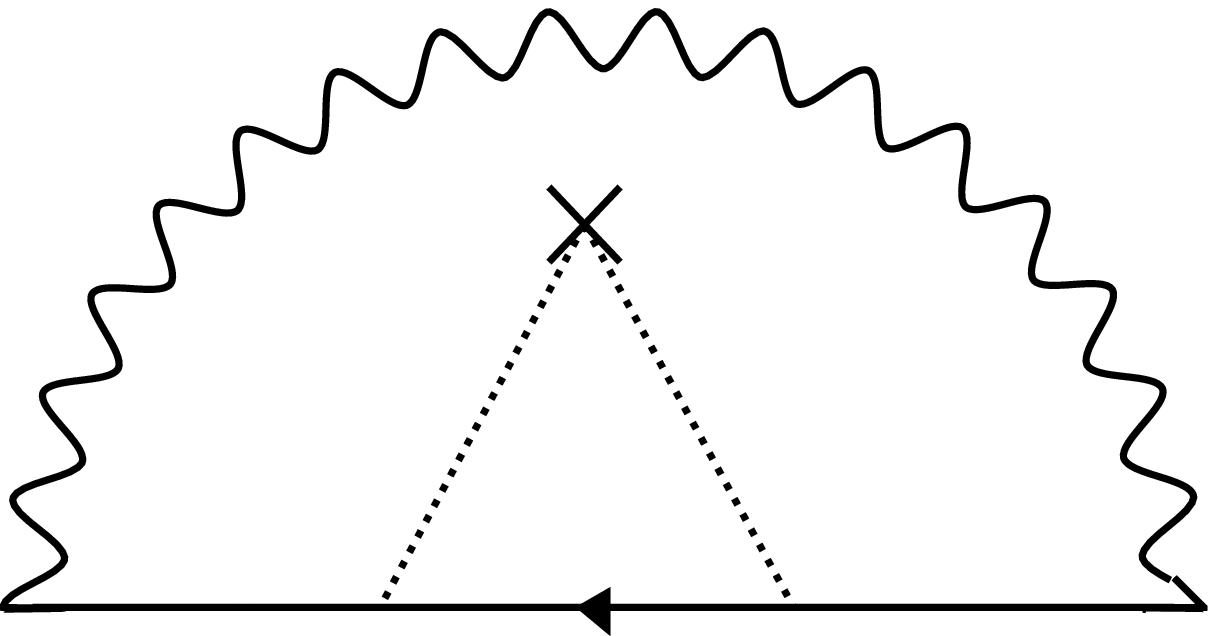}
\end{center}
\caption{The example of the self-energy diagram 
         included in the self-consistent calculation. 
         The wavy line means SC fluctuation interaction 
         and the dotted lines with 
         the cross mark the impurities scattering averaged over on 
         their positions.
         This type of diagram makes weak the contribution of SC fluctuation 
         to the self-energy.
         }
\label{impdia2}
\end{figure}

Comparing Fig. \ref{SEAs1_A}(a) and Fig. \ref{SEAs1_B}(a) for 
$n_\mathrm{imp}=0.00$, we find that the spectral weight at
hot spot has a short peak but the one at 
cold spot has a sharp and high peak. 
This difference owes to the form factor 
$\varphi_\mathbf{k}^2$ which has a large value at hot spot and a small value 
at cold spot. 
Physically, it can be seen that because of SC fluctuation, there is 
almost no coherence as the well-defined quasi-particle 
at hot spot even in the clean case. 
Let us introduce impurities. We find the remarkable feature that whereas 
the spectral peak at cold spot is largely suppressed by impurities, 
the one at hot spot is hardly affected by impurities and keeps the almost 
same form as that in the clean case. 
Owing to the direct connection between the spectral weight and 
the imaginary part of the self-energy, 
this kind of feature also can be seen in 
Fig. \ref{SEAs1_A}(b) and Fig. \ref{SEAs1_B}(b) which show the 
quasi-particle damping, $|\mathrm{Im}\Sigma|$. As impurity concentration 
increases, while the lifetime of the quasi-particle at cold spot becomes 
short, the one at hot spot hardly changes. That is, impurities seem 
to mainly affect the quasi-particles at cold spot and have no contribution 
to the ones at hot spot, 
although the contribution of impurities to the quasi-particle damping 
should not depend on momentum $\mathbf{k}$. 
The reason that such things arise is 
clarified by taking accout of Fig. \ref{SEAs1_A}(c) and Fig. \ref{SEAs1_B}(c) 
which shows $\mathrm{Im}\Sigma_\mathrm{sc}$, the contribution of 
SC fluctuation to the quasi-particle damping. These figures have the 
common feature that $|\mathrm{Im}\Sigma_\mathrm{sc}|$ becomes small 
by impurities. 
It means that as impurities increase, the quasi-particles tend not to be 
affected by the SC fluctuation. 
In terms of our formulation, this is due to the diagram such as 
Fig. \ref{impdia2}, which arises from the process of the 
self-consistent calculation. 
Impurities interfere in the scattering by the SC fluctuation. 
From these understandings, we can now appreciate
the behavior that $|\mathrm{Im}\Sigma|$ at 
hot spot keeps the almost constant value at Fermi level. 
This is because the contribution of $\mathrm{Im}\Sigma_\mathrm{sc}$ 
cancels out the one of $\mathrm{Im}\Sigma_\mathrm{imp}$, and 
the cancellation makes 
$\mathrm{Im}\Sigma
=\mathrm{Im}\Sigma_\mathrm{sc}+\mathrm{Im}\Sigma_\mathrm{imp}$ 
nearly constant. 
At cold spot, due to the smallness of the suppression of 
$|\mathrm{Im}\Sigma_\mathrm{sc}|$, the cancellation is not complete, 
and impurities effectively contribute there. 

\begin{figure}[t]
\begin{center}
\includegraphics[width=7cm]{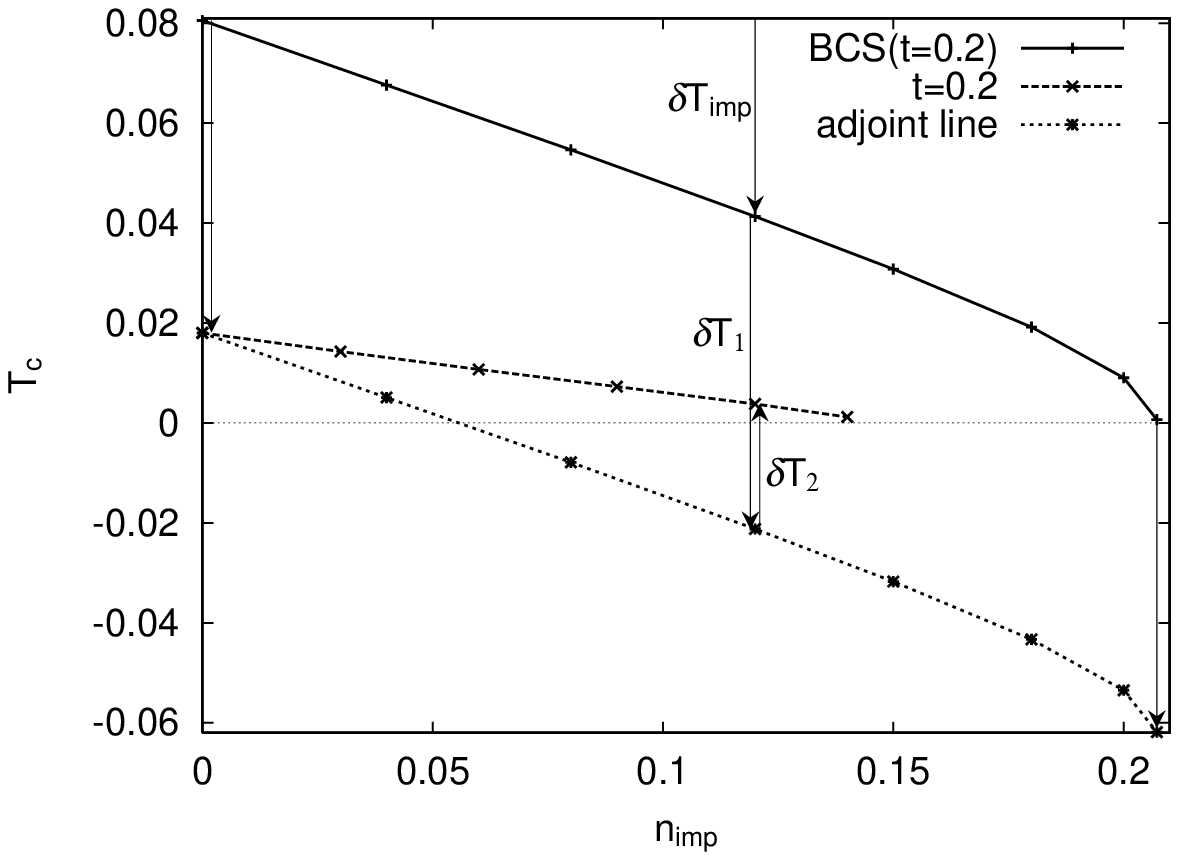}
\end{center}
\caption{The figure for the rough understanding that $\Tc$ is suppressed 
         almost linearly. $\delta T_\mathrm{imp}$ means the magnitude of the 
         $\Tc$ reduction form $T_\mathrm{c0}^\mathrm{MF}$ 
         in the BCS mean-field theory. 
         $\delta T_\mathrm{1}$ means the difference of 
         $T_\mathrm{c0}^\mathrm{MF}$ and $T_\mathrm{c0}$ in the 
         self-consistent 1-loop theory, that is, the magnitude of 
         suppression by SC fluctuation without impurities. 
         The 'adjoint line' is drawn by shifting the 'BCS($t=0.2$)' data 
         down by the amount of $\delta T_\mathrm{1}$. 
         $\delta T_\mathrm{2}$ means the decline in the the magnitude 
         of $\Tc$ reduction by SC fluctuation.
         }
\label{adjoint}
\end{figure}

The idea of the suppression of $|\mathrm{Im}\Sigma_\mathrm{sc}|$ allows 
us to tackle the above issue. 
On the basis of the idea, we have drawn Fig. \ref{adjoint} 
which gives us a rough schematic understanding for 
the linear variation of $\Tc$.
From the figure, the value of $\Tc$ in the self-consistent 1-loop theory 
can be expressed by the following form:
\begin{equation}
\Tc
=T_\mathrm{c}^\mathrm{MF}-\delta T_\mathrm{1}
-(\delta T_\mathrm{imp}-\delta T_\mathrm{2}),
\end{equation}
where $\delta T_\mathrm{imp}$ $etc$. are explained in the caption 
of Fig. \ref{adjoint}. 
As impurity concentration increases, not only $\delta T_\mathrm{imp}$ but also 
$\delta T_\mathrm{2}$ increases via the suppression of 
$|\mathrm{Im}\Sigma_\mathrm{sc}|$. 
Then, if $\delta T_\mathrm{imp}-\delta T_\mathrm{2}$ were in proportion to 
$n_\mathrm{imp}$, $\Tc$ would decrease linearly according as 
$n_\mathrm{imp}$ picks up. 
In this sense, our parameters used to draw Fig. \ref{agfit_t02uni} is 
such ones that keep the delicate balance between 
$\delta T_\mathrm{2}$ and $\delta T_\mathrm{imp}$ which leads to 
the nearly linear variation of $\Tc$ as the function of $n_\mathrm{imp}$. 
This is the answer for the above issue.

\begin{fullfigure}[t]
  \begin{center}
    \begin{tabular}{ccc}
      \resizebox{50mm}{!}{\includegraphics{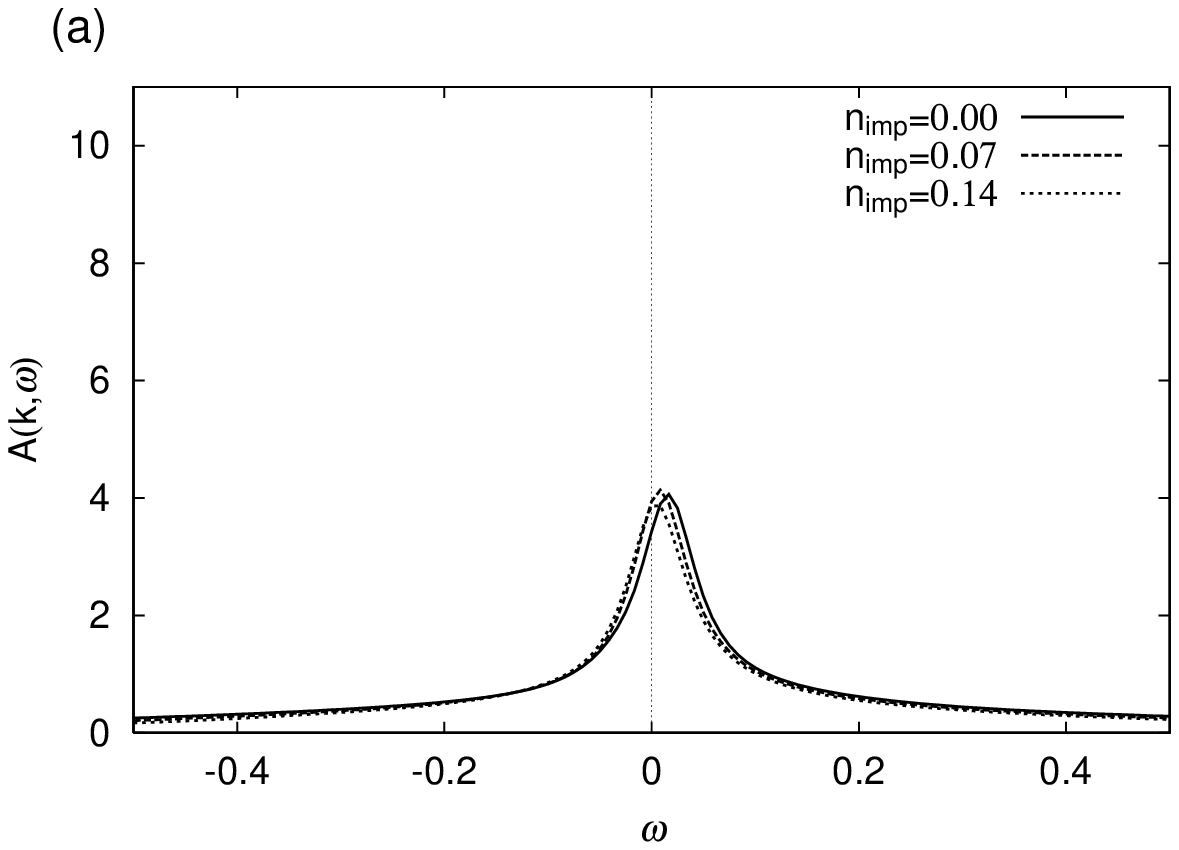}} &
      \resizebox{50mm}{!}{\includegraphics{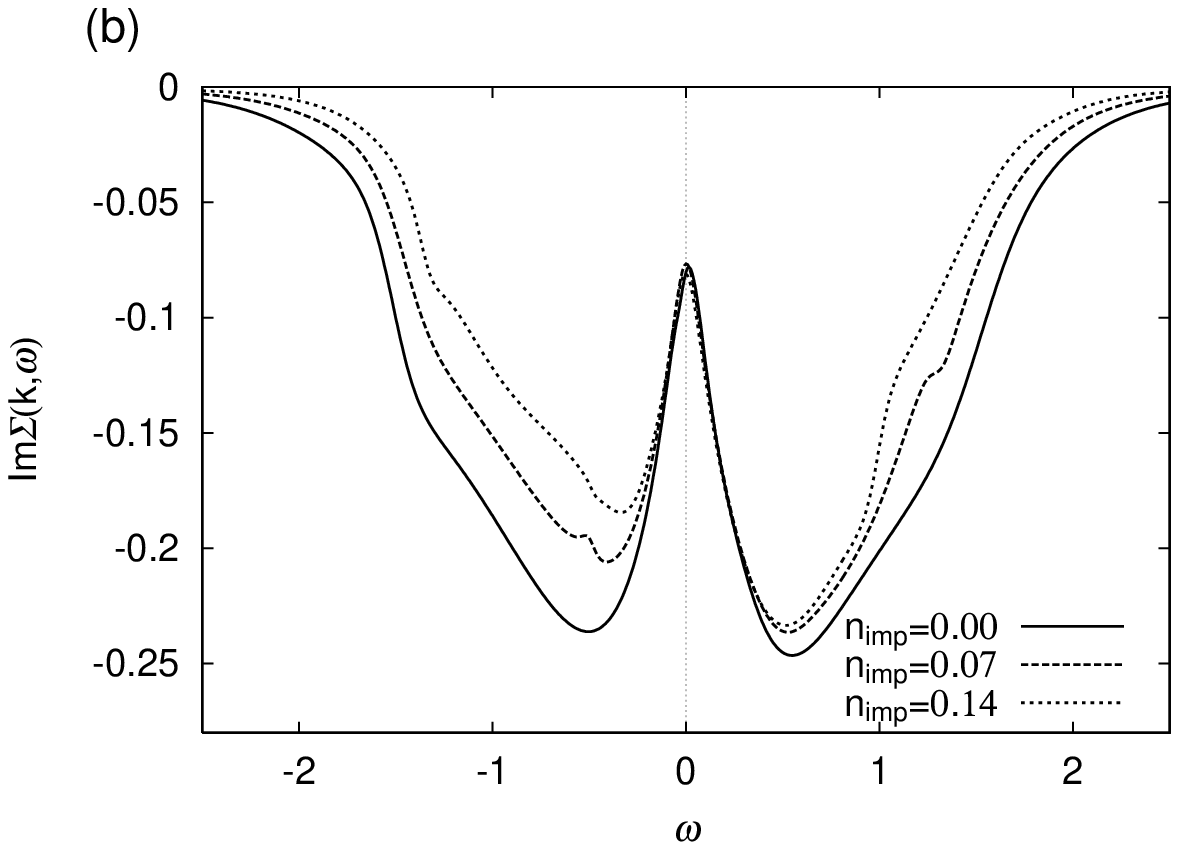}} &
      \resizebox{50mm}{!}{\includegraphics{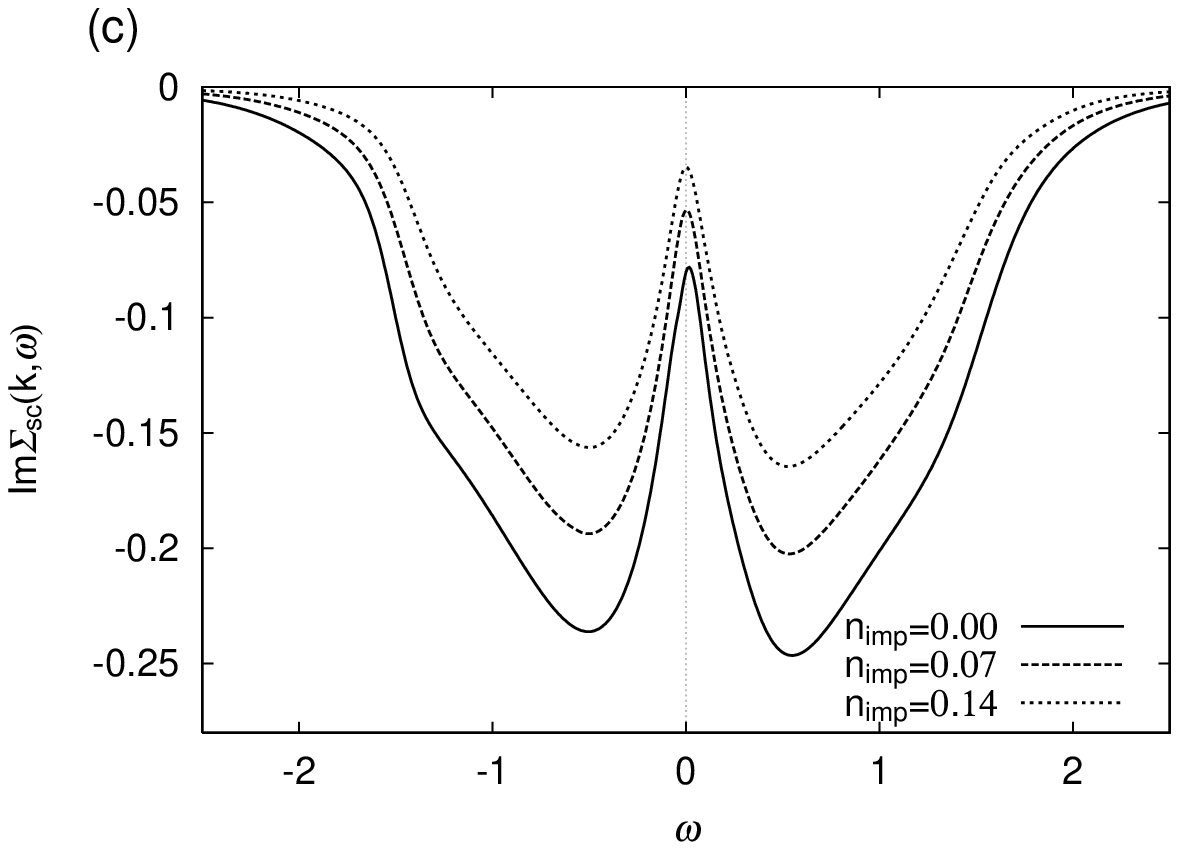}} 
    \end{tabular}
    \caption{The spectral weight and the imaginary part of 
             $\Sigma$ and $\Sigma_\mathrm{sc}$ as the function of 
             real frequency $\omega$ obtained by the self-consistent 
             1-loop theory with impurities. 
             Here, $t=0.2$, $g=0.4$, $T=0.02$, and 
             $\mathbf{k}$ is at {\bf A} in Fig. \ref{AB}.
             The impurity concentrations $n_\mathrm{imp}$ are chosen as 
             $n_\mathrm{imp}=0.00$, $0.07$ and $0.14$. }
    \label{SEAs1_A}
  \end{center}
\end{fullfigure}

\begin{fullfigure}[t]
  \begin{center}
    \begin{tabular}{ccc}
      \resizebox{50mm}{!}{\includegraphics{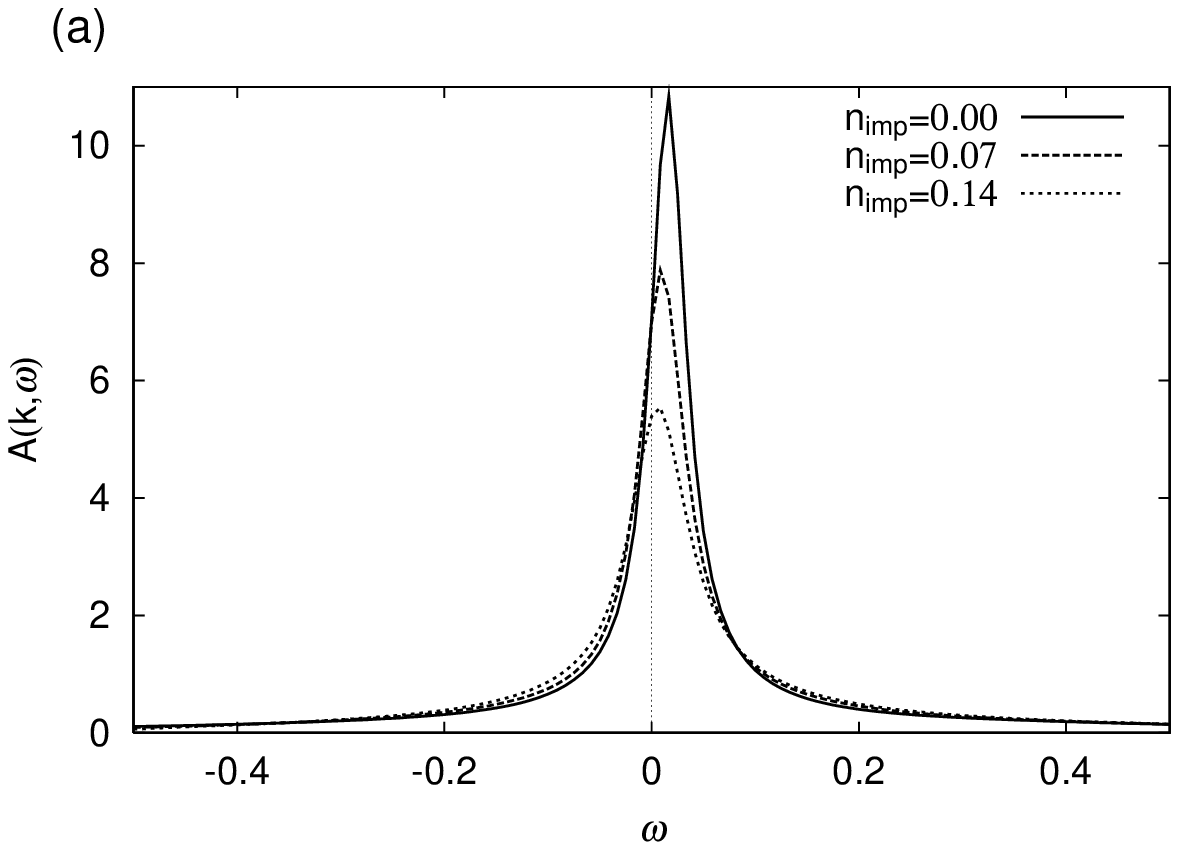}} &
      \resizebox{50mm}{!}{\includegraphics{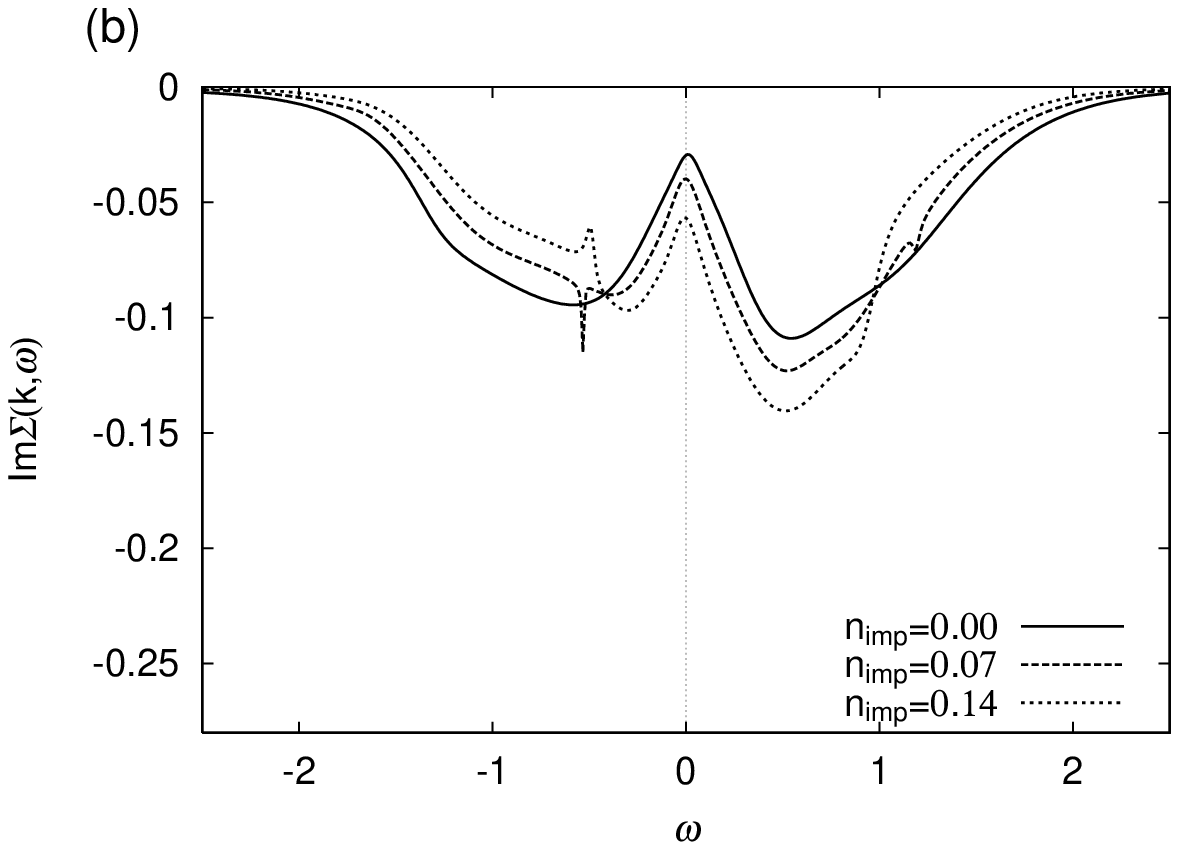}} &
      \resizebox{50mm}{!}{\includegraphics{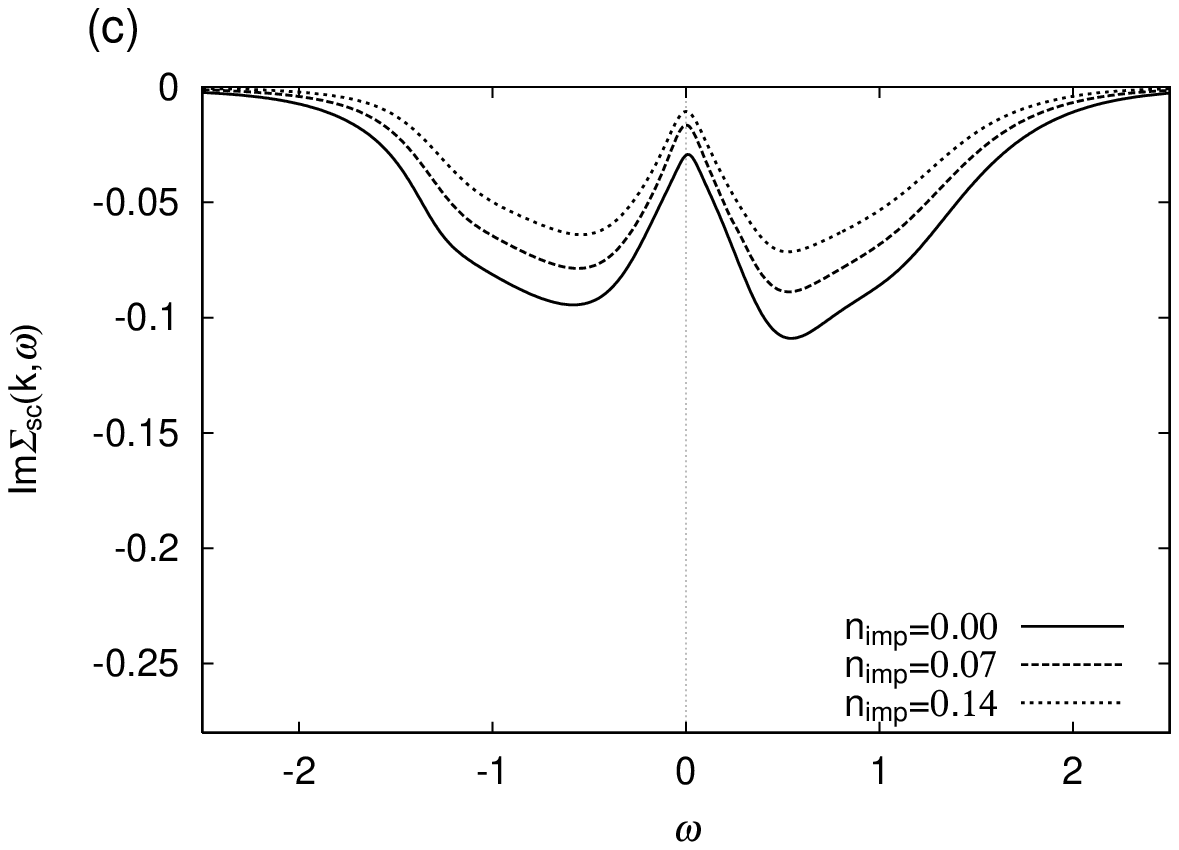}} 
    \end{tabular}
    \caption{The spectral weight and the imaginary part of 
             $\Sigma$ and $\Sigma_\mathrm{sc}$ as the function of 
             real frequency $\omega$ obtained by the self-consistent 
             1-loop theory with impurities. 
             Here, $t=0.2$, $g=0.4$, $T=0.02$, and 
             $\mathbf{k}$ is at {\bf B} in Fig. \ref{AB}.
             The impurity concentrations $n_\mathrm{imp}$ are chosen as 
             $n_\mathrm{imp}=0.00$, $0.07$ and $0.14$.  }
    \label{SEAs1_B}
  \end{center}
\end{fullfigure}

The essential and important point of the discussion up to here was 
the idea that $|\mathrm{Im}\Sigma_\mathrm{sc}|$ is suppressed by impurities. 
We call the idea 'psuedogap breaking'. We may doubt that 
the word is not consistent with the idea because there is no gap-like 
behavior in the spectral weight of Fig. \ref{SEAs1_A} or Fig. \ref{SEAs1_B}. 
However, the pseudogap is indeed destructed by impurities as we will explain 
in $\S4.3$.

\begin{figure}[t]
\begin{center}
\includegraphics[width=7cm]{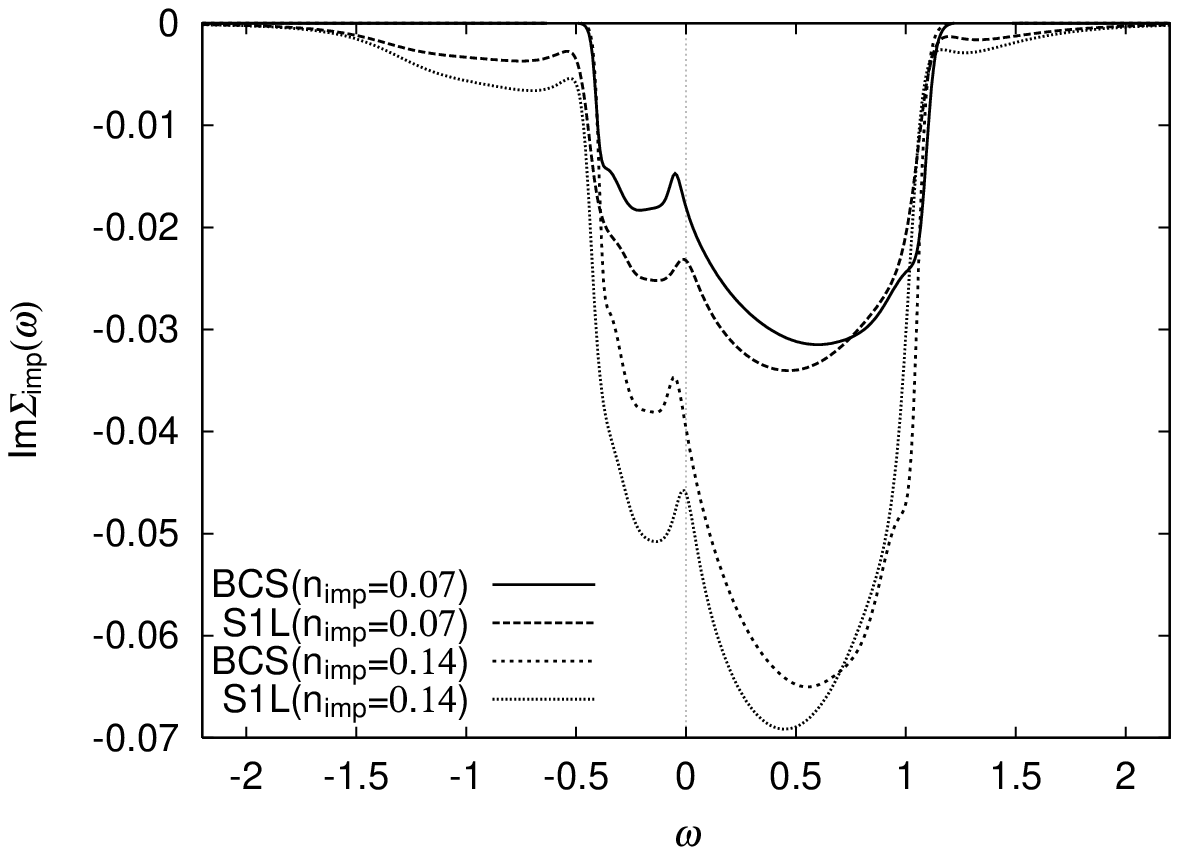}
\end{center}
\caption{The impurities contribution to the imaginary part of self-energy as
         the function of $\omega$ in the self-consistent 1-loop theory(S1L) 
         and in the mean-field theory(BCS). 
         The parameters are chosen as $t=0.2$, $g=0.4$, $n=0.9$, and 
         $T=0.02$.}
\label{comp_bcs}
\end{figure}

Another important result we have to point out is included in 
Fig. \ref{comp_bcs} which shows the comparison of the quasi-particle 
damping by impurities in the self-consistent 1-loop theory and that in 
the mean-field theory. From the figure, 
we can see that the former is larger than the latter. 
As we mentioned in $\S4.1$, SC fluctuation results in the reduced DOS. 
In our formulation, impurities are treated in the unitarity limit, then 
the reduced DOS leads to a large $\mathrm{Im}\Sigma_\mathrm{imp}$ 
as we can understand by Eq.(\ref{uniimp}). 
Therefore, the behavior shown in Fig. \ref{comp_bcs} can be interpleted 
by the reasoning 
that the contribution of impurities to the quasi-particle damping becomes 
effectively large by the effect of the SC fluctuation.

\subsection{Effective 1-loop model}

In this section, we explain the concept of 
the 'pseudogap-breaking' by impurities. 
To the aim, we introduce an effective model for the self-consistent 
1-loop theory with impurities. 
The feature of this model is the maintenance of the gap-like 
behavior as can be seen in the 1-loop order theory 
even though the calculation to determine $\Tc$ is done self-consistently.

\subsubsection{Derivation of the effective 1-loop model}

Since $T(Q)$ has the sharp peak around $Q=(\mathbf{q},i\nu_m)\cong 0$ 
for $T\cong \Tc$, we can expand $T(Q)^{-1}$ around $Q\cong 0$. 
Moreover, we use the static approximation, i.e. we only take into account 
$\nu_m=0$ for $T(\mathbf{q},i\nu_m)$. 
Then 
\begin{equation}
T(\mathbf{q},0)=\frac{-g}{1-g\chi_\mathrm{pp}(\mathbf{q},0)}
\label{Tq0}
\end{equation}
can be written as the following form:
\begin{equation}
T(\mathbf{q},0) \cong \frac{-g}{r+b\mathbf{q}^2},
\label{effT}
\end{equation}
where $b$ is a constant and 
\begin{equation}
r=1-g\chi_\mathrm{pp}(\mathbf{0},0).
\end{equation}
By making use of Eq.(\ref{effT}), we can rewrite Eq.(\ref{SEsc}) as follows:
\begin{equation}
\begin{split}
& \Sigma_\mathrm{sc}(\mathbf{k},i\omega_n) \\
& =\varphi_\mathbf{k}^2 \frac{1}{\beta N} \sum_\mathbf{q}
\frac{-g}{r+b\mathbf{q}^2}
\frac{1}{-i\omega_n-\xi_\mathbf{q-k}-\Sigma_\mathrm{imp}(-i\omega_n)} \\
& =\varphi_\mathbf{k}^2 \frac{g}{\beta r \xi^2} \frac{1}{N}
\sum_\mathbf{q} \frac{1}{\xi^{-2}+\mathbf{q}^2}
\frac{1}{i\omega_n+\xi_\mathbf{q-k}+\Sigma_\mathrm{imp}(-i\omega_n)} \\
& \cong \Delta^2 \varphi_\mathbf{k}^2 
\frac{1}{N} \sum_\mathbf{q}
\frac{2 \xi^{-1}}{\xi^{-2}+q_x^2}\frac{2 \xi^{-1}}{\xi^{-2}+q_y^2} \\
& \hspace{1.5cm} \times \frac{1}{i\omega_n+\xi_\mathbf{k}+v_xq_x+v_yq_y
+\Sigma_\mathrm{imp}(-i\omega_n)}.
\label{SEsc2}
\end{split}
\end{equation}
Here, $\xi^2 \equiv b/r$, and we have replaced $\frac{1}{\xi^2+\mathbf{q}^2}$ 
with $\frac{\xi^{-1}}{\xi^{-2}+q_x^2}\frac{\xi^{-1}}{\xi^{-2}+q_y^2}$. 
Moreover, we have used that $\xi_\mathbf{q-k}$ around 
$\mathbf{q} \cong \mathbf{0}$ can be written as
\begin{equation}
\xi_\mathbf{q-k}=\xi_\mathbf{k}+v_xq_x+v_yq_y,
\end{equation}
where 
\begin{equation}
v_x(\mathbf{k})
=-2\sin k_x (t-2t' \cos k_y),
\end{equation}
and
\begin{equation}
v_y(\mathbf{k})
=-2\sin k_y (t-2t' \cos k_x),
\end{equation}
and defined $\Delta^2$ as follows:
\begin{equation}
\Delta^2
=-\frac{1}{\beta N}\sum_\mathbf{q}T(\mathbf{q},0).
\label{Gapp}
\end{equation}
By using this $\Delta^2$, we can express $\xi$ as
\begin{equation}
\xi=\sqrt{\frac{g}{4 \beta r \Delta^2}}.
\label{length}
\end{equation}
Integrating the last expression of Eq.(\ref{SEsc2}), we finally obtain
\begin{equation}
\begin{split}
& \Sigma_\mathrm{sc}(\mathbf{k},i\omega_n) \\
& =\frac{\Delta^2 \varphi_\mathbf{k}^2}
{i\omega_n+\xi_\mathbf{k}+i(|v_x|+|v_y|)\xi^{-1}\mathrm{sgn}\omega_n
+\Sigma_\mathrm{imp}(-i\omega_n)}.
\label{SEscfinal}
\end{split}
\end{equation}
The renormalized Green function is given by 
\begin{equation}
G(\mathbf{k},i\omega_n)
=\frac{1}{i\omega_n-\xi_\mathbf{k}-\Sigma_\mathrm{sc}(\mathbf{k},i\omega_n)
-\Sigma_\mathrm{imp}(i\omega_n)},
\label{Geff}
\end{equation}
where $\Sigma_\mathrm{imp}(i\omega_n)$ is as same as Eq.(\ref{uniimp}).

We solve Eqs.(\ref{chipp}), (\ref{Tq0}), (\ref{Gapp}), 
(\ref{length}), (\ref{SEscfinal}), (\ref{uniimp}) and (\ref{Geff}) 
self-consistently choosing the chemical potential $\mu$ so as to keep 
$n=0.9$.

\subsubsection{Results of calculation and 'pseudogap breaking'}

\begin{figure}[t]
\begin{center}
\includegraphics[width=7cm]{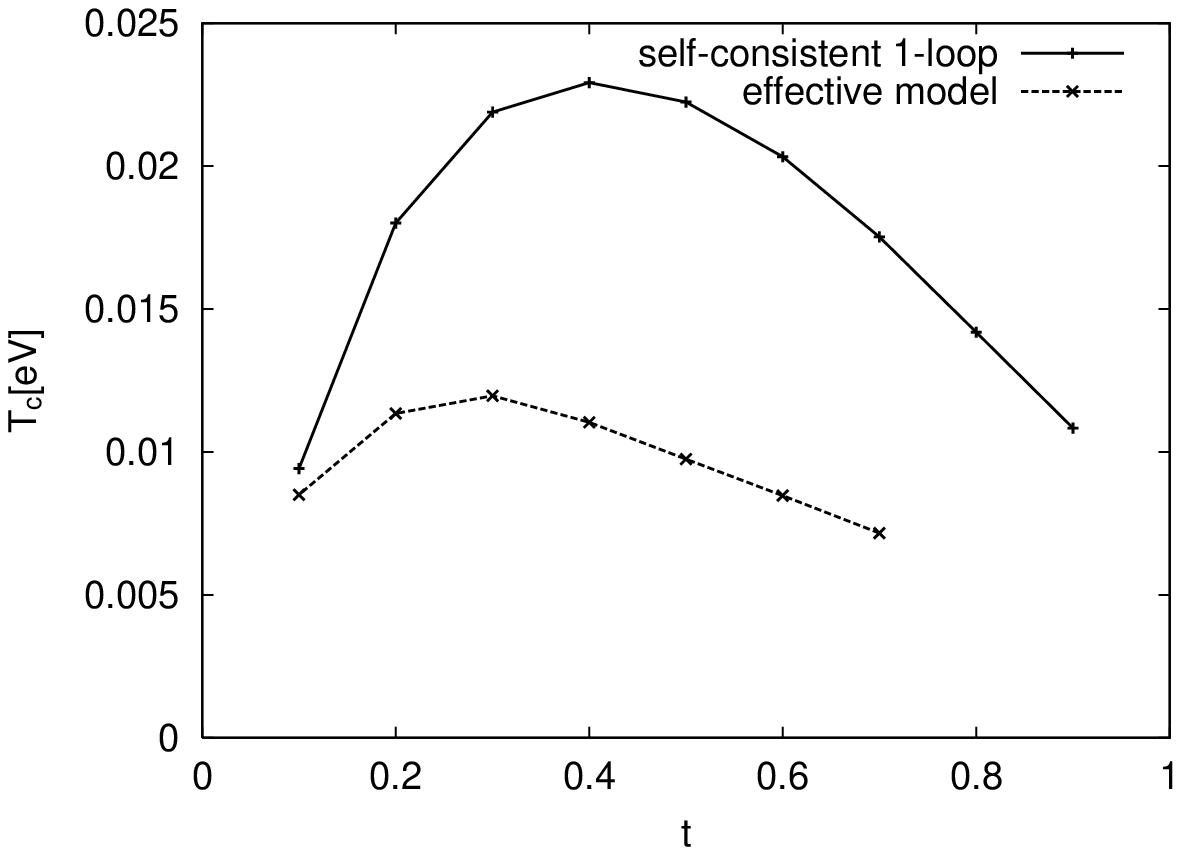}
\end{center}
\caption{$t$ vs. $\Tc$ in the self-consistent 1-loop theory and in the 
         effective 1-loop model. The parameters are chosen as 
         $g=0.4$ and $n=0.9$. The data of 'self-consistent 1-loop' is 
         as same as the data in Fig. \ref{tT1}. }
\label{tTeff}
\end{figure}

Fig. \ref{tTeff} shows $t$-$\Tc$ phase diagram in the 
effective 1-loop model without impurities 
and for comparison that in the self-consistent 1-loop 
theory we have already displayed in Fig. \ref{tT1}. 
In the effective 1-loop theory, the larger suppression of $\Tc$ 
from the BCS value than that in the self-consistent 1-loop theory 
can be seen. That may be because we neglect the components of $\nu_m \neq 0$ 
in $T(\mathbf{q},i\nu_m)$ and then the SC fluctuation as a thermal fluctuation 
is large in the effective model. 

\begin{figure}[t]
\begin{center}
\includegraphics[width=7cm]{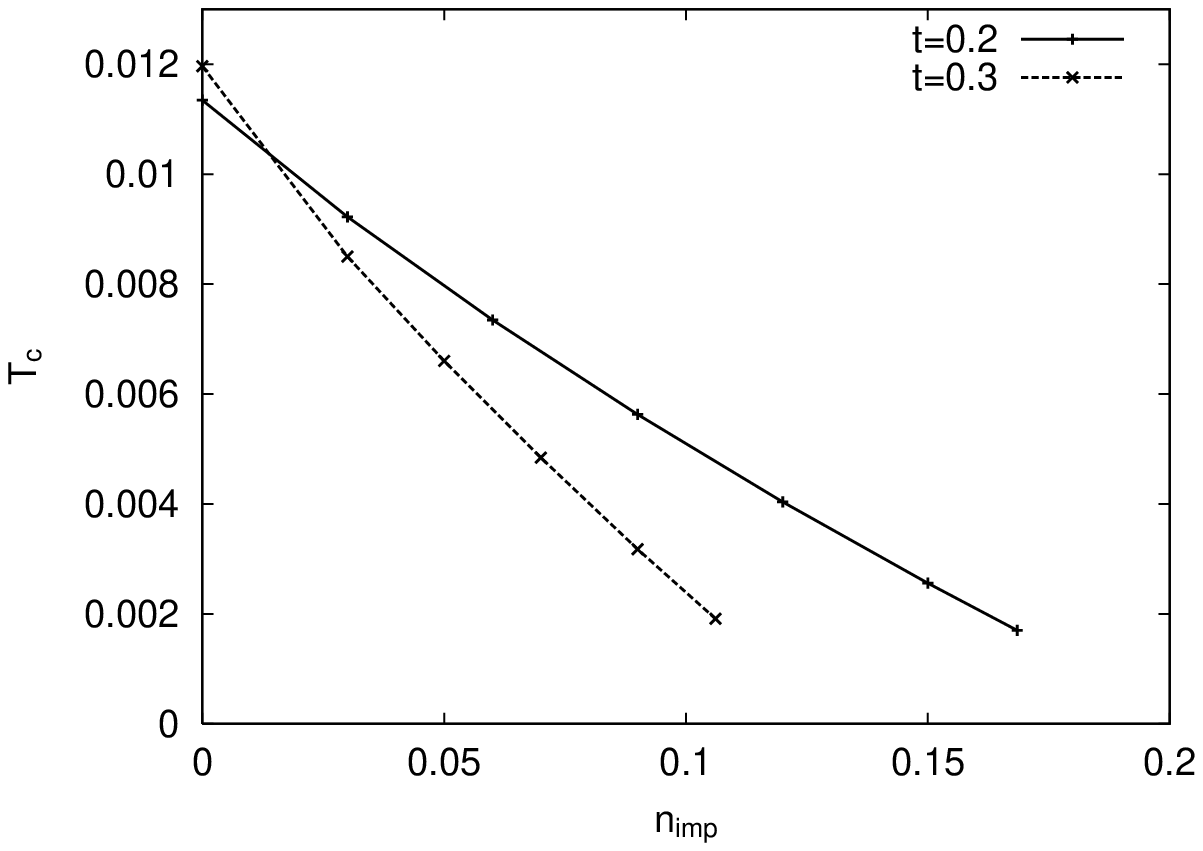}
\end{center}
\caption{$n_\mathrm{imp}$ vs. $\Tc$ in the effective 1-loop model for the 
         case of $t=0.2$ and $0.3$. The parameters are chosen as 
         $g=0.4$ and $n=0.9$.}
\label{impTeff}
\end{figure}

Fig. \ref{impTeff} shows $n_\mathrm{imp}$-$\Tc$ phase diagram in the 
effective 1-loop theory with the parameter is the same as in 
Fig. \ref{bcs_self}. The rough behavior of $\Tc$ is identical to 
that in the self-consistent 1-loop theory. 

The obvious difference between the effective model and 
the self-consistent 1-loop theory is reflected in the spectral 
peak at hot spot which is shown in Fig. \ref{ASEeff}. 
The gap-like behavior in the spectral weight, that is, the 
pseudogap, can be seen in the data for $n_\mathrm{imp}=0.00$. 
It is notable that as $n_\mathrm{imp}$ increases, 
$|\mathrm{Im}\Sigma_\mathrm{sc}|$ becomes small, the pseudogap is suppressed, 
and then the single peak in the spectral weight emerges. 
This is nothing but 'pseudogap breaking'. 
Now, we have understood that the suppression of 
$|\mathrm{Im}\Sigma_\mathrm{sc}|$ 
by impurities is closely related to the pseudogap breaking.

\begin{fullfigure}[t]
  \begin{center}
    \begin{tabular}{ccc}
      \resizebox{50mm}{!}{\includegraphics{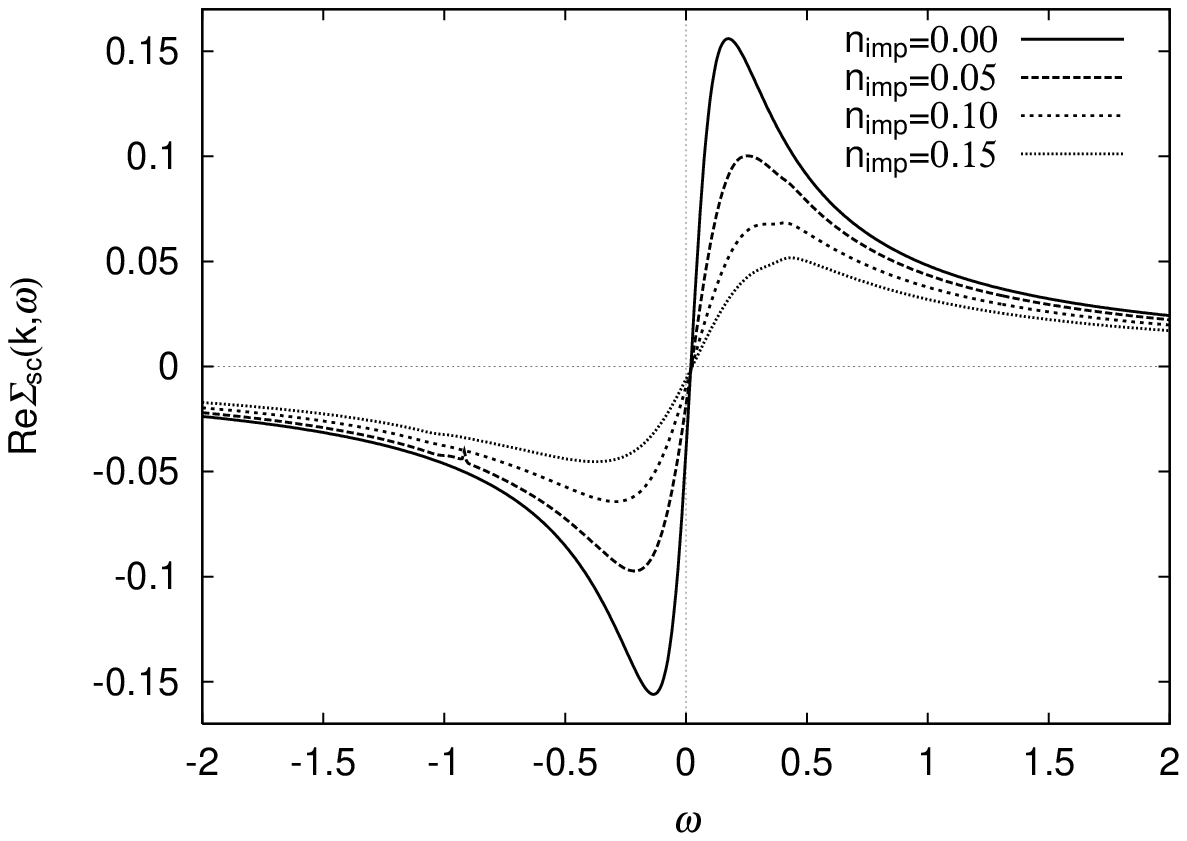}} &
      \resizebox{50mm}{!}{\includegraphics{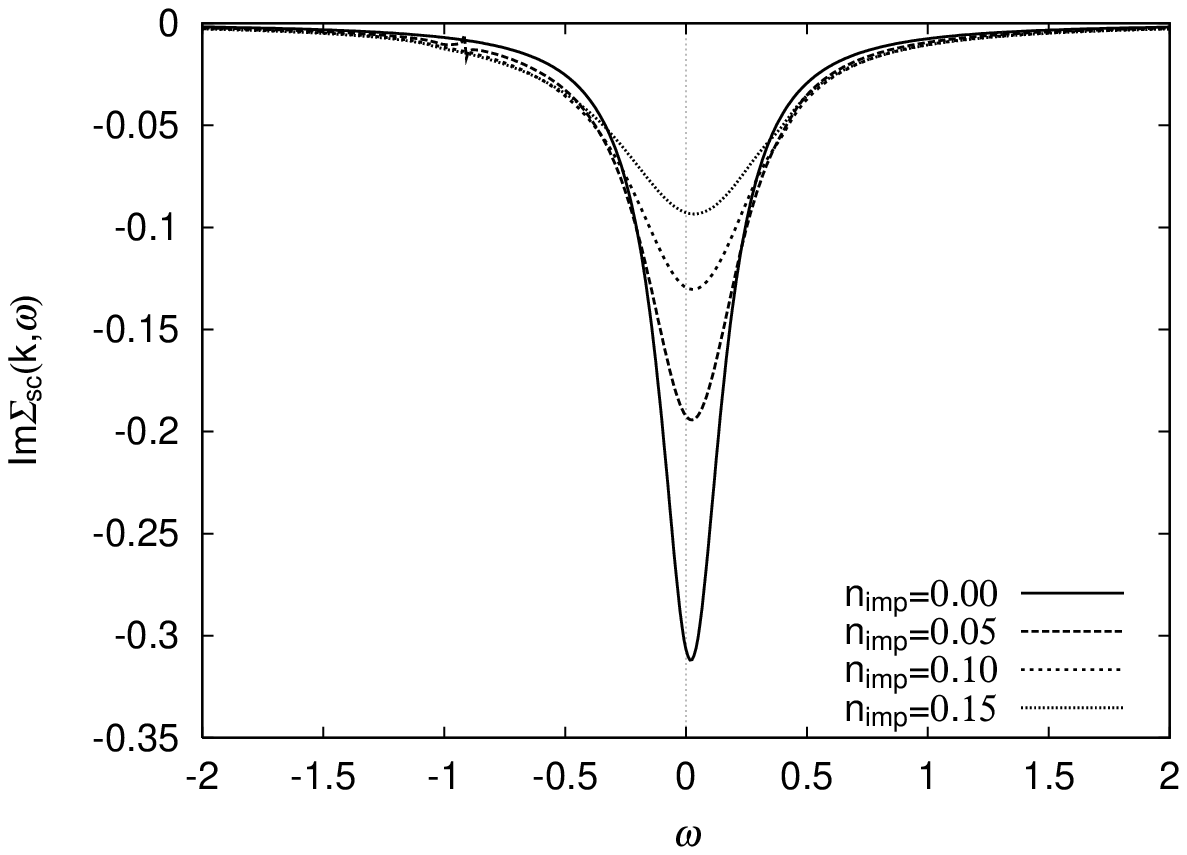}} &
      \resizebox{50mm}{!}{\includegraphics{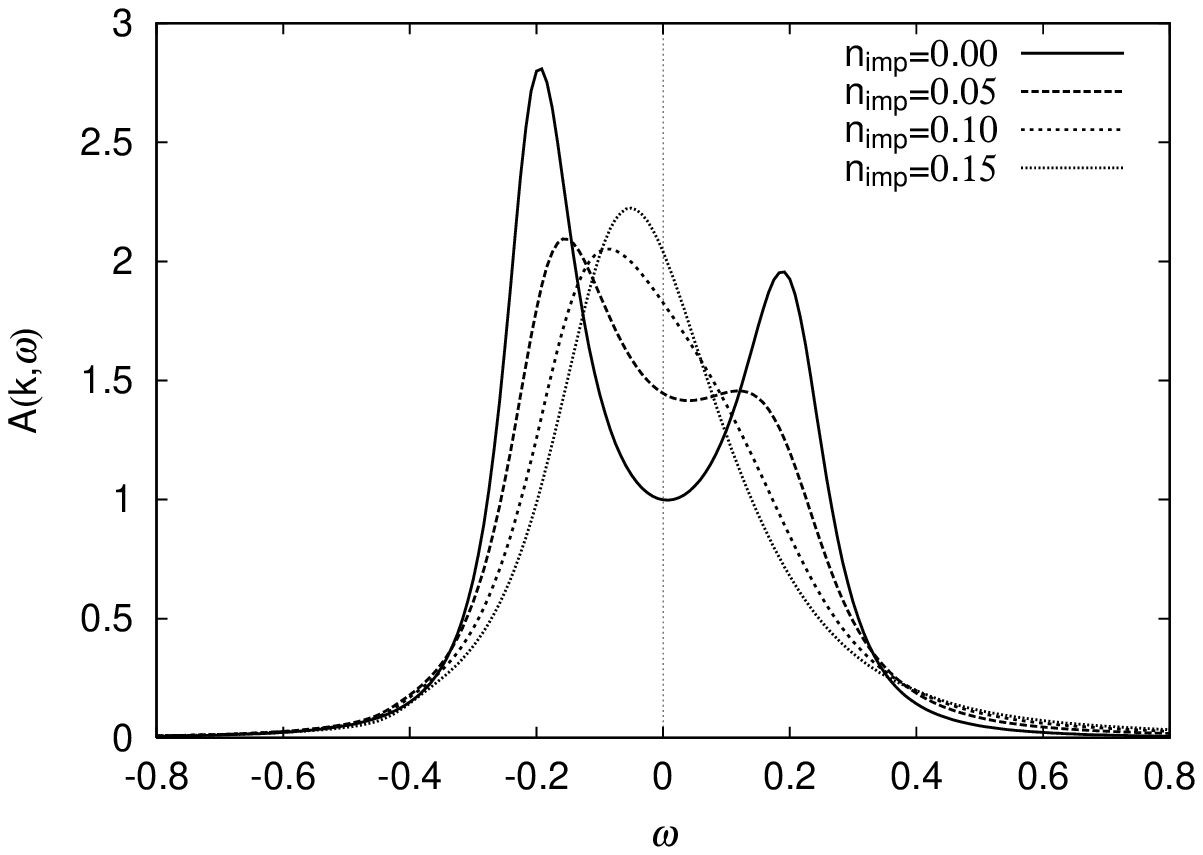}} 
    \end{tabular}
    \caption{The real and imaginary part of self-energy and the spectral 
             weight as the function of $\omega$ obtained by the effective 
             1-loop theory with impurities. Here, $t=0.2$, 
             $g=0.4$, $T=0.014$, and $\mathbf{k}$ is at {\bf A}. 
             The impurity concentrations $n_\mathrm{imp}$ are chosen as 
             $n_\mathrm{imp}=0.00$, $0.05$, $0.10$, and $0.15$.
             }
    \label{ASEeff}
  \end{center}
\end{fullfigure}

As is already mensioned, the suppression of 
$|\mathrm{Im}\Sigma_\mathrm{sc}|$ i.e. the pseudogap breaking, 
results from the self-energy diagram with 
such a form as Fig. \ref{impdia2}. 
Because the causal relationship is very clear in the effective 
1-loop theory, let us consider the relation again. 

For simplicity, we adopt the following form of $\Sigma_\mathrm{imp}$, 
\begin{equation}
\Sigma_\mathrm{imp}(i\omega_n)
=-i \frac{1}{2 \tau} \mathrm{sgn} \omega_n.
\end{equation}
Here, $\tau(\propto n_\mathrm{imp}^{-1})$ is the lifetime 
of the quasi-particle by impurity scttering. 
By using this expression, we can rewrite Eq.(\ref{SEscfinal}) as 
\begin{equation}
\Sigma_\mathrm{sc}(\mathbf{k},i\omega_n)
=\frac{\Delta^2 \varphi_\mathbf{k}^2}
{i\omega_n+\xi_\mathbf{k}+i \Gamma_\mathbf{k} \mathrm{sgn} \omega_n}, 
\label{RedSE}
\end{equation}
where 
\begin{equation}
\Gamma_\mathbf{k}=
\frac{|v_x|+|v_y|}{\xi}+\frac{1}{2 \tau}.
\end{equation}
It is notable that the expression of Eq.(\ref{RedSE}) is nothing but 
the diagram of Fig. \ref{impdia2} itself within the approximation 
by which we have used to derive the effective 1-loop model. 
By analitical continuation, $i\omega_n \to \omega+ i \delta$, 
in Eq.(\ref{RedSE}), we obtain 
\begin{equation}
\mathrm{Im}\Sigma_\mathrm{sc}(\mathbf{k},\omega)
=-\frac{\Delta^2 \varphi_\mathbf{k}^2 \Gamma_\mathbf{k}}
{(\omega+\xi_\mathbf{k})^2+\Gamma_\mathbf{k}^2}.
\end{equation}
From this equation, we find that 
$|\mathrm{Im}\Sigma_\mathrm{sc}(\mathbf{k},\omega)|$ has a peak at 
$\omega=0$ for the $\mathbf{k}$ on the Fermi surface and its top value is 
$\frac{\Delta^2 \varphi_\mathbf{k}^2}{\Gamma_\mathbf{k}}$. 
Therefore, as $n_\mathrm{imp}$ increases, $\Gamma_\mathbf{k}$ becomes 
large, and then the peak of 
$|\mathrm{Im}\Sigma_\mathrm{sc}(\mathbf{k},\omega)|$ lowers. 
This tendency coincides with the behavior of 
$\mathrm{Im}\Sigma_\mathrm{sc}$ in Fig. \ref{ASEeff}. 
As the result, we recognize that the scattering process such as 
Fig. \ref{impdia2} is essential for pseudogap breaking. 
That is, the broadening of the one-particle state by impurities 
makes weak the scattering due to SC fluctuation and 
it leads to the pseudogap breaking.

\subsection{Comments on the quantum dynamics of SC fluctuation}

\begin{figure}[t]
\begin{center}
\includegraphics[width=7cm]{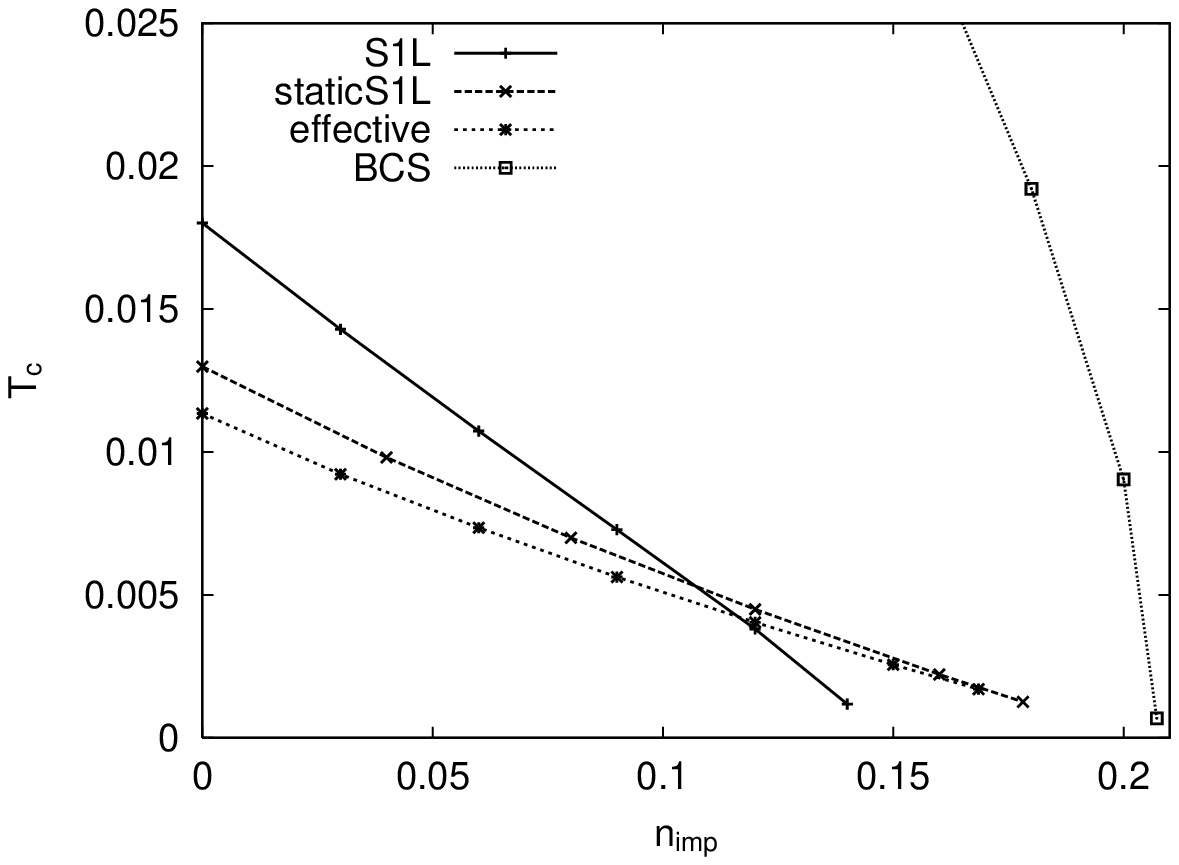}
\end{center}
\caption{$n_\mathrm{imp}$ vs. $\Tc$ in the self-consistent 1-loop theory(S1L), 
         S1L within the static approximation(staticS1L), 
         the effective 1-loop model(effective) 
         and the BCS mean-field theory(BCS). 
         The parameters are chosen as $t=0.2$, $g=0.4$ and $n=0.9$ in all 
         theories. 
         }
\label{SCcomp}
\end{figure}

In $\S4.2$, we have mensioned that the difference between 
$n_\mathrm{imp}^\mathrm{c}$ in the BCS theory and that in the self-consistent 
1-loop theory. 
If the SC fluctuation were perfectly thermal, 
$n_\mathrm{imp}^\mathrm{c}$'s in the both theories would coincide 
with each other, 
because the thermal fluctuation should vanish at $T=0$. 
In the self-consistent 1-loop theory, however, the quantum dynamics of 
SC fluctuation is included. Actually, in Eq.(\ref{SEsc}), all $\nu_m$'s 
are included in the summation not just only $\nu_m=0$. 
Therefore, we expect that the above difference originates from 
the quantum dynamics. 
The expectation is supported by Fig. \ref{SCcomp} which shows 
$n_\mathrm{imp}$ vs. $\Tc$ in some theories. 
The 'staticS1L' in the figure means that we have applied the static 
approximation to Eq.(\ref{SEsc}), that is, taken just only $\nu_m=0$ 
into account in the self-consistent 1-loop theory. 
In Fig. \ref{SCcomp}, it can be seen that 
$n_\mathrm{imp}^\mathrm{c}$ in 
the three theories except the self-consistent 1-loop theory 
almost coincide with each other and $n_\mathrm{imp}^\mathrm{c} \sim 0.21$. 
The reason would be that in 'staticS1L' and 'effective', only the thermal 
fluctuation is taken into account. 
On the other hand, $n_\mathrm{imp}^\mathrm{c}$ 
in the self-consistent 1-loop theory is smaller than 
that in the other theories and $n_\mathrm{imp}^\mathrm{c} \sim 0.15$. 
Then, we consider that the quantum dynamics of SC fluctuation 
plays a role to reduce $n_\mathrm{imp}^\mathrm{c}$. 
Farther understanding about the quantum fluctuation is 
one of our future problems.

\section{Conclusion}

The reduction of $\Tc$ due to impurities in the FLEX theory 
is almost same as that in the weak-coupling $d$-wave BCS model. 
This is because the cancellation between two effects by impurities; 
one is the pair-weakening, and another is the reduction of the pair-breaking 
produced by electron correlation. 
Therefore, even in the HTS, where the pairing interaction is caused by 
$U$, AG formula still can be used for rough estimation of 
$\Tc$ reduction due to impurities. However, the following points 
should be noticed. 
By the numerical calculation, we have found 
the deviation of the critical impurity concentration from the 
$n_\mathrm{imp}^\mathrm{c}$ expected by AG formula. It depends on 
the fine structure of the dispersion and the Fermi surface via DOS. 
The larger is DOS, the larger the deviation of $n_\mathrm{imp}^\mathrm{c}$ is. 
In order to analyze the hole-doping dependence of 
$\Tc$ reduction by impurites in detail, we should 
notice the form of Fermi surface not just only adopting AG formula.

As for the dependence of impurity effect on $\mathbf{k}$-point, 
impurities mainly affect the quasi-particles at cold spot. 
This is because impurities reduce the inelastic scattering 
by spin fluctuation which is effective to the quasi-particles at hot spot. 
This is the good example to indicate that we ought to take 
into account the reduction of the damping rate of quasi-particle arising from 
electron interaction.

In $\S4$, we have studied how pseudogap phenomena affect 
the reduction of $\Tc$ due to impurities. 
The variation of $\Tc$ as the function of $n_\mathrm{imp}$ is 
too linear to fit by AG formula 
in the region of rather large SC fluctuation. 
The linear variation is caused by 
the suppression of the quasi-particle damping arising from 
SC fluctuation i.e. the pseudogap breaking.

\section*{Acknowledgment}

Numerical computation in this study was carried out at the Yukawa Institute 
Computer Facility.

\end{document}